\documentclass[12pt]{article}

\pdfoutput=1

\usepackage{amsmath,amssymb,amsfonts,amscd,mathrsfs}
\usepackage{xcolor}
\definecolor{darkblue}{rgb}{0.1,0.1,.7}
\usepackage[colorlinks, linkcolor=darkblue, citecolor=darkblue, urlcolor=darkblue, linktocpage]{hyperref} 
\usepackage{geometry}
\usepackage{tablefootnote,colortbl}

\usepackage{braket}

\usepackage{cancel}
\usepackage{arcs}

\usepackage{arydshln}
\usepackage{booktabs,multirow}
\usepackage{amssymb}
\usepackage{fancyvrb}

\usepackage{tikz-feynman}
 \tikzfeynmanset{compat=1.0.0}
 
 \usepackage[square, comma, compress,numbers]{natbib}
\usepackage{subcaption}

\definecolor{myorange}{RGB}{199,146,32}

\definecolor{Gray1}{gray}{0.97}
\definecolor{Gray2}{gray}{0.9}
\definecolor{LightCyan}{rgb}{0.88,1,1}
\definecolor{blu}{rgb}{0,0,1}

\usepackage{ragged2e}
\usepackage{array}
\newcolumntype{L}[1]{>{\raggedright\let\newline\\\arraybackslash\hspace{0pt}}m{#1}}
\newcolumntype{C}[1]{>{\centering\let\newline\\\arraybackslash\hspace{0pt}}m{#1}}
\newcolumntype{R}[1]{>{\raggedleft\let\newline\\\arraybackslash\hspace{0pt}}m{#1}}

\usepackage{dsfont} 
\usepackage{tcolorbox}
\usepackage{booktabs}

\usepackage{titlesec}
\titleformat*{\section}{\large\bfseries}
\titleformat*{\subsection}{\normalsize\bfseries}
\titleformat*{\subsubsection}{\normalsize\it}
\titleformat*{\paragraph}{\normalsize\bfseries}
\titleformat*{\subparagraph}{\normalsize\bfseries}

\newcommand{\reef}[1]{(\ref{#1})}

\def\eps{\epsilon}

\newcommand{\beq}{\begin{equation}} 
\newcommand{\eeq}{\end{equation}}

\def\geq{\geqslant}
\def\leq{\leqslant}
\newcommand{\diffop}[2]{\ifthenelse{\equal{#2}{1}}{\frac{\mrm{d}}{\mrm{d} #1}}{\frac{\mrm{d}^#2}{\mrm{d} #1^#2}}}

\newcommand{\hs}{\bar{s}}

\newcommand{\mrm}[1]{{\mathrm #1}}

\newcommand{\im}{\text{Im}\, }
\newcommand{\re}{\text{Re}\, }

\def\l{\ell} 

\usepackage[normalem]{ulem}

\newcommand{\be}{\begin{equation}}
\newcommand{\ee}{\end{equation}}
\def\bea#1\eea{\begin{align}#1\end{align}}

  \def\th{\theta}

\usepackage{accents}
\newlength{\dhatheight}

\newcommand{\diagM}{
  \begin{minipage}[h]{0.12\linewidth}\begin{tikzpicture}
  [
  roundnode/.style={circle, draw=black!60, fill=black!30, very thick, 
  inner sep=2.1pt,
  text width=3mm},
roundnode3/.style={circle, draw=black!60, fill=black!6, very thick, 
  inner sep=2.5pt,
  text width=3mm},
]
\begin{feynman}[small]
\vertex (X) at (0,0);
\vertex (x1l) at (-1.,.4);
\vertex (xnl) at (-1.,-.4);
\vertex (x1r) at (+1,.4);
\vertex (x2r) at (+1,-.4);
   \vertex (t7) at (-1.2,-.45){$\scriptsize{^{p_1}}$};
   \vertex (t7) at (-1.2,+.35){$\scriptsize{^{p_2}}$};
   \vertex (t7) at (+1.25,-.45){$\scriptsize{^{p_4}}$};
   \vertex (t7) at (+1.25,+.35){$\scriptsize{^{p_3}}$};
   \diagram*{
   (x1l) -- [thick, quarter left, looseness=.8] (X) --[ thick, quarter left, looseness=.8] (x1r) ,
         (xnl) -- [thick, quarter right, looseness=.8] (X) --[ thick, quarter right, looseness=.8] (x2r) ,
 };
\node[roundnode3] (X1) at (0,0){\tiny{$\vspace{-.05cm}\hspace{-.0cm}M$}};
  \end{feynman}
\end{tikzpicture}
  \end{minipage} 
  }

\newcommand{\diagUVtwo}{
  \begin{minipage}[h]{0.12\linewidth}\begin{tikzpicture}
  [
roundnode/.style={circle, draw=black!60, fill=black!6, very thick, 
  inner sep=2.1pt,
  text width=3mm},
roundnode3/.style={circle, draw=black!60, fill=black!6, very thick, 
  inner sep=2.5pt,
  text width=3mm},
]
\begin{feynman}[small]
\node [dot] (X) at (-.4,0);
\vertex (x1l) at (-1.,.4);
\vertex (xnl) at (-1.,-.4);
\node [dot] (Y) at (0+1.1-.5,0);
\vertex (X1l) at (1.8-.5,.4);
\vertex (Xnl) at (1.8-.5,-.4);
   \diagram*{
   (x1l) -- [scalar, thick, quarter left, looseness=.5] (X) --[ scalar, thick, quarter left, looseness=.5] (xnl) ,
      (Y) -- [very thick] (X) ,
     (X1l)   -- [scalar, thick, quarter right, looseness=.5] (Y)  -- [scalar, thick, quarter right, looseness=.5] (Xnl)  ,
};
 \end{feynman}
\end{tikzpicture}
  \end{minipage} 
  }

\newcommand{\diagIRone}{
  \begin{minipage}[h]{0.12\linewidth}\begin{tikzpicture}
  [
roundnode/.style={circle, draw=black!60, fill=black!6, very thick, 
  inner sep=2.1pt,
  text width=3mm},
roundnode3/.style={circle, draw=black!60, fill=black!6, very thick, 
  inner sep=2.5pt,
  text width=3mm},
]
\begin{feynman}[small]
\node [dot] (X) at (-.4,0);
\vertex (x1l) at (-1.,.4);
\vertex (xnl) at (-1.,-.4);
\node [dot] (Y) at (0+1.1-.5,0);
\vertex (X1l) at (1.8-.5,.4);
\vertex (Xnl) at (1.8-.5,-.4);
   \diagram*{
   (x1l) -- [scalar, thick, quarter left, looseness=.5] (X) --[ scalar, thick, quarter left, looseness=.5] (xnl) ,
      (Y) -- [scalar, thick, half left, looseness=1.2] (X) -- [scalar, thick,  half left, looseness=1.2] (Y) ,
     (X1l)   -- [scalar, thick, quarter right, looseness=.5] (Y)  -- [scalar, thick, quarter right, looseness=.5] (Xnl)  ,
};
 \end{feynman}
\end{tikzpicture}
  \end{minipage} 
  }

\numberwithin{equation}{section}
\usepackage{tocloft}
\begin{document}

\vspace*{-.6in} \thispagestyle{empty}
\begin{flushright}
\end{flushright}
\vspace{1cm} {\Large
\begin{center}
\textbf{Bridging Positivity and  S-matrix Bootstrap Bounds
 }
\end{center}}
\vspace{1cm}
\begin{center}

{\bf  Joan Elias Mir\'o$^{a}$, Andrea Guerrieri$^{b,c,d}$, Mehmet As{\i}m G\"{u}m\"{u}\c{s}$^{e,f}$} \\[1cm] 
 {$^a$ The Abdus Salam ICTP,    Strada Costiera 11, 34135, Trieste, Italy  \\ 
 $^b$   School of Physics and Astronomy, Tel Aviv University, Ramat Aviv 69978, Israel  \\ 
 $^c$ Dipartimento di Fisica e Astronomia, Universita degli Studi di Padova, \& Istituto Nazionale di Fisica Nucleare, Sezione di Padova, via Marzolo 8, 35131 Padova, Italy. \\
 $^d$ Perimeter Institute for Theoretical Physics, Waterloo, Ontario N2L 2Y5, Canada \\ 
 $^e$SISSA, Via Bonomea 265, I-34136 Trieste, Italy \\
$^f$INFN, Sezione di Trieste, Via Valerio 2, 34127 Trieste, Italy
}
\vspace{1cm}

\abstract{ 

\bigskip
\noindent

The main objective of this work is to isolate Effective Field Theory scattering amplitudes in the space of non-perturbative two-to-two amplitudes, using the S-matrix Bootstrap. We do so by introducing the notion of Effective Field Theory cutoff in the S-matrix Bootstrap approach. 
We introduce a number of novel numerical techniques and improvements both for the primal and the linearized dual approach. 
We perform a detailed comparison of the full unitarity bounds with those obtained using positivity and linearized unitarity.
Moreover, we discuss the notion of Spin-Zero and UV dominance along the boundary of the allowed amplitude space by introducing suitable observables.
Finally, we show that this construction also leads to novel bounds on operators of dimension less than or equal to six.

}

\vspace{3cm}
\end{center}

 \vfill
 {
  \flushright
 \today 
}

\newpage 

\setcounter{tocdepth}{1}

{
\tableofcontents
}
 
 

\section{Introduction}

Quantum Effective Field Theory  is very much universal and has a wide range of application and flexibility.
 Nevertheless, the principles of unitary evolution and causality imply constraints  on the space of feasible Effective Field Theories (EFTs), that is on EFTs with a consistent UV completion. In other words, not \emph{ anything goes} and the coupling strengths of the interactions are subject to inequality constraints.

 A widely known  example is  the positivity bound: while a priori Wilson coefficients can take any real value,  the two-to-two forward scattering amplitude satisfies the positivity constraint $\text{Im}M>0$, implying that  certain Wilson coefficients are positive~\cite{Adams:2006sv} -- see also  studies in the context of the chiral Lagrangian~\cite{Pham:1985cr,Pennington:1994kc,Ananthanarayan:1994hf}.
 Several works have since then  exploited positivity,  leading to constraints  on renormalization group flows and the phenomenology of EFTs~\cite{Manohar:2008tc,Low:2009di,Komargodski:2011vj,Luty:2012ww,Bellazzini:2016xrt,Cheung:2016yqr,Distler:2006if,Englert:2019zmt,Bellazzini:2017fep,Alberte:2020bdz,Bellazzini:2019bzh,Gu:2020ldn,deRham:2018qqo}. See also~\cite{Arkani-Hamed:2020blm,Green:2019tpt,Bellazzini:2020cot,Bellazzini:2021oaj,Bellazzini:2021shn,Tolley:2020gtv,Caron-Huot:2020cmc,Komatsu:2020sag,Caron-Huot:2021rmr,Bern:2021ppb,Alberte:2021dnj,Chiang:2021ziz,Henriksson:2021ymi,Sinha:2020win,Knop:2022viy,Creminelli:2022onn, Haring:2022cyf,Li:2022rag} for interesting recent developments.

 Another  realisation of these principles 
  is the recent   version of the   S-matrix Bootstrap~\cite{Paulos:2016fap,Paulos:2016but,Paulos:2017fhb,Homrich:2019cbt}. In this approach  unitarity is not linearised  and it is treated non-perturbatively. 
These ideas have been used in  a number of theoretically and phenomenologically interesting  theories  such as two-dimensional flux-tube effective field theories \cite{EliasMiro:2019kyf, EliasMiro:2021nul}, four dimensional (pseudo)-Goldstone bosons \cite{Guerrieri:2018uew, Guerrieri:2020bto}, Majorana fermion scattering~\cite{Hebbar:2020ukp}, and higher dimensional supergravity \cite{Guerrieri:2021ivu,WhereIsMTheory}.

In this work we  obtain new  S-matrix Bootstrap bounds on the space of two-to-two scattering amplitudes in $d=4$ spacetime. 
More concretely, we consider two examples:  the  two-to-two scattering amplitude of a massive scalar singlet particle,
and a massive scalar particle with internal global $O(n)$ symmetry. 
These amplitudes  can be characterised by their Taylor expansion around the crossing symmetric point  $(s,t,u)=4/3(1,1,1)m^2$ in the centre of the Mandelstam triangle.
For the singlet theory, the first few terms of this expansion are given by
\be
M(\bar s, \bar t, \bar u) =
-c_0 + c_2 (\bar{s}^2 + \bar{t}^2 + \bar{u}^2) + c_3  (\bar{s}\bar{t}\bar{u}) + O(\bar s^4,\bar t^4, \bar u^4) \, ;
\label{lowamp4dnoflav}
\ee
while the s-channel amplitude of the $O(n)$ scalar theory is 
\be
M   (\bar s | \bar t, \bar u)  =  - c_0 + c_H  \bar{s} + O(\bar s^2,\bar t^2, \bar u^2) \, ,
\label{ammmmp}
\ee
where $(\bar s, \bar t, \bar u)=(s,t,u)/m^2-4/3(1,1,1)$.
We will first show that unitarity, crossing, and analyticity of the amplitude imply non-perturbative bounds on the  $c_i$'s.
We will also characterise these amplitudes  with extremal values of the $c_i$ coefficients, and study a number of observables such as UV/IR dominance or Low/High spin dominance.

For weakly coupled EFTs we may interpret  the $c_i$'s as Wilson coefficients of operators. 
For instance consider the free $O(n)$ scalar theory  perturbed by the dimension-six  operator $  \Delta{\cal L}=  g_H  \partial^\mu( \vec\phi \cdot   \vec\phi)  \partial_\mu(\vec \phi \cdot \vec \phi)/(4\Lambda^2) $.
 Then,  $c_H  =  2  \, g_H  \, m^2/\Lambda^2+\dots$ at tree-level. 
This is a priori very suggestive, because it is generally hard to 
set bounds on dimension-six operators using positivity methods, or linearised unitarity on the imaginary part.  This stems from the Froissart-Martin bound and the fact that scattering amplitudes satisfy double subtracted dispersion relations. The dispersive representation of the dimension six operators involve a real subtraction constant that cannot be bounded unless we access the real part of the amplitude too.
Nevertheless dimension-six operators  are of  physical importance because they parametrise at leading order  generic deviations from  the Standard Model  predictions (barring the Weinberg operator for neutrino masses). 
Therefore it is quite interesting that using the  S-matrix Bootstrap one is able to bound these dimension-six operators --  as well as to characterise the amplitudes achieving such extremal values. 

However the extremal values of the $c_i$'s are often achieved by strongly coupled amplitudes and therefore the weakly coupled EFT interpretation is not accurate. Namely   $|c_H-  2  \, g_H  \, m^2/\Lambda^2|  >O(1)$  and as a consequence the bound on $c_H$ does not  translate simply into a bound on $g_H$. 
After finding the  bounds on the  $c_i$'s, one of our main objectives  is precisely to amend this problem. That is we will 
 show that     \textbf{A)} the   S-matrix Bootstrap can output min/max values of  Wilson coefficients for theories that are described by a weakly coupled field theory  for energies  below a  physical cutoff $\Lambda$,
and \textbf{B)} this construction provides  min/max values of dimension-six operators as well. 
 

In section~\ref{sec:full_space} we set the stage by determining precisely the space of amplitudes with maximal $c_i$ values in the singlet case. 
In order to get this result we introduce a number of numerical improvements that allow us to achieve a faster convergence of the Bootstrap algorithm. We also compare in great detail the S-matrix Bootstrap bounds with a rigorous positivity approach.
In section~\ref{singeft} we carve out the space of amplitudes with extremal $c_i$'s that are weakly coupled in the IR.
In section~\ref{ddbounds} we derive new dual bounds using linearised unitarity, and compare with the results in section~\ref{singeft}.
In section~\ref{dimsix} we begin the exploration of the extremal values for the $O(n)$ theory. In section~\ref{medio} we address the role of the EFT cutoff in the bounds on the $c_i$'s.
Finally we conclude in section~\ref{conc}.


\section{The space of QFTs in $3+1$-dimensions}
\label{sec:full_space}

We first study the space of QFTs in $3+1$ dimensions that contain in the IR a single stable scalar particle of mass $m$, even under field parity.~\footnote{This $\mathds{Z}_2$-symmetry implies poles in the amplitude below threshold $0<s<4m^2$ are forbidden because of the absence of the trilinear coupling. Our analysis could be easily generalised by relaxing this assumption.}
We analyse a particular  \emph{slice}  of this space by determining the possible values of the $2\rightarrow 2$ on-shell scattering amplitude with momenta $p_1+p_2\rightarrow p_3+p_4$.
\be
\diagM\,   \nonumber
\ee
As we shall see below this observable is very rich, containing a wealth of information about the spectrum and properties  of the theory.

It is possible to describe 
the $2\rightarrow 2$ scattering  amplitude   as a function of the three Mandelstam invariants $M(s,t,u)$.~\footnote{Recall that the Mandelstam invariants are $s=(p_1+p_2)^2$, $t=(p_1-p_3)^2$ and $u=(p_1-p_4)^2$, where $p_i$ are Lorentz four-momenta.} Due to \emph{crossing symmetry}, $M$ is invariant under any permutation of its arguments $s\leftrightarrow t \leftrightarrow u$.
Momentum conservation $s+t+u=4m^2$ further reduces the number of independent variables to two: e.g.
$M(s,t)\equiv M(s,t,4m^2-s-t)$.

The amplitude $M(s,t)$ is further constrained by the two particle sector of the unitarity condition $S^\dagger S \preceq 1$ where, as usual, the $2\to 2$ S-matrix is given by $S= \mathds{1}+i(2\pi)^4\delta^{(4)}(p_1+p_2-p_3-p_4)M(s,t)$.
It is useful to diagonalise unitarity by projecting onto partial waves 
$
f_{\l}(s) \equiv 1/(32 \pi) \int_{-1}^{1} dx \, P_{\l}(x) \, M(s, t(s,x), u(s,x)) \, , 
$
 for $\ell\in \mathds{N}$, and $x=1+2t/(s-4m^2)$. The unitarity condition then takes the simpler form $2\im f_\ell(s) \geq \sqrt{(s-4m^2)/s}|f_\ell(s)|^2$ for $s>4m^2$. \footnote{This is also consequence of \emph{real analyticity} $M(s^*,t)=M(s,t)^*$. This is always true for the scattering processes we are considering.}
 The inequality is saturated in the \emph{elastic region} $4m^2\leq s< 16m^2$, due to the absence of multi-particle processes.

\begin{figure}[t] \centering
 \includegraphics[scale=.35]{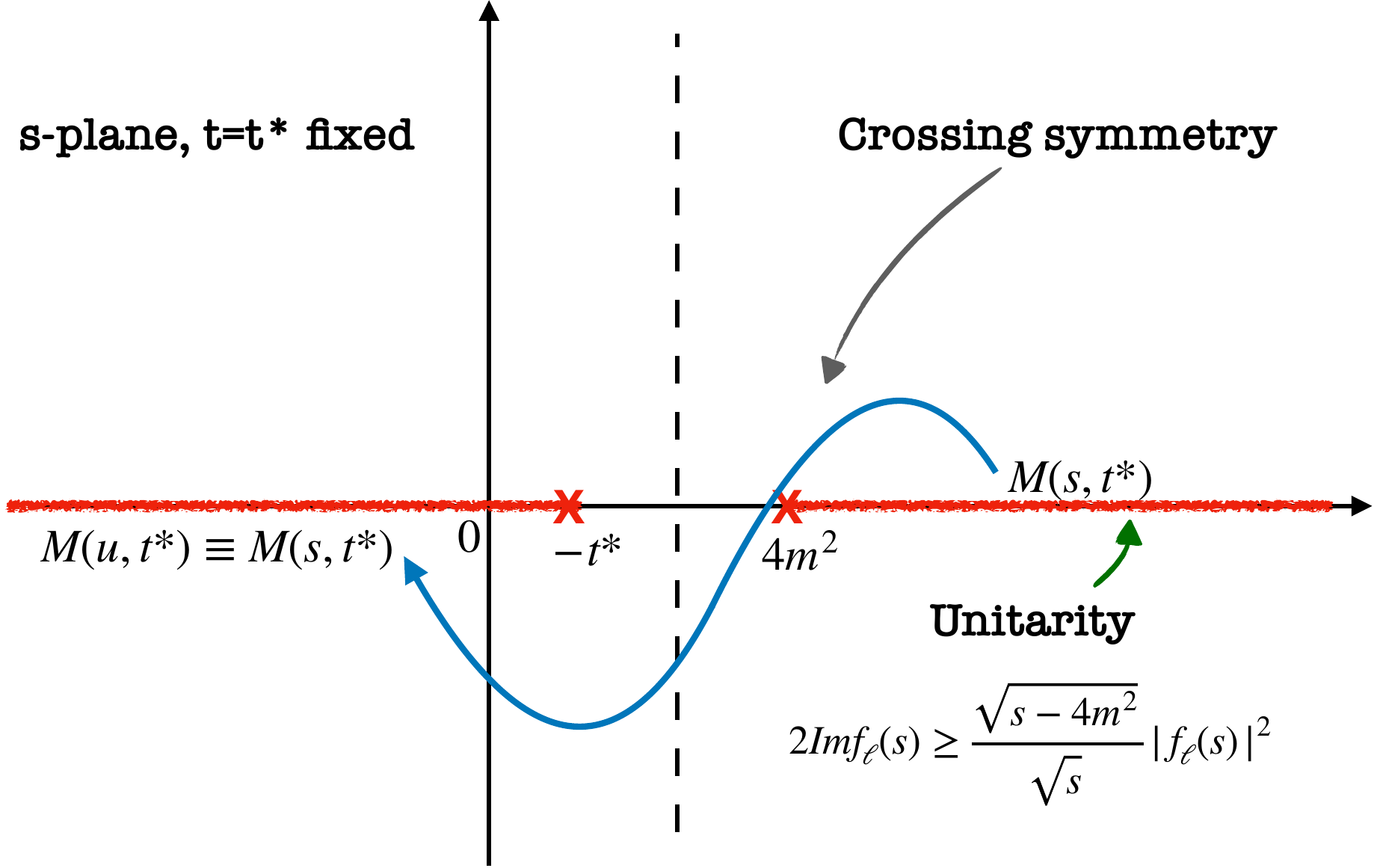} 
 \caption{Analytic properties of the amplitude $M(s,t)$ in the complex $s$-plane for a fixed value of $-4m^2<t^*<4m^2$. We denote in blue the crossing path continuing $M(s,t^*)$ into $M(u,t^*)$. The black dashed vertical line passes through the $s-u$ crossing symmetric point $(4m^2-t^*)/2$. Due to real analyticity, the amplitude is real in between the two cuts and along the dashed line.
The right-hand cut is subject directly to the unitarity constraints. }
  \label{4d_complex_plane} \end{figure}

We also assume \emph{maximal analyticity} (or  Mandelstam analyticity): the scattering amplitude is an analytic function in the $(s,t)$ complex planes everywhere except on the unitarity cuts $s, t,u>4m^2$, and possible bound state poles for $0<s,t,u<4m^2$.~\footnote{The validity of Maximal analyticity of the scattering amplitude is a long-standing conjecture that has resisted all the perturbative checks, see \cite{Correia:2020xtr} for a review, and the references therein. In \cite{Correia:2021etg} it has been pointed out that even for the simple case of the scattering of the lightest particles, its perturbative proof would require the cancellation of an infinite number of intricate physical sheet Landau singularities. }
In our setup, we assume the absence of bound states below threshold $0<s,t,u<4m^2$, but generalising our results to include these cases is possible. 
In Fig. \ref{4d_complex_plane} we summarise the assumed analytic structure in the complex $s$-plane at fixed $-4m^2<t^*<4m^2$. 
The right-hand cut starting at the two-particle threshold $s=4m^2$ is a consequence of unitarity.
The left hand-cut is due to the physical $u$-channel process starting at $s=-t^*$. The $u$-channel cut moves as we move $t^*$, and overlaps the  $s$-channel cut for $4m^2<s<-t^*$ when $t^*<-4m^2$.

All in all, the $2\to 2$ scattering amplitude is a function of two variables satisfying crossing-symmetry, unitarity, and analyticity. These properties are summarised in    Fig.~\ref{4d_complex_plane}.
Any such amplitude can be defined by its 
 Taylor expansion around the crossing symmetric point 
\be
M(\bar s,\bar t,\bar u)=\sum_{n,p,q=0}^\infty \tilde c_{(npq)} \bar s^n \bar t^p \bar u^q,  \label{exp}
\ee
with 
coefficients $\tilde c_{(npq)} \in\mathds{R}$, 
and with the  shifted Mandelstam variables being $m^2 \bar{x} = x - 4m^2/3$. 
In this new variables the crossing symmetric point is at  the origin $(\bar s, \bar t, \bar u)=(0,0,0)$ and  momentum conservation is given by $\bar{s} + \bar{t} + \bar{u} = 0$.
Given the values of all the coefficients  $\{ \tilde c_{(npq)}\}$ in \reef{exp},   we can reconstruct the whole amplitude $M(s,t)$ by analytic continuation.
Therefore, we parametrise  the space of amplitudes by the  values of  these Taylor coefficients. 

The first few terms of \reef{exp} can be simply written as
\reef{lowamp4dnoflav}
after imposing $\bar{s} + \bar{t} + \bar{u} = 0$, and a straightforward linear redefinition of the coefficients $\{ \tilde c_{(npq)}\}$.

In perturbation theory the  $c_i$'s have a simple interpretation in terms of couplings or Wilson coefficients in the EFT. 
For instance, consider the field theory   Lagrangian
\be
{\cal L}[\partial \phi,\phi]
=   \frac{1}{2} (\partial_\mu \phi)^2 - \frac{1}{2} m^2 \phi^2 - \frac{1}{24} g_0 \phi^4 + \frac{1}{2} \frac{g_2}{\Lambda^{4}} [(\partial_\mu \phi)^2]^2 + \frac{1}{3} \frac{g_3}{\Lambda^{6}} (\partial^\mu \partial_\rho \phi) (\partial^\nu \partial_\mu \phi) (\partial^\rho \partial_\nu \phi) \phi  
+ \cdots 
\label{lagone}
\ee
where $\Lambda$ is  the cutoff of the EFT,  the dots 
 $\cdots$ involve higher order corrections $O(\Lambda^{-8})$ in the derivative expansion and operators with more than four fields $O(\phi^6)$.
If the theory is weakly coupled, i.e. $p_i ,m \ll \Lambda$ with $g_i\lesssim O(1)$, 
it is straightforward to compute the amplitude $M(s,t)$ from \reef{lagone}, leading to
\bea
c_0 &= g_0   -  4/3\, g_2  \eps^2   +   \dots  \  , \label{appr}\\[.2cm]
c_2 &= g_2 \eps^2+ \dots \, ,  \label{appr2} \\[.2cm]
c_3 &=  g_3 \eps^3 + \dots    \, ,   \label{appr3}
\eea
where $\eps=m^2/\Lambda^2$.
The first  term of these equations  is based on  standard field theory  analysis: the coupling $c_i=g_i \eps^i$ at tree level  but it can be renormalised by $O(g_0)$ loops involving marginal interactions. 
This explanation however is a bit too naive, which is demonstrated by the presence of the second term $4/3 g_2\eps^2$ in  \reef{appr}. 
This piece also  arises from a tree-level correction because we are Taylor expanding the amplitude around the crossing-symmetric point. 
Corrections to \reef{appr}-\reef{appr3} are either loop  suppressed or involve further powers of $\eps$.

A priori, the  coefficients in \reef{lowamp4dnoflav}  are only required to be real $c_i \in\mathds{R}$. However,  the analytic, crossing,  and unitarity constraints imply that the admissible values of these coefficients lie in a compact and convex subspace.
This optimization problem can be addressed with the S-matrix Bootstrap, which we will describe next.

In the first part of this work we will not make reference to perturbative physics or effective field theory analysis. We will instead  take \reef{lowamp4dnoflav} as our definitions of couplings $c_0,c_2,c_3,\dots$.
We emphasize that we shall not expect (\ref{appr}-\ref{appr3}) to hold for the
results of this section. In the next section we will address  how to interpret  the bounds on $c_i$'s as bounds on the Wilson coefficients  of EFTs.

 In the rest of the paper we work in units where
$
 m^2=1$. 
 We will however often reintroduce factors of $m^2$ where it adds clarity.

\subsection{S-matrix Bootstrap in $d=3+1$}

An optimization problem can be viewed from two complementary perspectives, the \emph{primal} and the \emph{dual} formulation.
Most of the recent works on the numerical S-matrix Bootstrap have been formulated in the primal approach, put forward in \cite{Paulos:2016but,Paulos:2017fhb}, and later on applied to a variety of physical systems in two \cite{Doroud:2018szp,Homrich:2019cbt,Paulos:2018fym,He:2018uxa,Cordova:2018uop,Karateev:2019ymz,Chen:2021pgx}, in four \cite{Guerrieri:2018uew, Guerrieri:2020bto,Hebbar:2020ukp,Bose:2020shm, Bose:2020cod,Karateev:2020axc, Karateev:2022jdb,Chen:2022nym}, and even in higher dimensions for the case of supergravity theories \cite{Guerrieri:2021ivu,WhereIsMTheory}. Alternative approaches in two dimensions based on dispersion relations are~\cite{Gabai:2019ryw,Tourkine:2021fqh}.
The \emph{dual} approach to the S-matrix Bootstrap has been first proposed in \cite{Cordova:2019lot} to study two dimensional systems of identical particles, and generalized to multi-particle systems \cite{Guerrieri:2020kcs}, massless EFTs \cite{EliasMiro:2021nul}, and in the presence of boundaries \cite{Kruczenski:2020ujw}. Formulating the \emph{dual} S-matrix Bootstrap in higher 
dimensions $d>2$ has turned out to be a challenging mathematical problem. Recently, it has been solved assuming either maximal analyticity \cite{He:2021eqn},  or  in~\cite{Guerrieri:2021tak}  by employing the rigorous analyticity domain derived by Martin~\cite{Martin:1965jj}~\footnote{
Ref.~\cite{Guerrieri:2021tak} employed fixed-$t$ dispersion relations and  connected to the old literature on the subject~\cite{Lopez:1974cq,Lopez:1975wf,Lopez:1975ca,Bonnier:1975jz,Lopez:1976zs}.}.

For concreteness, suppose that we are interested in finding the minimal  (min) and maximal (max) values of the $c_i$'s, under the constraint that the $2\rightarrow 2$ scattering amplitude $M(s,t)$ satisfies   unitarity, analyticity, and crossing-symmetry. 
The \emph{primal} strategy proceeds by filling the space of allowed S-matrices ``from inside", 
i.e. producing feasible scattering amplitudes while extremising $c_i$. It is thus a constructive approach. 
Instead,  the \emph{dual} approach excludes those values of $c_i$ that cannot be attained by a unitarity, analytic and crossing-symmetric amplitude, and therefore produces bounds on the values of $c_i$'s. 
A number of problems have been studied with both the primal and the dual~\cite{Cordova:2019lot,Guerrieri:2020kcs,Kruczenski:2020ujw,He:2021eqn,Guerrieri:2021tak,EliasMiro:2021nul}, 
finding that both approaches lead to the same optimal values for the objectives (i.e. the  values for the $\text{min/max}(c_i)$ values in our example above).
 \begin{figure}[h] \centering
 \includegraphics[scale=.5]{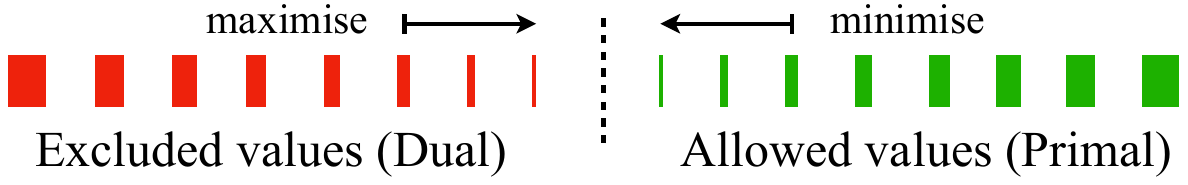} 
\end{figure}

In this section  we use the primal formulation of the S-matrix Bootstrap. 
Our point of view is that if interesting min/max values of the $c_i$'s 
are found
by using the primal approach, then it will be worth to nail  further down  those interesting bounds by developing a specific dual formulation. 

We next briefly describe the primal approach~\cite{Paulos:2017fhb}, with few modifications adapted to our problem. 
In the primal formulation one introduces  a crossing-symmetric and (real) analytic ansatz for the amplitude, 
for instance
\be
M^\text{ans}(s,t,u)=\sum_{a,b,c=0}^{N}\alpha_{(abc)} \rho_s^a \rho_t^b \rho_u^c,
\label{generic_ansatz}
\ee
where the permutation symmetry of the Taylor coefficients ensures crossing-symmetry, and  $\rho_x$  is a map ensuring maximal analyticity by mapping the complex plane with a cut $(4m^2,\infty)$ into the unit disk. 
The ansatz depends on a parameter $N$ counting the number of Taylor coefficients 
. For each $N$ we have a different ansatz, that defines a finite-dimensional truncated space of amplitudes. 
In the limit $N\rightarrow \infty$ the ansatz \reef{generic_ansatz}  describes any analytic and crossing-symmetric amplitude. 

The coefficients  in \reef{lowamp4dnoflav}  are computed from \eqref{generic_ansatz} by  taking derivatives of $M^\text{ans}$ at the crossing-symmetric point. These coefficients are   linear functionals  $c^\text{ans}_i[\alpha_{(abc)}]$ of the ansatz's Taylor coefficients, 
\be
c_0^\text{ans} = -M^\text{ans} (0,0,0)\, , \ 
c_2^\text{ans}  = \frac{1}{4} \left[ \frac{\partial^2}{\partial \bar{s}^2} M^\text{ans} (\bar{s},0,-\bar{s}) \right]_{\bar{s}=0}\, , \
 c_3^\text{ans}  = - \frac{1}{2} \left[ \frac{\partial^2}{\partial \bar{s}^2} \frac{\partial}{\partial \bar{t}} M^\text{ans} (\bar{s},\bar{t},-\bar{s}-\bar{t}) \right]_{\bar{s}, \bar{t}=0} \, .
\label{lowamp4dnoflav2}
\ee

The min/max value of the $c_i$'s is then searched for by scanning the values of the Taylor coefficients $\alpha_{(abc)}$ subject to the unitarity constraints. 
The unitarity constraints  are obtained by projecting  
the amplitude \reef{generic_ansatz} into partial waves
\be
f^\text{ans}_{\l}(s) = \frac{1}{32 \pi} \int_{-1}^{1} dx \, P_{\l}(x) \, M^\text{ans}(s, t(s,x), u(s,x)) \, , 
\ee
and imposing
$
|S^\text{ans}_\l(s)|^2 \equiv  \left|1 + i \sqrt{s-4m^2}/\sqrt{s} f^\text{ans}_{\l}(s)\right|^2 \leq 1 \, ,
$
for any $\ell\in\mathds{N}$ and for any physical value of the energy, i.e. for   $s>4m^2$. 
The number of unitarity constraints is infinite. 
Therefore,  to set up a numerical algorithm
we need to introduce two cutoffs in order to deal with finitely many constraints: 
 a cutoff $L$ in spin, so that we bound all partial waves  for 
$\ell \leq L$;
 and choose a grid of points  in $\{s_1,\dots s_n\}\in[4m^2,\infty)$    where we  impose unitarity point-wise for each spin partial wave. We denote by $1/a$ the density of points in the grid, so that $1/a \rightarrow \infty$ corresponds to the continuum limit.~\footnote{In practice we do not use a uniform grid with same spacing for all points. We employ a Chebyshev grid on the boundary of the $\rho$ disk. See also Appendix~\ref{numerics} for more details on the numerics.}
 All in all 
 we set the unitarity constraints
 \be
|S^\text{ans}_\l(s_i)|^2 \equiv  \left|1 + i \frac{\sqrt{s_i-4m^2}}{\sqrt{s_i}} f^\text{ans}_{\l}(s_i)\right|^2 \leq 1 \quad \text{for} \quad \ell=0,2,\dots L \quad \text{and} \quad i=1,2,\dots n \, . 
\label{unitarity_constraints}
 \ee

We are thus left with a finite dimensional optimization problem:
\be
c^{\text{Max}}_i(N,L,1/a)\equiv \text{Max}\big[ \, c_i^{\text{ans}}(N)  \text{ ; subject to  \reef{unitarity_constraints} }  \big] \, , 
\label{routine1}
\ee
and analogously for the minimal values. 

Finding  optimal values requires taking $N$ arbitrarily large, in order to explore the whole space of analytic and crossing-symmetric amplitudes.
It is important however to take first the number of constraints to infinity, that is the number of spins $L$ and the density of points $1/a$ to infinity first; and only then extrapolate $N\rightarrow \infty$.
In this way we are lead to numerically optimal min/max values of the  coefficients
\be
c_i^{\text{opt Max}}=\lim_{N\to\infty}\lim_{L \to \infty} \lim_{1/a \to \infty} c^{\text{Max}}_i(N,L,1/a) \, ,
\ee
and analogously for the minimal values.

Taking large values of both  the number of unitarity constraints \reef{unitarity_constraints} and $N$   is crucial in order to obtain accurate results.
In this work we  introduce a number of  numerical advances to improve the convergence in both  $L$ and $N$, which are key to reproduce our numerical results. 
We include  in   appendix~\ref{numerics} all the necessary details, and proceed now to  briefly explain these numerical advances:  
\begin{itemize}
\item \emph{Subtracted positivity constraints. }
Positivity is a necessary condition that any unitary amplitude must satisfy. Unlike unitarity, many positivity conditions can be imposed at the level of the amplitude not requiring partial wave projections. Moreover, constraints imposed at the level of the amplitude involve all infinite partial wave projections, in particular those with $\ell >L$ that are not bounded. In ref.~\cite{Guerrieri:2021ivu} it has been first shown the power of additional positivity constraints in accelerating $L$ convergence for massless particles.
In the gapped case we are studying, there are infinite different positivity constraints that can be implemented, since $\im M(s,0\leq t<4m^2)\geq 0$.
These constraints can be further improved by subtracting the contributions from the spins $\ell \leq l$ that we are already bounding by unitarity, leading to the inequalities of the form
\be
\im \tilde M^L(s,t) =\im M(s,t) -16\pi\sum_{\ell=0}^L (2\ell+1) \im f_\ell(s)P_\ell(1+2\tfrac{t}{s-4m^2})\geq 0,
\label{sub_pos}
\ee
for $0\leq t<4m^2$.~\footnote{We are grateful to Harish Murali who first made this observation for the amplitude of massless particles~\cite{WhereIsMTheory}.}
 
\item \emph{Wavelet basis. }
The naive ansatz~\reef{generic_ansatz} must be taken with care. Not all the terms in the triple sum are independent because of momentum conservation, and the redundant terms must be removed. Moreover, it has an unphysical triple discontinuity. Both issues can be simply solved considering an ansatz of the form $M^\text{ans}(s,t,u)=\bar M(s,t)+\bar M(s,u)+\bar M(t,u)$.
Instead of Taylor expanding as in eq.~\reef{generic_ansatz}, we propose a new different ansatz for $\bar M(s,t)$ inspired by the so called \emph{wavelet} expansion~\cite{wavelet} 
\bea
\bar M(s,t)&=\alpha_0+\sum_{\sigma\in \Sigma} \alpha_\sigma (\rho_s (\sigma)+\rho_t(\sigma))+\sum_{(\sigma,\tau) \in \Sigma^2} \alpha_{\sigma,\tau} ((\rho_s(\sigma)\rho_t(\tau)+\rho_s(\tau)\rho_t(\sigma)) ),
\label{wavelet_ansatz}
\eea
with $\alpha$'s free coefficients, and $\Sigma$ a set of points in the interval $(4m^2,\infty)$. 
The map
\be
\rho_s(\sigma)=\frac{\sqrt{\sigma-4m^2}-\sqrt{4m^2-s}}{\sqrt{\sigma-4m^2}+\sqrt{4m^2-s}},
\ee
maps the complex plane excluded the cut $[4m^2,\infty)$ into the unit disk, with the point $s=8m^2-\sigma$ sent to the origin $\rho=0$. We call $\sigma$ the scale parameter.
There exist a one-to-one conformal transformation that maps any $\rho$ variable into another with different scale parameter. Indeed, each element of the new basis can be viewed as a shifted and rescaled version of another -- see Appendix \ref{Appendix:wavelet} for more details.
The analogy with wavelets stops here. It would be interesting to explore the possibility of making this connection more precisely as it might lead to novel improved ansatzes.

\item \emph{High energy. }
The ansatz \reef{wavelet_ansatz}, goes to a constant when $s$ goes to infinity for fixed $t$. This is of course compatible with double-subtractions, but not very general. In particular, this behavior is far from saturating the Froissart bound $\im M(s,t=0) \sim s \log^2(s)$ \cite{Froissart:1961ux}.
We have explored the possibility of adding growing terms to the ansatz of the form $(4m^2-s)^{3/2} \rho_\sigma(s)$. 
We observe some improvement in the $N$ convergence in specific regions of the amplitude space, but we haven't used this improvement systematically in our numerics.
The reason is because these growing terms turn out to be numerically redundant, implying huge cancellations among different terms of the ansatz.
We think that making the construction of growing ansatzes more systematic would be an important direction to explore, especially in the case of $d>4$ S-matrix Bootstrap studies -- see also Appendix~\ref{Appendix:HighEnergy}.  

\item \emph{Approximating Karplus. }
Finally, we have also explored the effect of imposing a more physical domain of the double discontinuity. As reviewed in \cite{Correia:2020xtr}, analytic continuation of elastic unitarity implies that the double discontinuity must be zero below the so-called Karplus curve. In all gapped Bootstrap numerics, the double discontinuity has generically support on a larger region $(s,t) \in [4m^2,\infty)^2$. To study the effect of the change of its domain, we have constructed an ansatz where the double discontinuity is exactly zero in the square $(s,t)\in [4m^2,16m^2]^2$, with the idea of developing a rectangle approximation of the Karplus region. However, we haven't observed any visible effect on the bounds of the $c_i$'s that we have studied in this paper. It would be interesting to understand which observables  are most sensitive to the precise form of the domain of the double discontinuity.
\end{itemize}

\subsection{Filling in the space of QFTs in $d=3+1$}
\label{Sec:filling}

\begin{figure}[t] 
\centering 
        \includegraphics[width=1
       \textwidth]{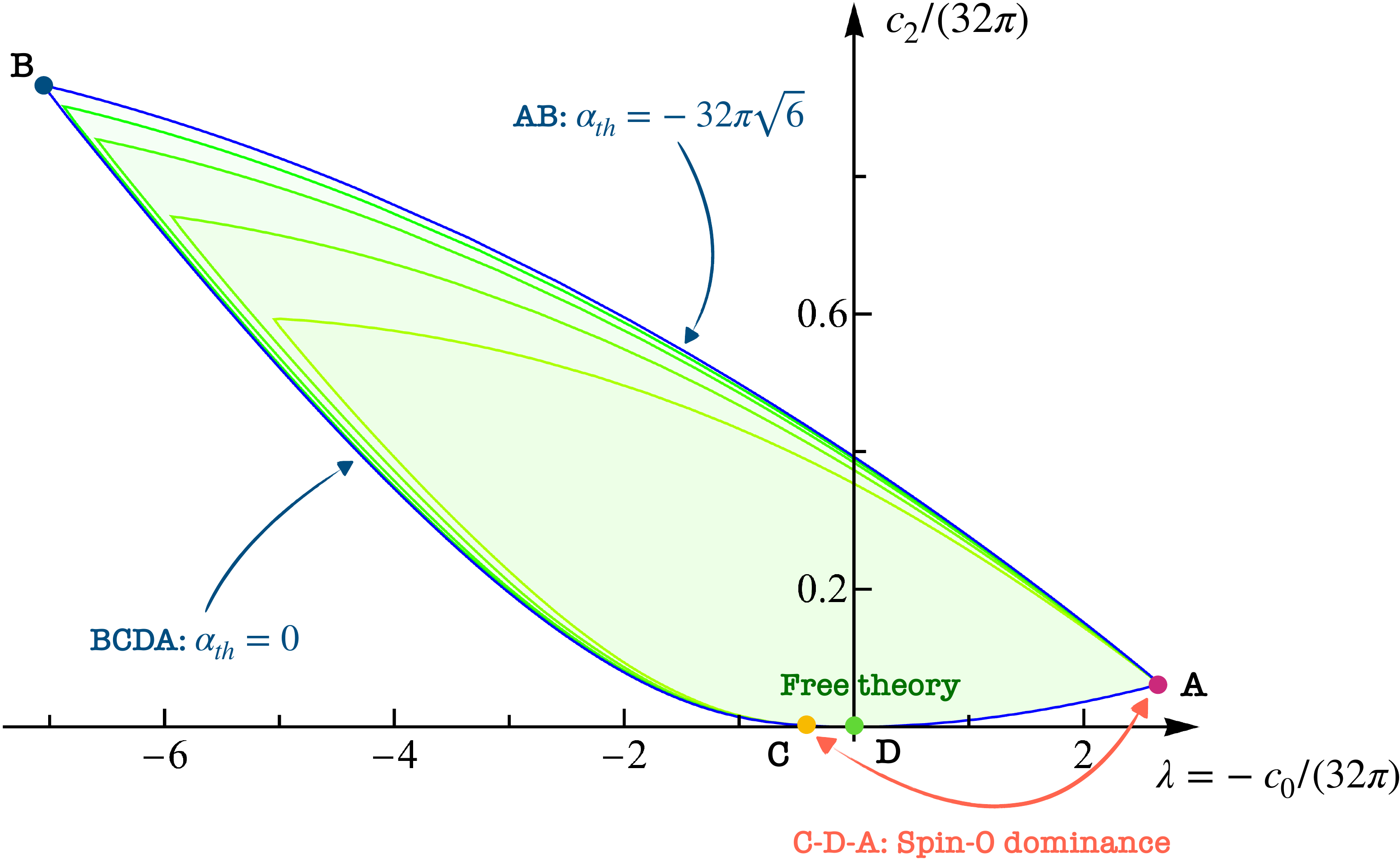} 
        \caption{ Allowed values of $(c_0,c_2)$. The different green lines correspond to different $N=6,\dots,12$, while $N=14$ is depicted in blue. The data are obtained at fixed $L=16$: the positivity constraints we impose~\eqref{sub_pos} are so efficient in constraining the large spin behavior that we can consider this value asymptotic.}
\label{butterfly_plot}
\end{figure}

In Fig.~\ref{butterfly_plot} we show the allowed space of the coefficients $(c_0,c_2)$.
All four dimensional scalar amplitudes take values of $(c_0,c_2)$ inside the green region whose approximate boundary is depicted in blue.
Next we discuss the numerical setup used to determine this plot, and the physical properties of the amplitudes saturating the bounds.

\subsubsection{Numerical aspects of Figure~\ref{butterfly_plot}}
The boundary of the allowed region has been obtained by solving a \emph{radial optimization} problem by setting $\lambda\equiv -c_0/(32\pi)=R\cos\theta$ and $c_2/(32\pi)=R_0+\kappa R\sin\theta$. For each fixed $\theta$, and for a suitable choice of $\kappa$ and $R_0$, we maximize $R$ within our numerical ansatz subject to unitarity.~\footnote{It is hard to estimate a priori the values of $R_0$ and $\kappa$ as they depend on the shape of the boundary that we can only know after solving the optimization problem. In our numerics we use $\kappa=1/10$ and $R_0=1/10$.} The result of each optimization problem determines a point $\{c_0(\theta),c_2(\theta)\}$ on the boundary. The collection of all these points obtained by choosing a discrete set of values of $\theta \in [0,2\pi)$ determines the closed regions in Fig.~\ref{butterfly_plot}. 

We use the wavelet ansatz in \eqref{wavelet_ansatz}, and add to the unitarity constraints the subtracted positivity conditions introduced in \eqref{sub_pos}.~\footnote{We add also few growing terms such that our ansatz at fixed $0<t<4m^2$ behaves as $s^{3/2}$, while in the physical region is bounded. See appendix~\ref{numerics}  for details on the parameters used in the numerics.} In order to accelerate convergence, we add a threshold singularity  term of the form $\alpha_\text{th}/(\rho_s-1)+\dots$ allowing for a square root singularity when $s\to 4m^2$ compatible with unitarity.~\footnote{In ref.~\cite{Paulos:2017fhb} it was observed that allowing for this freedom helps with convergence of the maximal coupling $c_0$ problem. }

The different green curves correspond to different values of $N$ ranging from $N=6$ to $N=12$ in steps of two following the color gradient, and  $N=14$ is depicted in blue. The inclusion of the positivity constraints allow us to work at fixed $L=16$, where we observe that the boundary does not change significantly by increasing or lowering the spin cutoff. We take the bounds at $N=14$ as a good approximation of the boundary of the allowed $(c_0,c_2)$ space. 
The position of the points around the cusp $B$ will change mildly as we increase $N$ further. 
However, in Appendix~\ref{minimum_coupling_numerics} we show that this change is small, of the order of 0.2. We have not attempted a systematic extrapolation of the whole boundary, but it would be interesting to derive an exclusion plot using a dual formulation.

\subsubsection{Properties of the amplitudes in the boundary}

At each point in the boundary, the solution of the optimisation problem determines  the values of the $\alpha_{(abc)}$ coefficients.
The solution not only provides the min/max values of $(c_0, c_2)$ but also a non-perturbative  $2\rightarrow 2$ scattering  amplitude of  a scalar particle. 
We must therefore look for observables that would help us characterise the physics along the boundary, thus giving us information about these extremal putative QFTs. 

We identify four special points. The points A and B are easily identified by looking at Fig. \ref{butterfly_plot} since they correspond to visible cusps.
At those points we attain respectively the maximum and minimum value of the coupling $\lambda$. 
The green curves show that convergence of the optimisation algorithm towards point A is  fast, while convergence is   slower towards point B. 
The  scattering amplitudes associated to points A and B have   been studied  in ref.~\cite{Paulos:2017fhb}. 
More recently,  precise numerical determinations of the maximum coupling $\lambda$, both from the primal and a dual approach, have been obtained in ref.~\cite{He:2021eqn}. 
The error on the minimum coupling is still  quite large. By combining our primal data and the dual bound in ref.~\cite{Guerrieri:2021tak}, we obtain the best estimate to date:
\be
-8.02 \leq \min \lambda \leq -7.0 \, . 
\ee
It is still unclear if one can construct a physical theory with a $\lambda$ close to the minimum value and why  numerical convergence is hard. It would be interesting to find some physical intuition behind this minimum coupling amplitude that might help designing a faster converging ansatz.

The point D at the origin $(c_0,c_2)=(0,0)$  is the free theory. It is simple to show that $c_2\geq 0$ by looking at its dispersive representation~\footnote{By using a similar dispersion relation it is easy to show that the second derivative of the amplitude around any point in the Mandelstam triangle is positive, i.e.   the amplitude   is a \emph{convex function} in the Mandelstam triangle.}
\be
c_2=  \frac{m^4}{\pi}\int_{4m^2}^\infty  \frac{\text{Im} M(z,\tfrac{4}{3}m^2)}{(z-\tfrac{4}{3}m^2)^3}dz \geq 0 \, . 
\label{dispc2one}
\ee
The lower bound $c_2=0$ is saturated when $\text{Im} M(z,t)=0$, hence by the free theory. Combining this analytic bound with our numerical results we obtain an estimate on the allowed values of $c_2$
\be
0\leq \frac{c_2}{32\pi}\lesssim 0.93 \, .  \label{absc2}
\ee

\begin{figure}[t] 
\centering 
        \includegraphics[width=1
       \textwidth]{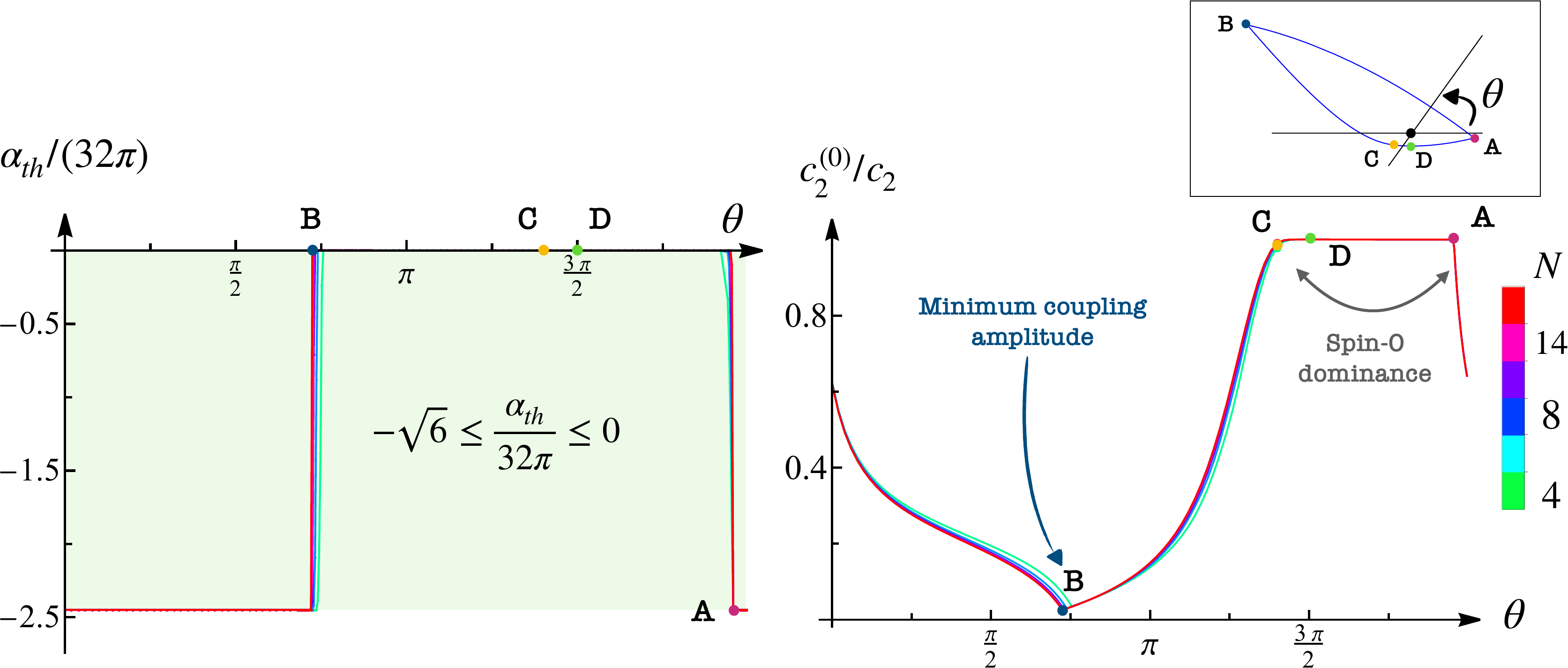} 
        \caption{
        On the left plot we show the residue of the pole at threshold $\alpha_\text{th}$ along the boundary of the allowed $(c_0,c_2)$ values in Fig.~\ref{butterfly_plot}.
        On the right we show the spin-zero dominance, also along the boundary of the allowed $(c_0,c_2)$ values.}
\label{threshold_and_ratios}
\end{figure}

Cusps like the points $A$ and $B$ must be associated to drastic discontinuities in the behavior of the amplitude. 
In Fig.~\ref{threshold_and_ratios} on the left we plot the value of the residue at threshold $\alpha_\text{th}$ of the amplitude as a function of the radial angle $\theta$. Unitarity at threshold determines the allowed region of the threshold residue $-32\pi \sqrt{6} \leq \alpha_\text{th} \leq 0$ depicted in green in the figure.
We clearly see that the points $A$ and $B$ divide the boundary in two regions: the upper arc $AB$ where the threshold residue $\alpha_\text{th}$ attains its minimum value, and the arc $BA$ where it is zero.

An interesting observable can be defined by projecting the dispersion relation \reef{dispc2one} into partial waves, $
c_2=\sum_{\ell=0}^\infty c_2^{(\ell)},
$ where 
\be
c_2^{(\ell)}=\frac{m^4}{\pi}(2\ell+1)16 \pi \int_{4m^2}^\infty \frac{\im f_\ell (z)P_\ell(1+\tfrac{8}{3(z-4m^2)})}{(z-\tfrac{4}{3}m^2)^3}dz \, 
\ee
is the contribution coming from the spin $\ell$ partial wave to $c_2$. 
We then define the ratio  $c_2^{(0)}/c_2$, which measures  the relative contribution of the spin-0 to the whole $c_2$ value.  
In Fig.~\ref{threshold_and_ratios} on the right we plot the ratio $c_2^{(0)}/c_2$ as a function of the radial angle $\theta$. Curves with different colors correspond to different values of $N$ (from $N=6$ in green, to $N=14$ in red).
First, we notice there is a region of the boundary where the spin-0 contribution dominates $c_2^{(0)}/c_2\sim 1$: it starts on the $BD$ arc before the free theory point, and it stretches all the way up to maximum coupling cusp $A$. We use the terminology introduced in \cite{Bern:2021ppb} and call this region Spin-zero dominance arc.
In all figures we have added a point $C$ signalling the starting point of the Spin-zero dominance. 
It is worth noticing that this is also the region that numerically converges faster.

Following the boundary anti-clockwise from point $A$ the Spin-zero dominance ratio attains its minimum at the minimum coupling cusp $B$.
The whole arc $ABC$ is saturated by scattering amplitudes with non trivial higher spin partial waves.
It would be interesting to study this ratio for higher spins and see whether it is possible to define a sequence of points $C^\prime$, $C^{\prime\prime}$, $\dots$ where the sum over the first $2,4,\dots$ spins dominates. This might serve as a way to shed light on these exotic amplitudes and hint at the physical theories that might appear close to the boundary of the allowed region.

We conclude with a comment on the minimum coupling cusp. 
It is interesting to notice that the point $B$ maximises also the value of $c_2$, and that features almost zero Spin-zero dominance.
This meets the intuition that the minimum coupling amplitude may be realized by the most strongly coupled QFT in four dimensions, justifying why convergence is so hard at  point $B$.

\subsection{Unitarity vs Positivity}
\label{projbounds}

In this section we compare 
the amplitudes in the boundary of the  allowed $(c_0,c_2)$-space
with the positivity bounds. 
Using positivity   we will  constrain ratios of the $c_i$ coefficients, like for instance $c_3/c_2$.
This is in contrast with the primal S-matrix Bootstrap results of the previous section where we find the absolute allowed space  e.g. for $c_2$ in 
equation~\reef{absc2}, in units of $m=1$. 

Recently refs.~\cite{Tolley:2020gtv,Caron-Huot:2020cmc}  exploited 
full crossing-symmetry to derive two-sided bounds on ratios of Wilson coefficients. 
Inspired by these  developments we will next introduce an alternative, but equivalent,  derivation of positivity
constraints exploiting full crossing symmetry. 
We will also need to adequate this derivation to our set up by including unitarity cuts for all physical energies $s\in [4m^2,\infty)$.

We start with the double-subtracted, fixed $t$, dispersion relation
\be
M(s,t)=M(s_0,t_0)+\frac{1}{\pi}\int_{4m^2}^\infty dz \, [M_z(z,t)K(z,s,t;t_0)+M_z(z,t_0)K(z,t,t_0;s_0)]
\label{fixed-t}
\ee
where the  kernel is given by
\be
K(z,s,t;t_0)=\frac{1}{z-s}+\frac{1}{z-4m^2+s+t}-\frac{1}{z-t_0}-\frac{1}{z-4m^2+t+t_0}  \, , 
\label{crossing_kernel}
\ee
and  we have defined the discontinuity 
 \be
 M_z(z,t) \equiv \frac{1}{2i}\text{disc}_z M(z,t) \,.
 \ee
This dispersion relation encodes $s\leftrightarrow u$ crossing-symmetry and analyticity in $s$ for fixed $t$.~\footnote{See for instance appendix A of ref.~\cite{Guerrieri:2021tak} for a recent derivation of \reef{fixed-t}. }
The validity of the fixed-$t$ dispersion relation \reef{fixed-t} has   been proved by Martin~\cite{Martin:1965jj}.
The   coefficients $c_i$ in \reef{lowamp4dnoflav} can be  obtained from  \reef{fixed-t} by taking suitable derivatives with respect to $s$ and $t$
\bea
-c_0&= M(s,t) -\frac{1}{\pi}\int_{4m^2}^\infty dz (M_z(z,t)K(z,s,t;\tfrac{4}{3}m^2)+M_z(z,\tfrac{4}{3}m^2)K(z,t,\tfrac{4}{3}m^2;\tfrac{4}{3}m^2))\,, \label{c0sr} \\[.2cm]
c_2&=\frac{m^4}{\pi}\int_{4m^2}^\infty dz \frac{M_z(z,\tfrac{4}{3}m^2)}{(z-\tfrac{4}{3}m^2)^3} \geq 0\,, \label{c2def}\\[.2cm]
c_3&=\frac{m^6}{\pi}\int_{4m^2}^\infty dz \left(3\frac{M_z(z,\tfrac{4}{3}m^2)}{(z-\tfrac{4}{3}m^2)^4}- 2\frac{\tfrac{\partial}{\partial t}M_z(z,t)|_{t=4/3m^2}}{(z-\tfrac{4}{3}m^2)^3}  \right)\,, \label{c3sr}
\eea
(for convenience we reproduce again the dispersion relation for $c_2$ in \reef{dispc2one}).
The sum rule for $c_0$ in \reef{c0sr} depends on an arbitrary analytic point $(s,t)$, within the region of convergence of the fixed-$t$ dispersion relations.
Since $c_0$ depends on an arbitrary subtraction constant $M(s,t)$, it is impossible to extract bounds without introducing constraints on the real parts of the amplitude.
The integrand in the dispersive representation of $c_3$ is not manifestly positive, and it is hard to claim bounds on this  coefficient  without further assumptions.

We will treat crossing symmetry 
starting from the \emph{crossing equation}
\be
M(s,t)=M(s,4m^2-s-t) \, . 
\label{crossing_basic}
\ee
The fixed-$t$ dispersion relation \reef{fixed-t} is only $s\leftrightarrow u$ invariant
and  therefore must be equipped with the explicit $s\leftrightarrow t$ crossing-symmetry, or the equivalent $t\leftrightarrow u$ crossing-symmetry constraint \reef{crossing_basic}. As first observed by Roy~\cite{Roy:1971tc}, 
using the fixed-$t$ dispersion relation \eqref{fixed-t} we can write the crossing equation \reef{crossing_basic} in a form that involves only the imaginary parts of the amplitude
\be
\int_{4m^2}^\infty dz \left[ M_z(z,t)K(z,s,t;t_0){-}M_z (z,u) K(z,s,u;t_0) {+}M_z(z,t_0)(K(z,t,t_0;s_0){-}K(z,u,t_0;s_0))\right]=0.
\label{crossing_eq_integral}
\ee
Next we expand the last equation  into partial waves
\be
\int_{4m^2}^\infty dz \sum_{\ell=2}^\infty (2\ell+1)\im f_\ell(z) F_\ell(z,s,t;t_0,s_0)=0 \, , 
\label{crossing_expanded}
\ee
where the kernel is given by 
\be
\begin{aligned}
&F_\ell(z,s,t;t_0,s_0)=\\
&P_\ell(1{+}\tfrac{2t}{z-4})K(z,s,t;t_0){-}P_\ell(1{+}\tfrac{2u}{z-4})K(z,s,u;t_0){+}P_\ell(1{+}\tfrac{2t_0}{z-4})(K(z,t,t_0;s_0){-}K(z,u,t_0;s_0)) \, ,
\label{blocks}
\end{aligned}
\ee
in $m^2=1$ units.  
If $(s,t)$ are inside the Mandelstam triangle (as we are interested in) then the partial wave expansion is convergent.
Notice that the sum over spins starts at $\ell=2$. This is a consequence of the double subtractions that leave the spin $\ell=0$ unconstrained.~\footnote{Equation \reef{crossing_expanded}   reminds us  the crossing equation used in the Conformal Bootstrap \cite{Rattazzi:2008pe,Poland:2018epd}.}
Next we can act on the crossing equation using different set of functionals, for instance, by applying any number of derivatives, or evaluating it at different points. 
We choose to take derivatives of \reef{crossing_expanded} around the crossing symmetric point $s=t=u=4/3$, and we are lead to 
\be
\mathcal{F}^{(l,k)}=\int_{4m^2}^\infty dz \sum_{\ell=2}^\infty (2\ell+1)\im f_\ell(z) F^{(l,k)}_\ell(z)=0.
\label{crossing_constraints}
\ee
where $
F^{(l,k)}_\ell(z)=\frac{\partial }{\partial s^l}\frac{\partial }{\partial t^k}F_\ell(z,s,t;\tfrac{4}{3},\tfrac{4}{3})\big |_{s=t=4/3}
$.~\footnote{
Schematically $M(s,t)\sim\sum_\ell f_\ell(s) P_l((u-t)/(u+t))$,  therefore $u\leftrightarrow t$ crossing-symmetry implies that the 
odd spin contributions to the amplitude  vanish. 
Thus an equivalent way to impose crossing symmetry is to first set to zero  the odd spin contributions to the imaginary parts $M_z(z,t)$ and  $M_z(z,t_0)$  in \reef{fixed-t}. Then   project into partial waves the fixed $t$ dispersion relation \reef{fixed-t}. 
This projection contains spin odd contributions because the Kernel is not $t\leftrightarrow u$ symmetric. 
Imposing that these spin odd projections vanish should be equivalent to  \reef{crossing_constraints}, see ref.~\cite{Guerrieri:2021tak}. }
The last equation is the final form of the $s\leftrightarrow t$ crossing-symmetry constraints that we will employ.

Although powerful, these constraints are still not sufficient to bound  $c_0$, $c_2$ or $c_3$ separately. 
Therefore we will use positivity to bound ratios of these coefficients. In particular, we  address the problem
\be
\begin{aligned}
&\min_{\{\im f_\ell\}\text{ in \reef{c3sr}}}\quad c_3\\
\text{subject to}\quad  &c_2=1, \quad \mathcal{F}^{(l,k)}=0, \quad \im f_\ell(s)\geq 0.
\end{aligned} \label{plinear} 
\ee
and analogously for the maximisation problem. 
Equation \reef{plinear} means minimise (min) the coefficient $c_3$ by varying the imaginary part of the partial waves $\text{Im}f_l$, subject to the constraints
$c_2=1$, the crossing-symmetry constraints  $ \mathcal{F}^{(l,k)}=0$ and  positivity  $ \im f_\ell(s)\geq 0$.
Since \reef{plinear} is a linear optimisation problem, its dual version is  given by 
\be
\begin{aligned}
&\max_{\lambda_2, \nu_{n,m}} \quad -\lambda_2\\
\text{subject to}\quad  (-)(&\lambda_2 K^{(2)}_\ell(s)+K^{(3)}_\ell(s)+\sum_{n,m}\nu_{n,m}F_\ell^{(n,m)}(s))\geq0\,,
\end{aligned}
\label{dualc3c2ratio}
\ee
where the kernels are given by 
\be
\begin{aligned}
&K^{(2)}_\ell(s)=\frac{16}{(s-4/3)^3}P_\ell(1+\tfrac{8}{3(s-4)})\\
&K^{(3)}_\ell(s)=48\left( \frac{P_\ell(1+\tfrac{8}{3(s-4)})}{(s-4/3)^4}-\frac{(\ell+1)\left((4{-}3s)P_\ell(1{+}\tfrac{8}{3(s-4)})+3(s{-}4)P_{\ell+1}(1{+}\tfrac{8}{3(s-4)})\right)}{4(s-4/3)^3(3s{-}8)}  \right)\, .
\end{aligned} \label{kkk}
\ee
The last equations can be obtained by expanding in partial waves the kernels  \reef{c2def} for $c_2$ and $c_3$ respectively.

\bigskip

\begin{figure}[t] 
\centering 
        \includegraphics[width=1 
       \textwidth]{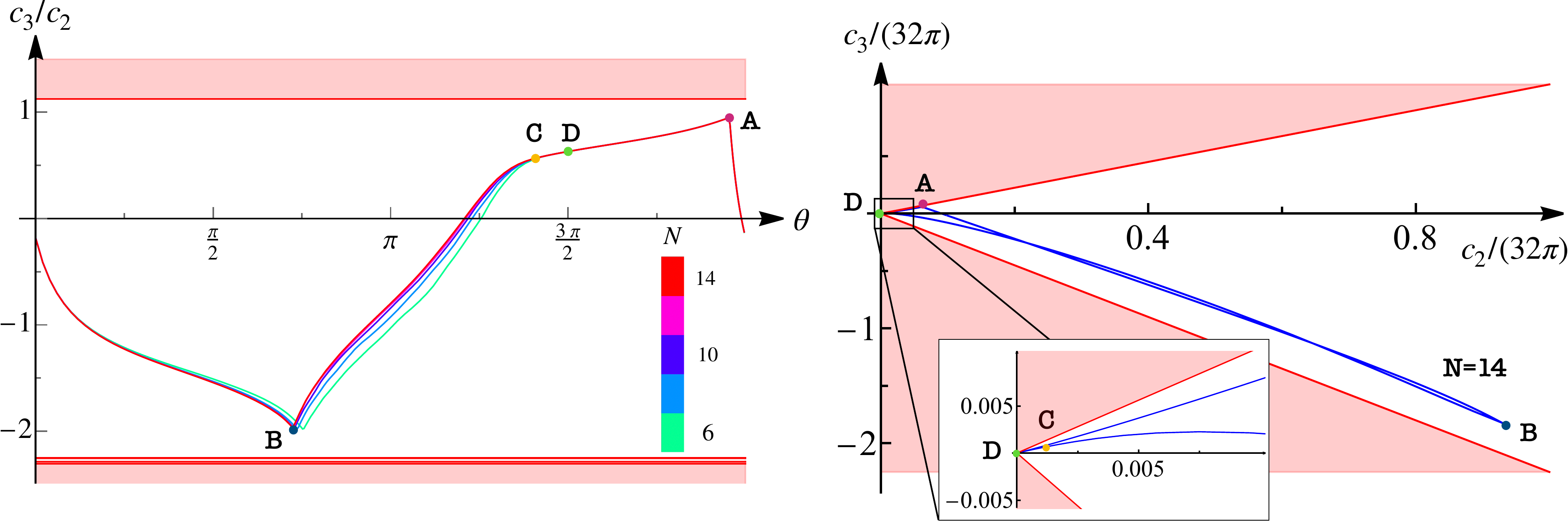} 
        \caption{On the left ratio of derivatives $c_3/c_2$ as a function of the radial angle $\theta$ used to parametrize the boundary of the allowed region in Fig.~\ref{butterfly_plot}. On the right the same plot in the $(c_2,c_3)$ space, for fixed $N=14$.}
\label{ratioc3c2}
\end{figure}

We solve numerically the problem  by considering all constraints with  $n+m \leq 7$. \footnote{Note that not all derivatives are independent, there is only one independent constraint up to $n+m\leq 6$. For $n+m<4$ all crossing constraints vanish.} 
Even in the dual approach, like in the primal, there is an extrapolation involved since we have to impose an infinite number of linear dual constraints. In practice, we impose constraints up to $\ell_\text{max} = 20$ and check that increasing the maximum spin does not change the bounds.
The results are given in Fig. \ref{ratioc3c2}, shown as the red excluded regions defined by the inequality 
\be
-2.250 \leq \frac{c_3}{c_2}\leq 1.125 \, . 
\label{c3c2ratiobounds}
\ee 

Note however that the upper bound is independent of  the number of crossing constraints imposed.
Indeed, it can be obtained analytically observing that from the definition of $c_3$ \reef{c3sr}
\be
c_3
\leq 
\frac{1}{\pi}\int_{4}^\infty dz  \frac{3 M_z(z,\tfrac{4}{3})}{(z-\tfrac{4}{3})^4} 
\leq \frac{1}{\pi} \int_{4}^\infty dz \frac{1}{(4-\frac{4}{3})} \frac{3 \, M_z(z,\tfrac{4}{3})}{(z-\tfrac{4}{3})^3} = \frac{9}{8} c_2 \, , 
\label{c3vsc2}
\ee
in $m^2=1$ units.
This bound is saturated by a function with $\tfrac{\partial}{\partial t} M_z(z,t)=0$, that is only possible  when $ M_z(z,t)$ contains only a spin-0 component.
The crossing constraints \eqref{crossing_constraints} involve only partial waves with spin $\ell\geq 2$, hence they cannot improve the bound.

The lower bound is sensitive to the crossing constraints. The different red lines  approaching the lower bound 
in the left plot of Fig.~\ref{ratioc3c2}
 correspond to increasing values of crossing constraints $n+m$, but they seem to converge fast. Our numerical result coincide with the analytic lower bound $c_3/c_2>-9/4$ derived in \cite{Haldar:2021rri} using crossing symmetric dispersion relations and the Geometric Function Theory.

We conclude this section by comparing the positivity bounds with the S-matrix Bootstrap data obtained in Sec. \ref{Sec:filling}.
The compatibility of the positivity bounds and the Bootstrap data has also  been observed in  refs. \cite{Haldar:2021rri,Zahed:2021fkp} in different systems.
The curves with different colours in Fig.~\ref{ratioc3c2} on the left show the ratio $c_3/c_2$ as a function of the radial angle $\theta$ used  in Fig. \ref{butterfly_plot}.
We notice that the maximum and minimum amplitudes correspond respectively to the maximum and minimum value for this ratio.
These values are close to saturate the positivity bounds, although there is still a gap. This is expected since we are not optimizing the S-matrix Bootstrap amplitudes  in the $(c_2,c_3)$ space.~\footnote{It has been checked in \cite{Chen:2022nym} that the bounds are  saturated using the S-matrix Bootstrap. See below a simple Quantum Field Theory that saturates the upper bound. } In the spin-0 dominance arc the ratio $c_3/c_2$ is positive, and therefore $c_3$ is positive.
This is expected since the negativity of $c_3$ comes from a term proportional to $\tfrac{\partial}{\partial t}\im M_z(z,t)$ that does not have any spin-0 contribution. We shall see other instances of this phenomenon in the next sections (see Fig. \ref{M2_ratio}  below where  $c_3$ is positivity along the boundary of the EFT region). In Fig.~\ref{ratioc3c2} on the right we show the trajectory in the $(c_2,c_3)$ plane as we move along the boundary, showing it is well inside the excluded region determined using positivity in red.

\subsubsection{Un-subtracted physics }

\begin{figure}[t] 
\centering 
        \includegraphics[width=1 
       \textwidth]{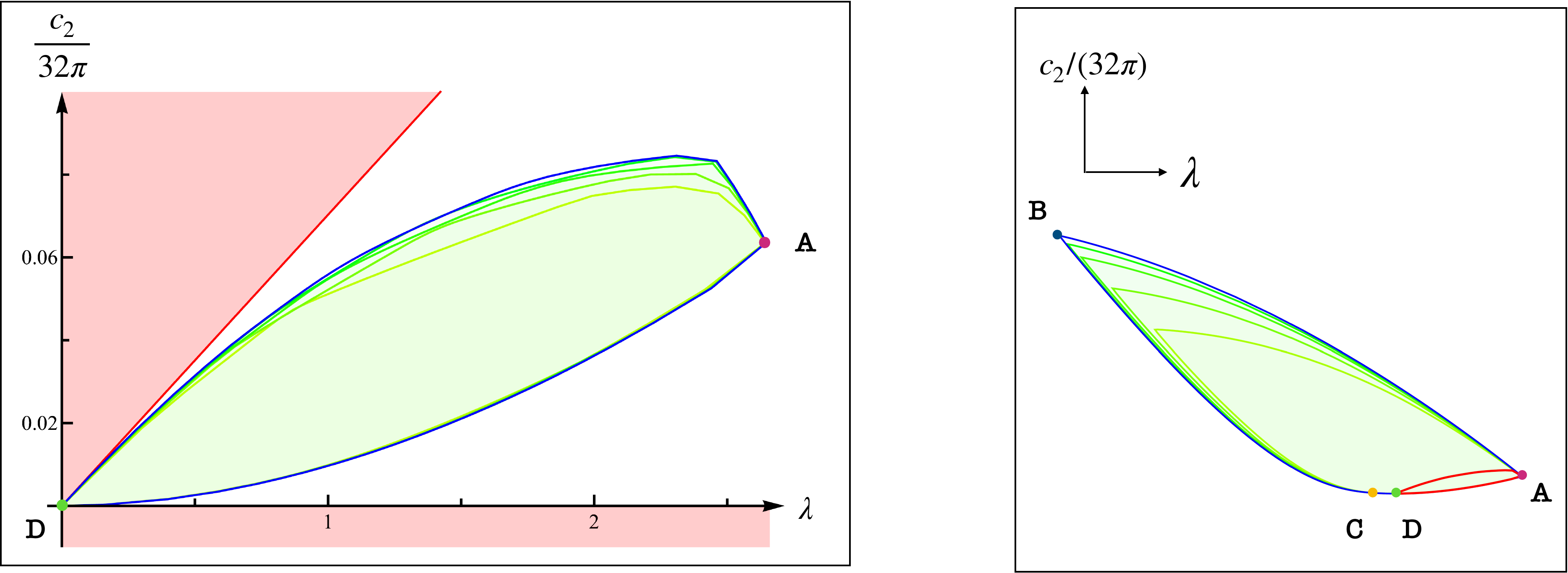} 
        \caption{Allowed values of $(c_0,c_2)$ in the un-subtracted physics scenario (left).
        On the right plot we show the dramatic effect of this constraint: the plot on the left is shown  there by a red curve  embedded in the plot of Fig.~\ref{butterfly_plot}. }
\label{decaying_butterfly}
\end{figure}

In this section we comment on the effect of subtractions on physical observables.
Let us assume that the amplitude decays at infinity at some fixed $t=t^*$
\be
\lim_{|s|\to\infty} M(s,t^*)=0.
\ee
We can write an un-subtracted dispersion relation and immediately see the dramatic effect of this condition on the quartic coupling by choosing $t^*=\tfrac{4}{3}$~\footnote{We have checked that imposing the same condition for any fixed angle, does not affect the bounds. It might be interesting to study the effect of different number of subtractions and conditions imposed at infinity. We leave these considerations for future studies.}
\be
\lambda \equiv \frac{1}{32\pi}M(\tfrac{4}{3},t^*)=\frac{1}{16\pi^2}\int_{4}^\infty dz \frac{\im M(z,t^*)}{z-\tfrac{4}{3}}\geq 0 \, . 
\label{g0disp}
\ee
Then, using \reef{c2def} and \reef{g0disp} we obtain an upper bound on the ratio $c_2/\lambda$
\be
c_2 
\leq \frac{1}{\pi} \int_{4}^\infty dz \frac{1}{(\frac{8}{3})^2} \frac{\im M(s,t^*)}{z-\tfrac{4}{3}} = \frac{9 \pi }{4}\lambda \, ,
\ee
where we have used the  inequality $\tfrac{1}{(s-4/3)^3}\leq (\tfrac{3}{8})^2 \tfrac{1}{s-4/3}$.
Thus, $c_2$ is bounded by 
\be
0 \leq \frac{c_2}{32\pi} \leq \frac{9}{128}\lambda \, .
\ee
This condition excludes negative values of $\lambda$ that would be compatible with weakly coupled $V(\phi)=+|g_0|\phi^4/4!$ theory.

In the left plot of Fig. \ref{decaying_butterfly} we show the allowed region in the $\{\lambda,c_2\}$ plane in green. As before, different lines correspond to different values of $N$. This region converges fast, so we use $N=4,...,8$, and $L=12$.
In the right plot of Fig. \ref{decaying_butterfly} we see how this region fits inside the full space in Fig.~\ref{butterfly_plot}, showing that part of the boundary is saturated by amplitudes that decays for large $s$ at fixed $t=4/3$.


\section{Weakly coupled EFTs in $3+1$ dimensions}
\label{singeft}

So far our discussion has been non-perturbative,  and based on the observation that below the discontinuities enforced by unitarity we can represent the amplitude by \reef{lowamp4dnoflav}.

For a limited range $|c_0|/(32\pi) \ll  1$, the boundary of the allowed  $(c_0,c_2)$  values in Fig.~\ref{butterfly_plot}  can be described by 
the effective field theory in \reef{lagone}.
However, most of the boundary involves large values of $(c_0,c_2)$, in $m^2$ units,  and thus cannot
 be simply interpreted as a bound on the coefficients of the  operators $\phi^4$ and $(\partial \phi)^4$. 
The higher order corrections  in the right hand side of (\ref{appr}-\ref{appr3}) are large; in other words,  the dots in $c_i= g_i \, (m^2/\Lambda^2)^i+\dots$ cannot be ignored.

The aim of this  section is to  formalise and give an answer to:  \emph{what are the extremal values of  $(c_0,c_2)$  in \reef{lowamp4dnoflav} for
theories that are weakly coupled  below the cutoff $\Lambda$ of a putative Effective Field Theory description?}
Namely, we will  isolate the region inside 
the  allowed $(c_0,c_2)$-plane of Fig.~\ref{butterfly_plot}
that at low energies can be described by a  weakly coupled field theory. 
In order to  achieve such purpose, we need to encode the notion of mass gap separation $m^2\ll \Lambda^2$ into the S-matrix Bootstrap problem. 

The imaginary part of the two-to-two scattering amplitude of quantum  theories that admit a weakly coupled field theory description at energies $s\ll \Lambda^2$   looks like
\be
\begin{minipage}{13cm}\centering
        \includegraphics[width=0.8
       \textwidth]{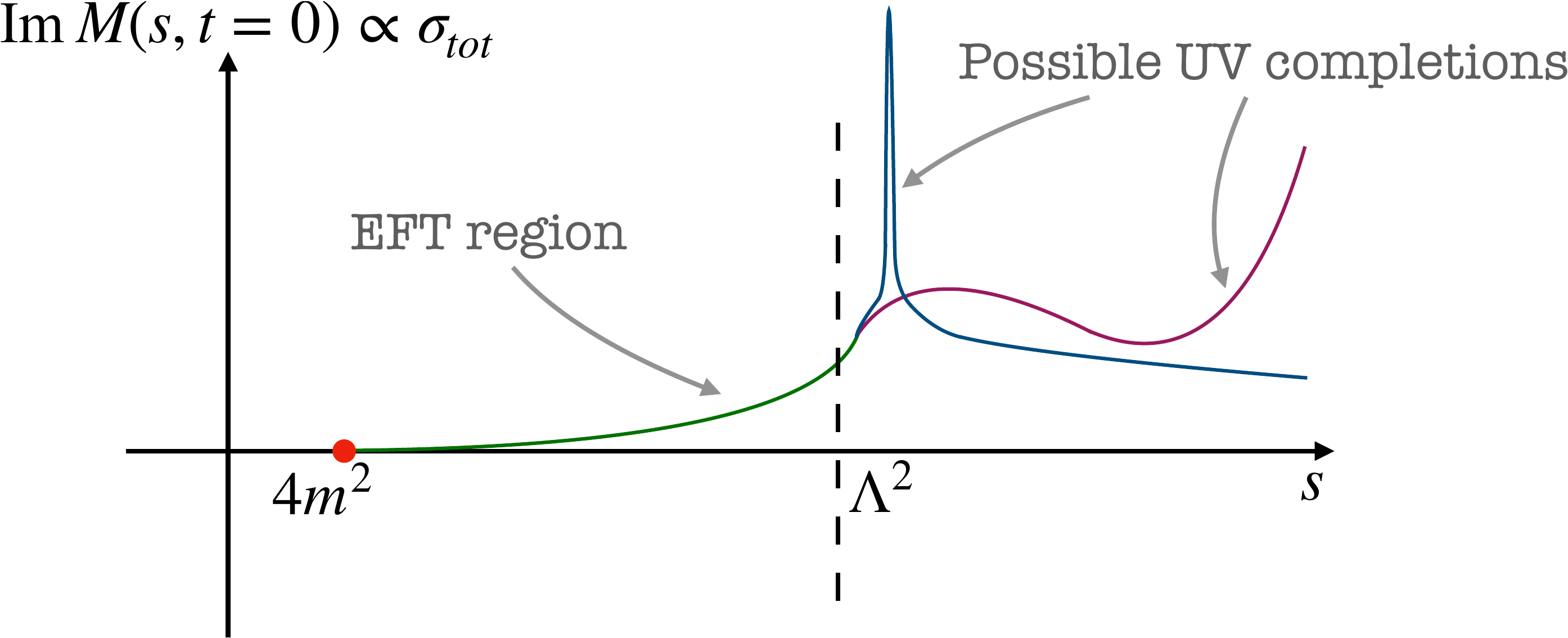} 
  \end{minipage} \label{ske} \nonumber
\ee
The imaginary part is controlled by loops. Therefore a sensible strategy is to require that $\text{Im}M(s)$ is small for energies below the physical cutoff. Thus we will set up a (primal) S-matrix Bootstrap approach  with  extra constraints  that ensure that  $\text{Im} M(s) \ll 16 \pi^2$ at low energies $s\leq \Lambda^2$.
We  make no assumption about the physics above the cutoff scale $\Lambda$, 
and allow strongly or weakly coupled  UV completions.
We will be interested in unitary amplitudes, and thus we will   require unitary  in the entire regime of physical values of $s>4m^2$. The assumption about maximal analyticity which we employed in the previous section is also   kept unchanged. 
Next we describe two   
methods, {\bf M1} and {\bf M2}, 
to approach this problem.

 \begin{enumerate}
\item[{\bf M1}] 
A first possible avenue is to require that the integral of the imaginary part of the amplitude, against some kernel,  is small in the low energy region $4\leq s \leq \Lambda^2$. In particular we will impose the extra constraint  
\be
\Delta a^\text{IR}_n \equiv  \int_{4}^{\Lambda^2} \frac{ M^\text{ans}_z(z,4/3)- M_z^{\text{EFT}}(z,4/3)}{(z-4/3)^{n+1}}dz \leq 0\quad  \label{M1_higherd}
\ee
where $n\geq 2$,  we set $m^2=1$, and $M^{\text{EFT}}$ is a phenomenological input based on the low energy theory of interest.
For instance, a simple minimalistic  choice  is to require  the  $g_0 \phi^4/4!$ theory in the IR. In this case the leading order  imaginary part of $M(s,t)$ is given by 
\be
M_z^{\text{EFT}}(s,t)=\frac{\sqrt{s-4}}{\sqrt{s}}\frac{g_0^2}{32\pi}  \, ,  \label{oneloopapp}
\ee
 which depends on  the EFT coupling $g_0 \ll 16 \pi^2$.
 Similar variables to $\Delta a_n^\text{IR}$ were analysed in \cite{Bellazzini:2020cot}, there called \emph{arcs}. Hence we  will call \reef{M1_higherd} \emph{arc constraints}. 
 All in all, {\bf M1} is a refinement of  \reef{routine1}
 to exclude theories featuring large cuts at low energy, and consists in the following optimisation problem 
\be
\text{Max}\big[ \, c_i^{\text{ans}}(N)  \text{ ; subject to  \reef{unitarity_constraints}  and \reef{M1_higherd}}  \big] \, .
\label{routine2}
\ee
\item[{\bf M2}] A perhaps more refined avenue is to impose  point-wise  constraints by requiring 
\be
\text{Im} f^\text{ans}_{\l}(s)  \leq  \text{Im} f^\text{EFT}_{\l}(s) \quad \text{for} \quad  4 \leq s \leq \Lambda^2 \, . \label{cooo}
 \ee
 where  $\text{Im} f^\text{EFT}_{\l}(s) $ is the phenomenological input, computed perturbatively  for the EFT we are interested in reproducing in the IR.
The real part is allowed to  freely vary in order to satisfy the unitary equations. 
In practice we will discretise  \reef{cooo} into a fine grid of points in the range $s\in[4,\Lambda^2]$.
We expect that   \reef{M1_higherd}  for $n=2,3,\dots n^*$ should be equivalent to \reef{cooo} for large values of $n^*$.~\footnote{Our intuition is also based on numerical experiments performed in two-dimensional toy models.}
In practice we will not achieve large values of $n^*$, and therefore it becomes convenient to employ this second method. 
All in all, {\bf M2} consists in  
\be
 \text{Max}\big[ \, c_i^{\text{ans}}(N)  \text{ ; subject to  \reef{unitarity_constraints}  and \reef{cooo}}  \big] \, .
\label{routine3}
\ee
Details on the   ansatz for $M(s,t)$ that we use for {\bf M2} are provided in  section~\ref{secm2}.
\end{enumerate}

\subsection{M1: bounding arcs}

\begin{figure}[t] 
\centering 
        \includegraphics[width=0.75
       \textwidth]{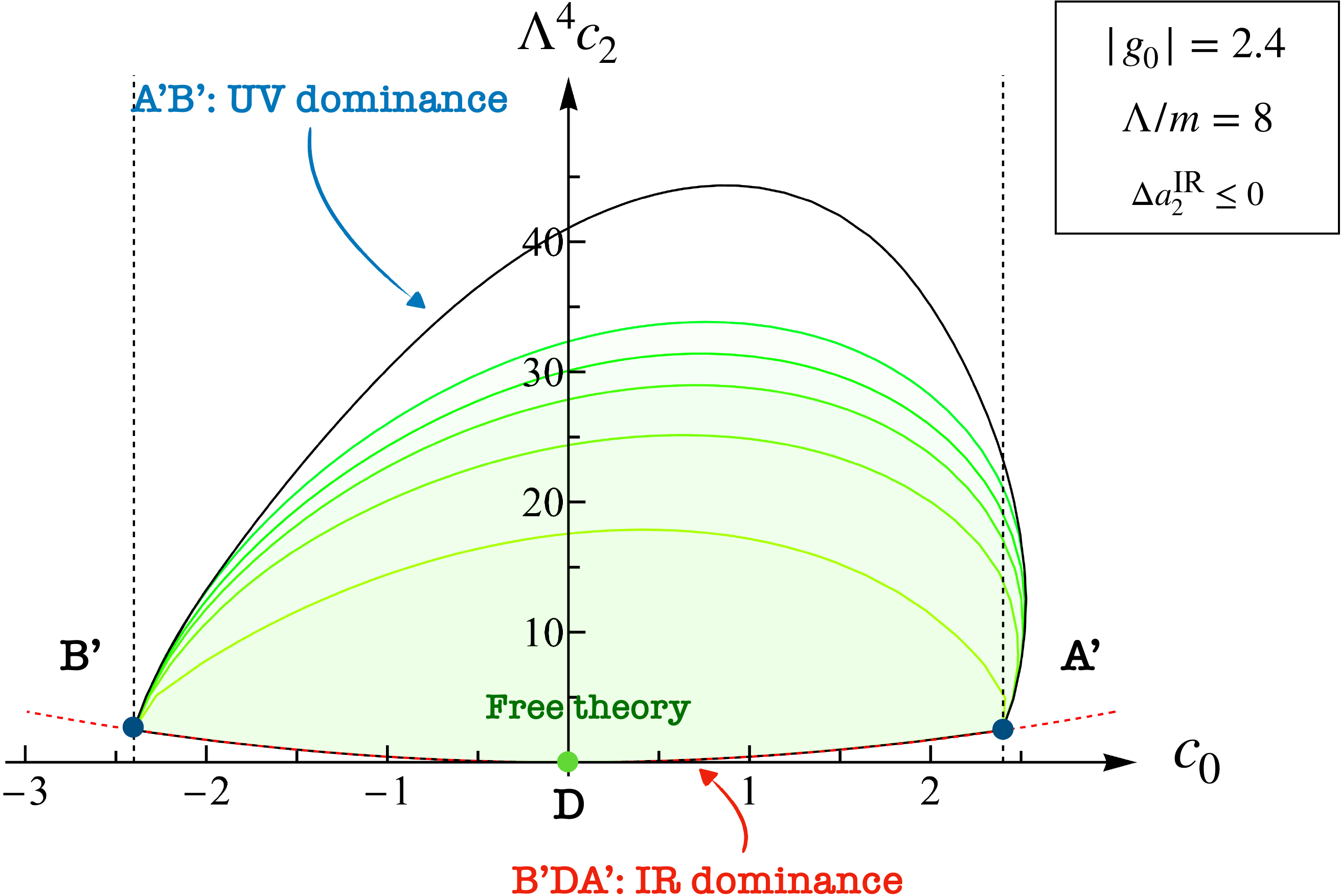} 
        \caption{Allowed values for $(c_0,c_2)$, using the method {\bf M1} with one arc constraint $\Delta a_2^\text{IR}\leq 0$. Different green curves correspond to different values of $N=6,8,\dots,14$, and the black solid line is our power law extrapolation. The bounds are obtained at fixed spin up to $L=34$ for the highest $N$. Unlike the numerics in Fig.~\ref{butterfly_plot} the bound in this case is more sensitive to the higher spins on the upper boundary. The dashed red line is the prediction from one-loop $g_0\phi^4$ theory. Black dashed vertical lines represent the coupling $|g_0|=2.4$ we input. }
\label{M1_bound}
\end{figure}

By imposing the inequality \reef{M1_higherd} for a fixed number of modes $n$ we make sure that the contributions  to the $c_i$'s coming from the IR physics are small, and  proportional to $g_0^2$.
The constraint depends on two phenomenological parameters the coupling $g_0$ and the cutoff $\Lambda$.
We take   $g_0=4! \times 0.1$ and $\Lambda^2=(8m)^2$. This value of the cutoff is  quite large because the EFT corrections are controlled by powers of $m^4/\Lambda^4=1/64^2$. We will comment on the cutoff dependence on sections~\ref{ddbounds} and \ref{medio}.~\footnote{For a loose analogy with Higgs physics, recall that the ratio $m_h/1~\text{TeV}\approx 1/8$ and the quartic $\lambda|H|^4$ is $\lambda\approx0.125$.}

In Fig. \ref{M1_bound} we plot the allowed region in the $(c_0,c_2)$ plane for different values of $N=6,\dots, 14$ with spin up to $L=34$ imposing 
only the constraint $\Delta a_2^\text{IR}$.  In black we show an extrapolation done with a power law fit with three free parameters. 

There are two cusps $A^\prime$ and $B^\prime$ corresponding respectively to the maximum and minimum allowed value of  $c_0$.
These values turn out to be approximately equal to the  effective coupling imposed in the arc constraint $g_0=2.4$.
 The point $D$ corresponds to the free theory, which is allowed by our constraints. 
 
 The lower part of the boundary corresponding to the arc $B^\prime D A^\prime$ converges fast with $N$, and  coincides with the lower bound in Fig.~\ref{butterfly_plot}. 
 As we will show  momentarily the arc $DA^\prime$  corresponds to  the EFT in \reef{lagone} with  a weakly coupled UV completion. 
 Instead, the segment $B^\prime D$ involves negative quartic coupling. Therefore the actual cutoff of the  EFT description in the $B^\prime D$ region might be smaller than $\Lambda$.~\footnote{
 For a field theory   intuition, consider  the potential $V(\phi){=} \phi^2(m^2 {  -} |\lambda_4|\phi^2 {+ }|\lambda_6|/\phi^4 /\Lambda^2{ +}\dots)$.  
 Taking  the  scaling $\lambda_6, \lambda_4^2{=}O(\delta)$, with $\Lambda^2{=}m^2$,  the  $O(\lambda_6)$ corrections to   $\text{Im}M \propto \lambda_4^2 + O(\delta^2)$ are negligible yet the $\phi{=}0$ vacua is stable. }
  On the other hand, the upper part of the boundary, the arc $A^\prime B^\prime$, 
  sets the maximal allowed values for $c_2$.
  
In order to characterise the theories in the boundary we introduce the following ratio
\be
r\equiv c_2^\text{UV}/c_2 \, ,  \label{ratio}
\ee
where 
\be
c_2^\text{UV}=\frac{1}{\pi}\int_{\Lambda^2}^\infty \frac{\im M(s,t=4/3)}{(s-4/3)^3}ds \, .
\ee
Theories that are weakly  coupled in the IR and strongly coupled in the UV have $r\approx 1$.
The opposite however is not true: $r$ may be arbitrarily close to $1$ for a weakly coupled UV completion as long as the contribution of  IR physics into $c_2$ is subdominant with respect to $c_2^\text{UV}$.

\begin{figure}[t] 
\centering 
        \includegraphics[width=1 
       \textwidth]{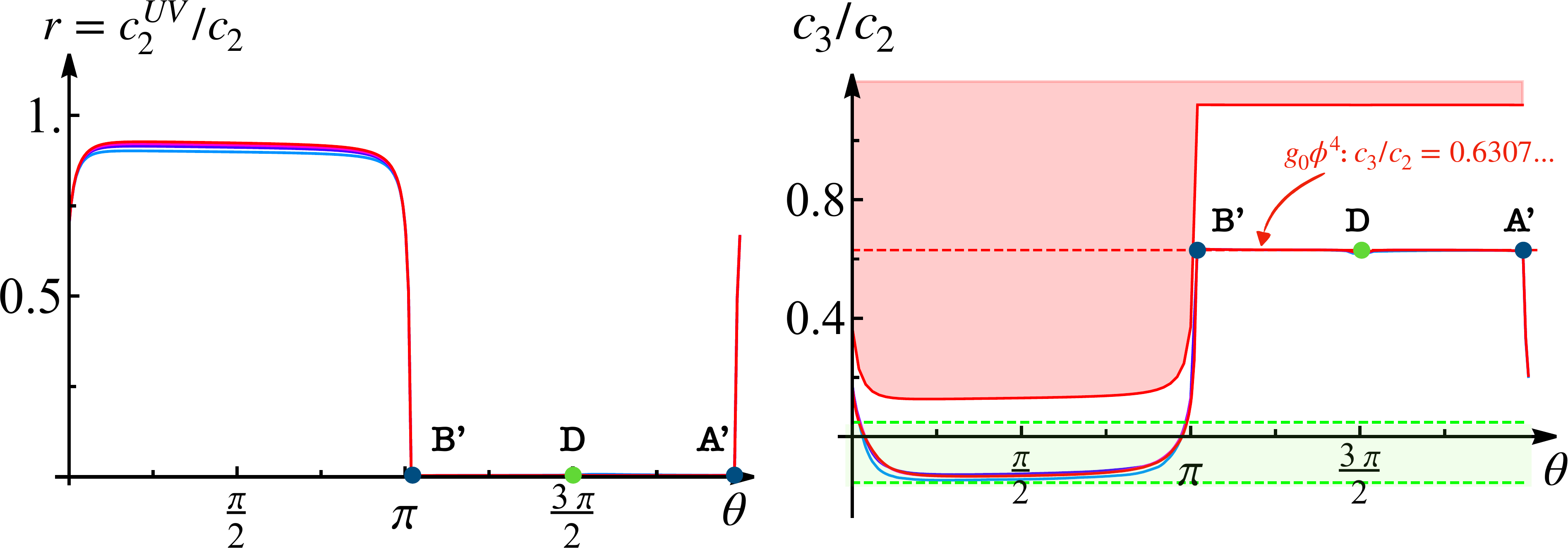} 
        \caption{ Value of $r$ (left) and $c_3/c_2$ (right) along the boundary of  Fig.~\ref{M1_bound}. Different colours denote different values of $N$ ranging from $N=8$ in light blue, to $N=14$ in red.  On the right, the green region determined by the green dashed lines represent the region allowed by the tree-level positivity constraints. The red dashed line is the value of the ratio $c_3/c_2$ in the $g_0\phi^4$ theory. The red excluded region has been obtained using the UV dominance ratio extracted from the left figure using eq.~\eqref{ic3vsc2}.}
\label{M1_ratios}
\end{figure}

In  Fig.~\ref{M1_ratios} on the left, we plot the value of $r$
along the boundary of Fig~\ref{M1_bound}, i.e. as a function of the  angle $\theta$ we used to determine the boundary of the allowed region 
of Fig~\ref{M1_bound}.  Different lines correspond to  different values of   $N=6,\dots, 14$, showing that this observable is nicely converged along the boundary. 
This observable presents an abrupt transition at the edges $A^\prime$ and  $B^\prime$, and shows that the upper 
branch of the boundary in  Fig~\ref{M1_bound} is UV dominated ($r\approx 1$) while the lower one is IR dominated ($r\approx0$).

In the limit of exact  UV domination  $r=1$, the 
  unitary cut along   $s\in[4,\Lambda^2]$ can be neglected. 
In this region  we can get an approximate estimate of the maximal $c_2$ measured at the cutoff scale simply by rescaling the plot in Fig.~\ref{butterfly_plot}, i.e. by reinterpreting $m^2$ as $\Lambda^2$ in that result. There we found $c_2 \lesssim 40$ for $|c_0|/(32\pi) \ll 1$,
which roughly coincides with  the maximal value of $c_2$ in Fig.~\ref{M1_bound}.
We remark however, that the amplitudes in Fig.~\ref{M1_bound} satisfy unitary in the whole range $s\in[4,\infty]$.

In Fig.~\ref{M1_ratios} on the right, we plot the ratio $c_3/c_2$ as function of the radial angle $\theta$, i.e. its value along the boundary of Fig~\ref{M1_bound}. 
The ratio $c_3/c_2$ is well inside the bounds  in \reef{c3c2ratiobounds}.
We remark that here we are not minimizing (or maximizing) the value of   $c_3/c_2$, but only providing a consistency check by showing that  the ratio $c_3/c_2$ is within the rigorous bounds in \reef{c3c2ratiobounds}. 
The plot shows two main behaviours of the amplitude depending on whether the value of $c_2$ is UV (which happens  for $\th$ approximately within the range $[0,\pi]$) or  $c_2$  is IR dominated  (i.e. $\th$ roughly in the range $[\pi,2\pi]$). 

In the IR dominated region, 
 we find that the  ratio is given by
 $
 c_3/c_2 \approx 0.6307
 $.
This value can be explained by a simple Effective Field Theory calculation. 
Consider a model with a light scalar with mass $m=1$, a heavy scalar with mass $M$, and a potential
\be
V(\phi,\Phi) =   \phi^2 (1/2+\eps M \Phi + \alpha \eps^2 \Phi^2) + M^2/2 \, \Phi^2+ \beta \eps^4 / 4! \, \phi^4 \, , \label{pot}
\ee
where we take $\alpha,\beta$ as $O(1)$ numbers, and $g_0\equiv \beta \eps^4$. 
One can readily compute the value of $c_3$ and $c_2$ from the diagrams
\be
\diagIRone \quad \quad  +\quad 
\quad 
\diagUVtwo \quad \quad  + \quad 
\quad \cdots 
  \label{diags}
\ee
 by expanding each diagram in powers of momenta around the crossing-symmetric point $(s,t,u)=(4/3,4/3,4/3)$ and matching the result into the expression   \reef{lowamp4dnoflav}.
Dashed lines indicate the propagation of the light scalar, while solid thicker lines the propagation of the heavier scalar. 
The first diagram represents the leading IR contribution, and the second  diagram is a   contribution from the heavy scalar. 
Next we will  discuss two interesting limits of this simple computation, 
 $1/M\rightarrow 0$ and $1/M\rightarrow 2$.
In the limit where one decouples the heavy physics  $1/M\rightarrow 0$  we get
\be
c_3/c_2= \frac{390-459 \sqrt{2} \cot ^{-1}\sqrt{2}}{-224+240 \sqrt{2} \cot ^{-1}\sqrt{2}} \approx 0.6307 \label{phi4ratio}
\ee
from the first diagram in \reef{diags}. This IR contribution    nicely matches the right plateau of the right plot in Fig.~\ref{M1_ratios},
something which  perhaps is not too surprising because   that region is IR dominated and can admit a weakly coupled EFT description.
In the second  limit we take the heavy particle as light as possible  $M\rightarrow 2$, i.e. at threshold $M^2=(2m)^2$ for $m=1$.~\footnote{
For values lighter than $M=2$ the second diagram would induce a pole in the $2\rightarrow 2 $ scattering of the light particle, and thus one needs to modify   the dispersion relations (\ref{c0sr}-\ref{c3sr}) accordingly.} 
We get
\be
c_3/c_2= 9/8+O(\eps^2) 
\ee
which  saturates the positivity bound \reef{c3c2ratiobounds} as $\eps\rightarrow 0$.~\footnote{Note that  as long as $\alpha>M^2/2$ the point $(\phi,\Phi)=(0,0)$ is the global minimum of $V$.} The value $9/8$ arises from the second diagram in  \reef{diags}. It is easy to understand why this theory saturates the bound:
the first inequality of \reef{c3vsc2} is saturated because the second term in the integrand of \reef{c3sr}  vanishes;
and the second inequality of  \reef{c3vsc2} is saturated because the discontinuity of the diagram is a delta function peaked at threshold $s=4$.~\footnote{A similar phenomena has been observed in ref.~\cite{Caron-Huot:2020cmc} and in other uses of positivity bounds. Positive moment constraints are saturated by integrating at tree-level heavy states at threshold~\cite{Bellazzini:2020cot}. } 

It is possible to derive an improved positivity bound taking into account the 
UV and IR contributions to the dispersion relation of $c_2$,
\be
\frac{c_3}{c_2}\leq     \frac{9}{8} (1-r)    +\frac{3}{\Lambda^2-4/3} r \label{ic3vsc2} \, . 
\ee
This is an optimal inequality, it is a weighted average of  IR and UV  threshold contributions,  and  is derived following similar steps to the derivation of \reef{c3vsc2}.
The upper bound \reef{ic3vsc2} is shown by the shaded red region in Fig.~\ref{M1_ratios} on the right, by taking $r$ from the left plot.
The amplitude that we get with the {\bf M1} method nicely agrees with the bound throughout. 
In the region where the value of $c_2$ is UV  dominated the IR cuts can be neglected,  and one has
\be
 -0.1564  \leq c_3/c_2 \leq  9/188 \approx 0.0479  \, , \label{chbound}
\ee
 after setting $\Lambda=(8m)^2$, and $m=1$. 
 The bound is  shown by the shaded green region in Fig.~\ref{M1_ratios}. Our amplitudes in the  UV dominated  closely saturate the lower bound \reef{chbound}.

The notion of UV domination is quite general, and does not necessarily apply to field theories with a weakly coupled description in the IR. 
We shall therefore verify that the theories that saturate the bounds in Fig.~\ref{M1_bound} do behave as EFTs.

\begin{figure}[t] 
\centering 
        \includegraphics[width=0.9
       \textwidth]{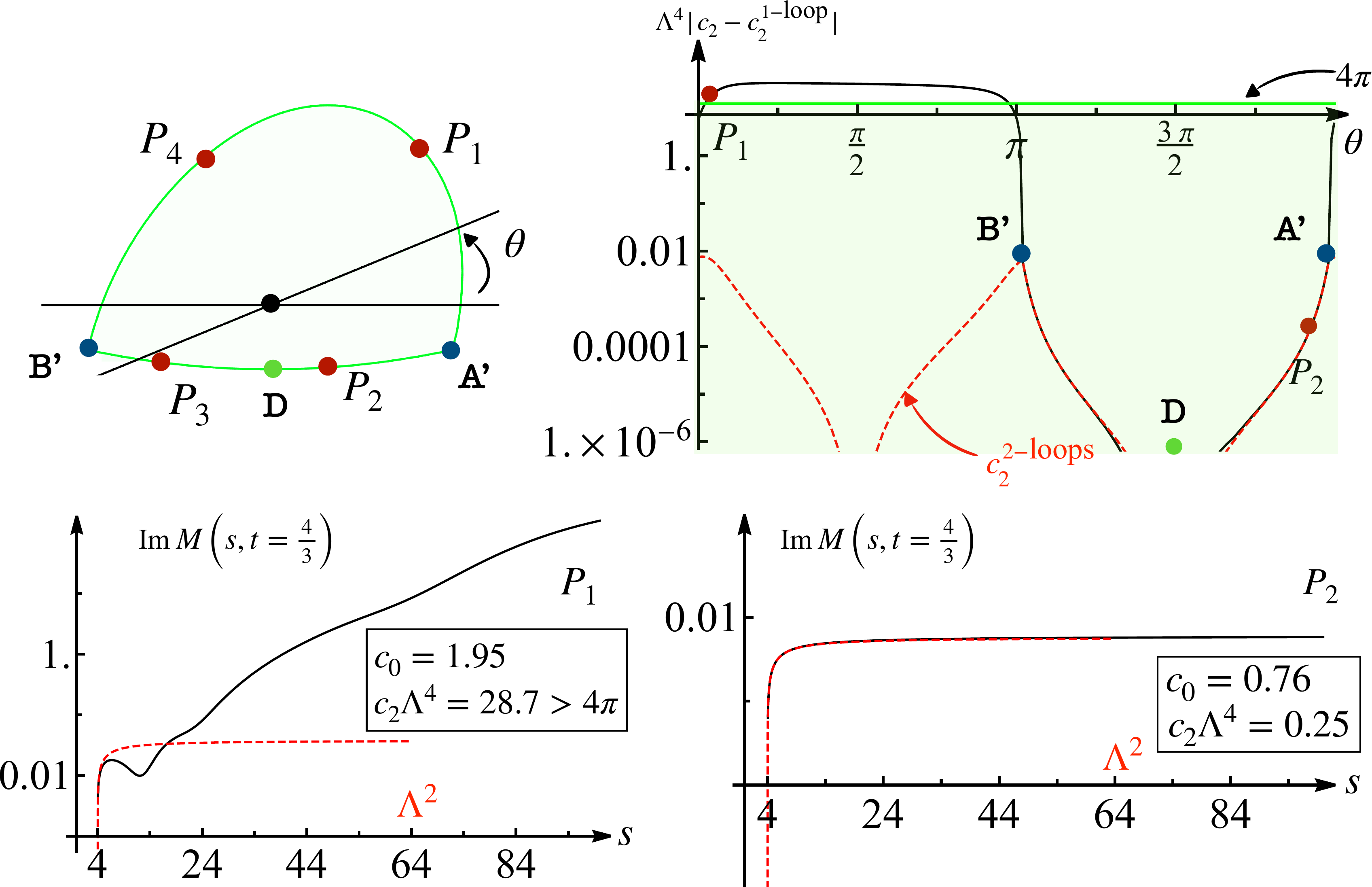} 
        \caption{Two bottom plots: $\im M (s,4/3)$ for the points $P_1$ and $P_2$ of Fig.~\ref{M1_bound} indicated in the upper left inset. 
        In the top right corner the difference $\Lambda^4 |c_2-c_2^\text{1-loop}|$ value along the boundary (solid black line), two-loops perturbation theory (red dashed line) and naive dimensional analysis (solid green line and the green region below it).}
\label{M1_plots}
\end{figure}

In the upper right plot of  Fig.~\ref{M1_plots}  we compute the difference $\Lambda^4 |c_2-c_2^{1-\text{loop}}|$ as we move along the boundary of the {\bf M1} region.
$c_2^{1-\text{loop}}$ is given in eq.~\eqref{c2_1loop}, and comes from the one-loop contribution to $c_2$ in the $g_0 \phi^4$ theory where we extract $g_0$ from the amplitude using \eqref{appr}. 
The green line is the naive dimensional analysis upper bound $4\pi$ on the UV contributions to $c_2$. 
We observe that in the UV dominance region, this bound is violated by a factor 2 at least, suggesting that there might be other non-perturbative contributions to $c_2$. 
The dashed red line is the two-loop contribution to $c_2$ given in   \eqref{c2_2loop}.
In the IR dominance region, after subtracting the one-loop contribution, what remains 
matches almost perfectly the two-loops curve, indicating that the UV terms are negligible and that the amplitudes come from a weakly coupled UV theory.

In virtue of the above discussion, we conclude this section by comparing 
the  imaginary part of the amplitude $\im M(s,t=4/3)$ for  different sample points along the boundary -- the two bottom plots in Fig.~\ref{M1_plots}. 
We  extract the quartic coupling by measuring the amplitude at the crossing symmetric point, and then use it to plot the red  dashed curves using the one-loop approximation \reef{oneloopapp}. 
We observe that indeed for the points $P_2$ or $P_3$ laying in the IR dominance region $B^\prime D A^\prime$, 
the one-loop approximation is valid up to the cutoff $\Lambda^2=64$, and even beyond that. 
Instead for the points $P_1$ or $P_4$, the one-loop approximation is valid in a smaller region well below the cutoff, showing that these theories receive still important contributions from higher derivative operators, and therefore our upper bound is too loose and could be improved.
In principle this can be done by adding further constraints $\Delta a_{n\geq 4}^{IR}$.
However we will not pursue further this route in the present work and instead next we switch to the more constraining   method {\bf M2}.

\subsection{M2: bounding the low energy ansatz}
\label{secm2}
The other possibility is to drastically approximate the amplitude in the IR in order to match the expected EFT behavior.
As we will see momentarily this approximation has to be taken with care.

Dispersion relations encode in a compact way the analyticity and crossing properties of the amplitude. 
The Mandelstam representation -- see  equation \reef{dispHDfinal}  in appendix~\ref{manrep}  -- determines the scattering amplitude everywhere in the maximal analyticity domain once we specify its single and double discontinuities, and eventually few subtraction constants i.e. $M[c_0,\sigma(s),\rho(s,t)]$.
In Sec. \ref{sec:full_space} we let these quantities vary arbitrarily constrained solely by unitarity.
Here, we discuss the possibility of bounding the single and double discontinuity in the IR, according to the predictions of the EFT we are interested in.
In particular, these new constraints will define a region in the space of the coefficients $\{c_0, c_2\}$, and we will provide evidence that our prescription nails down precisely the space of EFTs.

\begin{figure}[t] 
\centering 
        \includegraphics[width=0.57
       \textwidth]{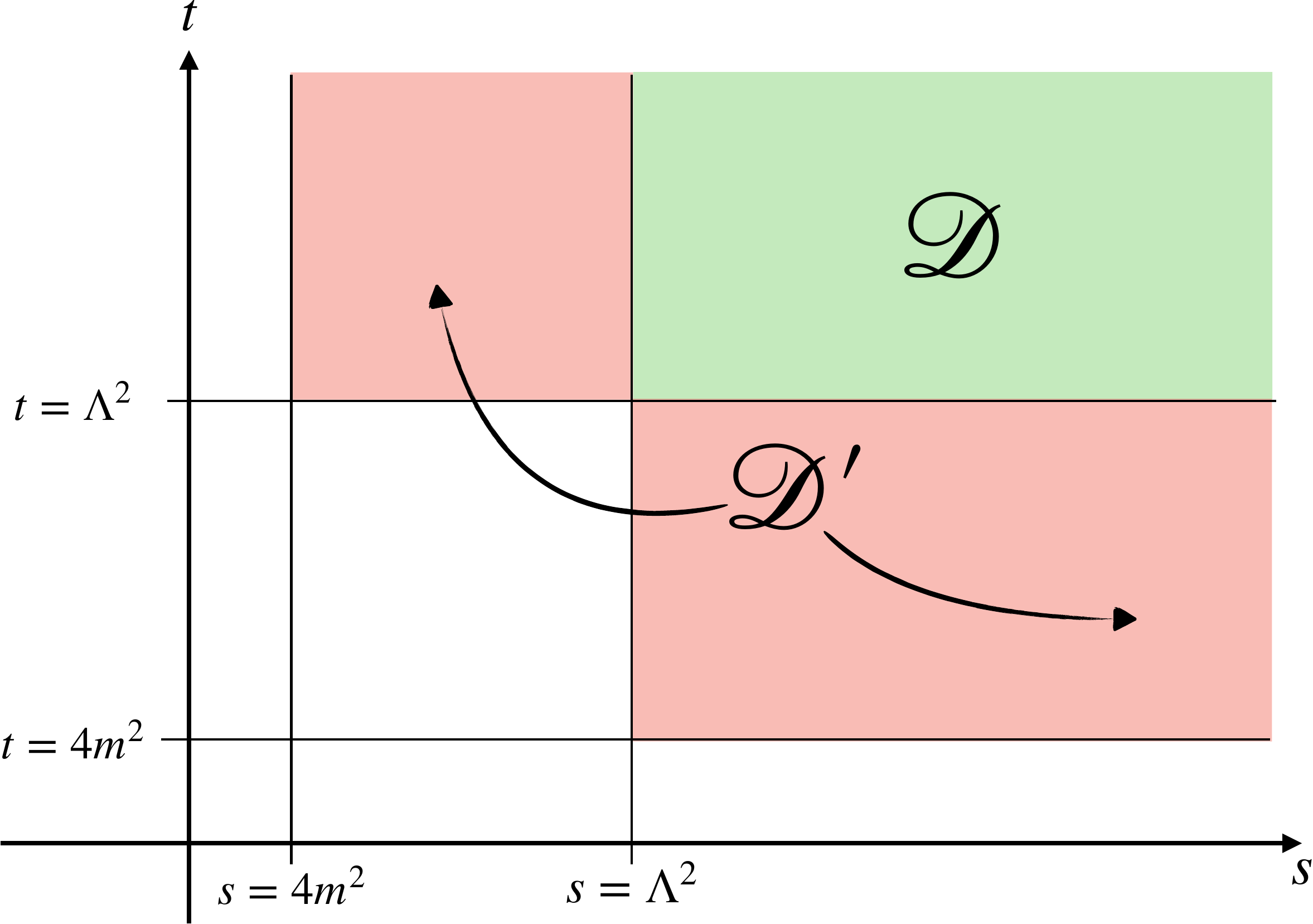} 
        \caption{ Domain of the double-discontinuity in the $\{s,t\}$ plane in the one-loop approximation. In the region $s,t<\Lambda^2$ we can approximate it by setting it to zero. The region $\mathcal{D}$ depicted in green is the domain used in our numerics. In red we denote the domain $\mathcal{D}^\prime$ that contributes to the higher spins in the IR region. }
\label{ddisc_plane}
\end{figure}

To make contact with perturbation theory, first we observe that the single discontinuity $\sigma(s)$ of the amplitude can be expressed in terms of the spin-zero imaginary part $\im f_0(s)$, and the double discontinuity, see \eqref{single_disc}.
We can straightforwardly compute the spin-zero imaginary in perturbation theory using the $g_0 \phi^4/4!$ interaction Lagrangian, and use it to bound the spin-zero projection of our numerical ansatz in the IR
\be
\im f^\text{ans}_0(s)\leq \frac{1}{2}\sqrt{\frac{s-4m^2}{s}}\left(\frac{g_0}{16\pi}\right)^2, \quad 4m^2<s<\Lambda^2 \, . 
\label{M2_constraint}
\ee
At the one-loop level, the imaginary part of the amplitude contains only the spin-zero contribution.
Using this fact and eq.~\eqref{higher_spins_equation} 
\be
\im f_{\ell \geq 2}(s)=\frac{1}{4 (s-4m^2)}\int_{}^\infty \rho(s,y) Q_\ell(1+\tfrac{2y}{s-4m^2})dy=0 \, , 
\label{zero_higher_spins}
\ee
we obtain a condition on the double discontinuity in the IR that can be automatically satisfied by setting $\rho^\text{ans}(s,t)=0$ for any fixed $t>4m^2$ and $4m^2<s<\Lambda^2$.
Using crossing symmetry, therefore, we shall consider the restricted support $\mathcal{D}= [\Lambda^2,\infty)\times  [\Lambda^2,\infty)$  for the double discontinuity-- see the green region in Fig.~\ref{ddisc_plane}.

\begin{figure}[t] 
\centering 
        \includegraphics[width=1
       \textwidth]{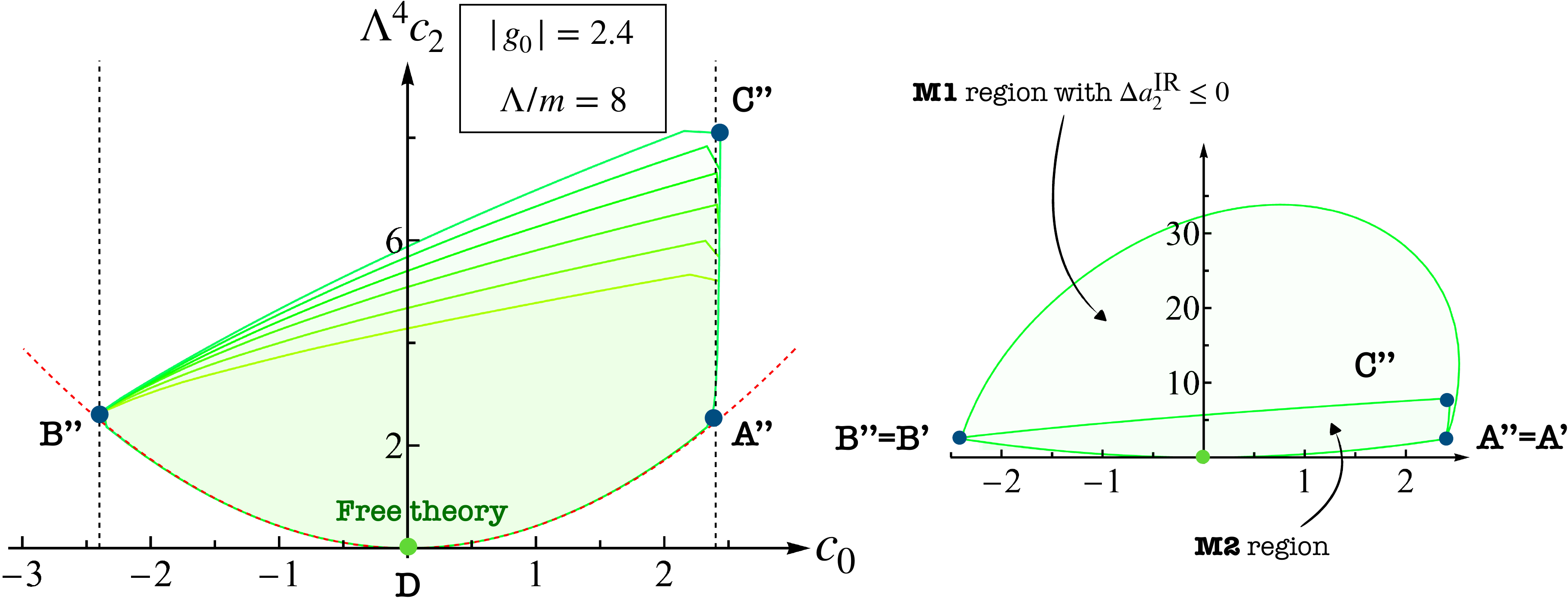} 
        \caption{On the left, allowed values for $(c_0,c_2)$, using the method {\bf M2}. Different green lines correspond to different values of $N$ ranging from $N=6$ to $N=16$; we keep $P=10$ and $Q=3$ fixed. For the highest $N$ the spin cutoff is $L=30$. On the right, comparison between the {\bf M1} and the {\bf M2} regions using the best numerics we have. }
\label{M2_bound}
\end{figure}

Of course, this is not the only option. At fixed $s$ in the IR and $t>\Lambda^2$, in principle, we have no right to set the double discontinuity to zero.
In particular, the double discontinuity in that region is not positive~\footnote{Indeed, there exists a region, called Mahoux region, where the double discontinuity is actually positive. The integral in \eqref{zero_higher_spins}, however, has support over the whole domain of the double discontinuity and it is not enough to mathematically justify our assumption. } and the equality \reef{zero_higher_spins} might still hold for all spins by requiring some nontrivial cancellations.
Therefore, another option is to turn on the double discontinuity on the domain $\mathcal{D}^\prime=[4m^2,\Lambda^2)\times [\Lambda^2,\infty) \cup [\Lambda^2,\infty) \times [4m^2,\Lambda^2)$ -- see red region in Fig.~\ref{ddisc_plane}, and add the additional constraints $\im f_{\ell \geq 2}\leq \max\{ \frac{m^4}{\Lambda^4},\lambda^3\}$. In this paper we will not consider this option as we will see that the allowed region for the coefficients $c_i$ obtained using the domain $\mathcal{D}$ is compatible with the EFT region defined by the arc bounds in Fig.~\ref{M1_bound}.

The ansatz we use to solve the bootstrap problem in this case is a mixed ansatz including both dispersive and power series terms.
In fact, since we effectively introduce a separation between the IR and the UV we think that is convenient to parametrize the two regions differently.
We use the dispersive part to parametrize the single discontinuity of the amplitude.
We choose a basis of Chebyschev polynomials for $4m^2<s<\Lambda^2$, and a simple power law series for the UV region $s>\Lambda^2$
\be
\frac{1}{\pi}\sigma^\text{disp}(s)=\theta((s-4)(\Lambda^2-s))\sqrt{s-4}\sum_{p=1}^P \delta^{IR}_p T_p(\tfrac{2s-\Lambda^2-4}{\Lambda^2-4})+\theta(s-\Lambda^2) \sum_{i=1}^Q \delta_i^{UV} s^{-i/2},
\label{single_disc}
\ee
where $T_p(x)$ are the Chebyshev polynomials.
The set of coefficients $\{ \delta_p^{IR}\}$ and $\{\delta_i^{UV}\}$ are subject to the constraint that $\sigma(s)$ is continuous at $s=\Lambda^2$.
Plugging the ansatz~\eqref{single_disc} into the dispersion relation in eq. \eqref{Mandelstam_rep} we obtain a crossing symmetric analytic continuation
\be
M^\text{disp}(s,t,u)=\int_4^\infty dx \, \sigma^\text{disp}(x)\, \left(\frac{s-s_0}{(x-s)(x-s_0)}+\frac{t-t_0}{(x-t)(x-t_0)}+\frac{u-u_0}{(x-u)(x-u_0)}\right).
\ee
Next we introduce an ansatz for the UV region that has both single and double discontinuity on the domain $\mathcal{D}$.
We introduce the generalized $\rho$ variable
\be
\rho_s(\Lambda,\sigma)=\frac{\sqrt{\sigma-\Lambda^2}-\sqrt{\Lambda^2-s}}{\sqrt{\sigma-\Lambda^2}+\sqrt{\Lambda^2-s}},
\ee
and write down the ansatz in  \eqref{wavelet_ansatz} where we replace $\rho_s(\sigma)\rightarrow \rho_s(\Lambda,\sigma)$.
The whole amplitude will be obtained by summing this power series ansatz and $M^\text{disp}(s,t,u)$.

\begin{figure}[t] 
\centering 
        \includegraphics[width=1
       \textwidth]{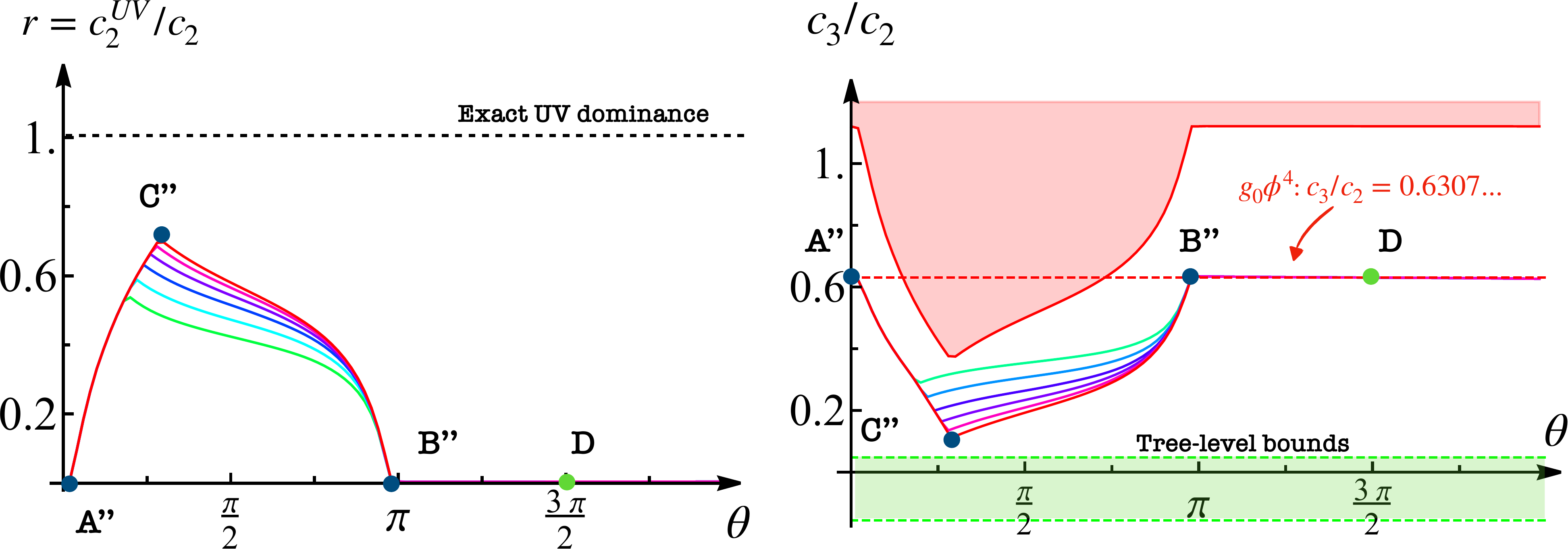} 
        \caption{ Value of $r$ (left) and $c_3/c_2$ (right) along the boundary of  Fig.~\ref{M2_bound}. Different colors correspond to different $N$ ranging from $N=6$ (green) to $N=16$ (red). See Fig.~\ref{M1_ratios} for a comparison and a detailed description.}
\label{M2_ratio}
\end{figure}

In the left plot of Fig. \ref{M2_bound} we present the bound obtained by solving the problem~\eqref{routine3}.
We show the allowed region in the $(c_0, c_2)$ plane for different values of $N=6,8,\dots 16$ and maximum spin cutoff  $L=30$. As for the previous case, the coupling $c_0$ is compatible with the perturbative coupling we input with the EFT constraint \reef{M2_constraint}. We identify four special points: the points $A''$ and $B''$ corresponding respectively to the maximum and minimum coupling amplitude, the edge $A''C''$ has a nearly degenerate coupling fixed at the maximum value $c_0 \sim g_0^{\text{input}}$, and the point $D$ is the free theory.

In the right plot of Fig.~\ref{M2_bound} we compare the two regions {\bf M2} and {\bf M1}.
As expected, {\bf M2} $\subset$ {\bf M1}, and the edge $B''DA''$ and $B'DA'$ of the two regions coincide.
This is yet another check that the amplitudes saturating the lower bound in both regions are insensitive to the presence of the double discontinuity in the IR, and that the higher spins approximately vanish $f_{\ell\geq 2}\sim 0$.

In the left plot of Fig. \ref{M2_ratio} we plot the UV dominance ratio $c_2^{UV}/c_2$ and on the right the ratio $c_3/c_2$  as a function of the radial angle $\theta$ used to parametrize the boundary. 
As in Fig.~\ref{M1_ratios} for the {\bf M1} constraints, along the lower edge we observe IR domination.
On the upper edge $A''C''B''$ there is UV dominance with maximum at the cusp $C''$, but the ratio $r$ is not close to one.
In turn, this implies that the tree-level bounds are always violated along the boundary of the EFT region -- the green strip in Fig.~\ref{M1_ratios}.
Moreover, our bound implies that along the boundary of the EFT region the coefficient of the dimension ten operator associated to $c_3 >0$ is always positive. 
We shall optimize the coefficient $c_3$ to make sure this holds in the EFT region, but it is interesting to find such non-trivial positivity to hold for this class of amplitudes.~\footnote{Looking at the dispersion \eqref{c3sr}, it is easy to see that in the IR the $c_3$ is always positive if we can neglect higher contributions than the one-loop to EFT amplitude. Only in the UV we have negative contributions that seems to not be able to change sign of the result.}

\begin{figure}[t] 
\centering 
        \includegraphics[width=1.0
       \textwidth]{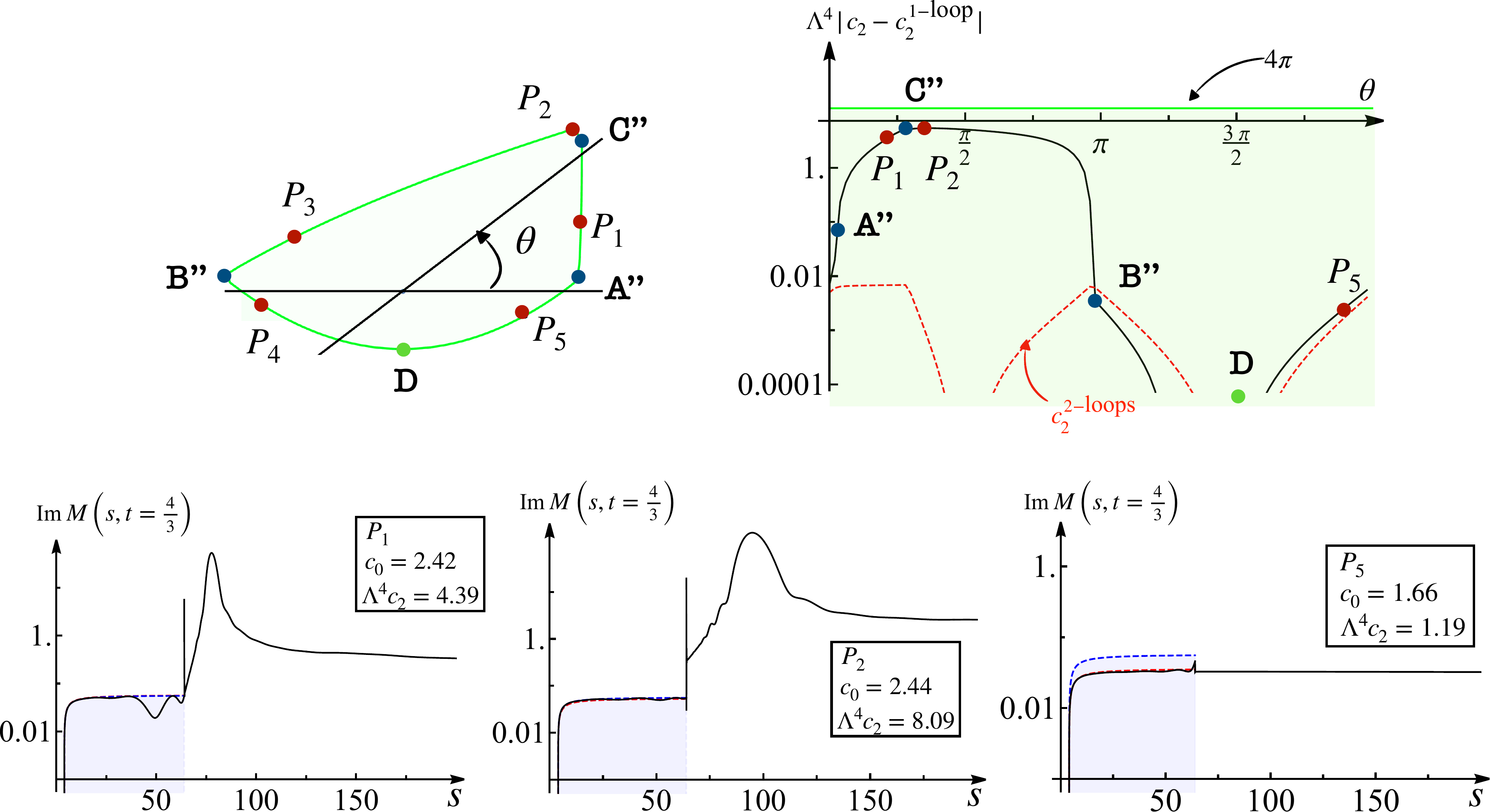} 
        \caption{ Lower three plots: $\im M (s,4/3)$ for the points $P_1$, $P_2$, and $P_5$ of Fig.~\ref{M2_bound} indicated in the top left inset. 
        Comparison of the $c_2$ value along the boundary against perturbation theory and naive dimensional analysis. }
\label{M2_plots}
\end{figure}

Lastly, we analyse the behaviour of the amplitude for different sample points along the boundary scattered respectively along the IR and UV dominance regions in Fig.~\ref{M2_plots}, top left plot.
In the right top plot we evaluate the quantity $\Lambda^4|c_2-c_1^{1-\text{loop}}|$ as a function of the radial angle $\theta$ that parametrizes the boundary (solid black line). As done in the previous section we compare this quantity with both the naive dimensional analysis bound $4\pi$ and the $c_2^{2-\text{loops}}$. Unlike the {\bf M1} region, here in the UV dominance region, the UV contributions to $c_2$ are below the $4\pi$ estimate, while in the IR dominance edge the agreement with the two-loops expectation (dashed red line) is good, especially for $c_0\geq 0$.

In the bottom plots we show the imaginary part of the amplitude at $t=4/3$ (solid-back) for three sample points along the boundary.
The dashed blue line and the depicted blue region represent the allowed values for the amplitude compatible with the constraint~\eqref{M2_constraint}.
All numerical amplitudes for $4<s<\Lambda^2\equiv 64^2$ must lie inside the blue region.
In the same plots we add the one-loop imaginary part computed using the coupling extracted from the numerical amplitude. For EFT amplitudes we should expect to follow the red dashed line up to the cutoff scale $\Lambda^2$. 

For the three sample points shown we find that the quartic extracted at the crossing symmetric point $c_0$ and the effective coupling entering the imaginary part (that we can see from the non-perturbative imaginary part of amplitude at low energies) agree well. For the point $P_1$ we observe the EFT seems to break at energies slightly lower than the cutoff. 
For the point $P_5$ we observe a slight difference between the $c_0$ we measure at the crossing symmetric point and the profile of the imaginary part that saturates the bound \eqref{M2_constraint}. The difference however is compatible with the threshold unitarity inequality $g_0^2 \geq c_0^2$, but there is no saturation. This means that most of the amplitudes along the upper branch are not perfectly elastic below $s=16m^2$; therefore, these are only approximate EFT amplitudes. 

All the amplitudes shown in Fig.~\ref{M2_plots} present an instability around $s=\Lambda^2$. We suspect this is due to the crude approximation we made of setting to zero the higher spins below the cutoff. We believe that this can be easily fixed, with the introduction of small double discontinuity in the domain $\mathcal{D}^\prime$ as mentioned above. However, we believe the bound and the conclusions will not change, although we plan to investigate this point in more detail in the future.


\section{Dual bounds}
\label{ddbounds}

\subsection{Linearised unitarity bounds}
\label{direct_bounds}

In this section we discuss the relation between the existence of absolute bounds on the Wilson coefficients, and the presence of a gap in the imaginary part.
The bounds derived will use crossing and boundedness of the imaginary part of the partial waves $0\leq \im f_\ell \leq 2/\rho^2(s)$,
where $\rho^2(s) = \sqrt{s-4}/\sqrt{s}$.

We will also show that it is possible to derive dual bounds in presence of IR cuts using additional constraints as the ones introduced in Sec.~\ref{singeft}.

\subsubsection{An analytic dual bound on $c_2$ with a \emph{gap} in the imaginary part }
\label{analytic_c2bound}

First, we derive an analytic expression for the bound in \cite{Caron-Huot:2020cmc} on $c_2$ for the scattering of massless particles in $3+1$ dimensions with the assumption that $\im M(s,0)=0$ for $0<s<\Lambda^2$. Since $\Lambda$ is the only scale in this problem, we set $\Lambda=1$.

The problem to solve can be formulated as follows
\be
\max  c_2 \, , \quad \text{subject to} \quad \mathcal{F}^{(n,m)}=0\, , \quad 0\leq \im f_\ell(s) \leq 2\, ,
\label{c2dual_gapless}
\ee
where
\be
c_2=\frac{1}{\pi}\int_{1}^\infty dz \frac{M_z(z,0)}{z^3}\, , \quad \text{and} \quad \mathcal{F}^{(n,m)}=\int_{1}^\infty dz \sum_{\ell=2}^\infty n_\ell^{(4)}\im f_\ell(z) F^{(n,m)}_\ell(z)\, ,
\ee
where we introduce the notation $n_\ell^{(4)}=16\pi (2\ell+1)$.
The $F^{(n,m)}_\ell(z)$ are computed from \eqref{blocks} taking the limit $m^2\to 0$, and setting $s_0=t_0=0$.
We can dualise this problem by means of the Lagrangian
\be
\mathcal{L}=c_2+\sum_{n,m} \nu_{n,m} \mathcal{F}^{(n,m)}+\sum_{\ell=0}^\infty  \int_{1}^\infty n_\ell^{(4)} \lambda_\ell(s) \im f_\ell(s)ds+\sum_{\ell=0}^\infty \int_{1}^\infty n_\ell^{(4)} \mu_\ell(s) (2-\im f_\ell(s))ds\, ,
\ee
where $\mu_\ell(s) \geq 0$ and $\lambda_\ell(s) \geq 0$ are arbitrary positive functions.
We obtain the dual formulation by  integrating out the primal variables $\text{Im}f_\ell$. After a bit of algebra we are led to 
\be
\begin{aligned}
& \min\, \sum_{\ell=0}^\infty \int_{1}^\infty 2n_\ell^{(4)} \mu_\ell(s)ds\\
\text{subject to} \quad &\frac{1}{\pi s^3} +\lambda_\ell-\mu_\ell + \sum_{n,m}\nu_{n,m} F_\ell^{(n,m)}(s) = 0 \, , 
\end{aligned}
\label{dualone}
\ee
The dual constraints can be recast into the  inequality
$
\frac{1}{\pi s^3} -\mu_\ell(s) + \sum_{n,m}\nu_{n,m} F_\ell^{(n,m)} = -\lambda_\ell(s) \leq 0
$, which in turn implies 
\be  \frac{1}{\pi s^3} + \sum_{n,m}\nu_{n,m} F_\ell^{(n,m)} \leq \mu_\ell(s) \, . 
\ee
The last inequality can be saturated by 
\be
\mu_\ell(s)=\bar \mu_\ell(s)\, \texttt{HeavisideTheta}  [\bar \mu_\ell(s)] \quad \text{where}\quad \bar \mu_\ell= \left(\frac{1}{\pi s^3} + \sum_{n,m}\nu_{n,m} F_\ell^{(n,m)}\right) \, .  
\ee
Next we should  plug this solution in  \reef{dualone} and minimise over the $\nu_{n,m}$.

Consider the problem where we add only the  first non-trivial null constraint
\be
F_\ell^{(1,3)}(s)=\frac{\ell(\ell+1)(\ell^2+\ell-8)}{s^5} \, . 
\ee
It is crucial to notice that $F_2^{(1,3)}<0$, while $F_{\ell \geq 4}^{(1,3)}>0$.
Then note that for  $\nu_{1,3}<0$ we have that   $\bar\mu_2 > 0$ for all $s>1$, while for  $\nu_{1,3}>0$ we have  $\bar\mu_{\ell \geq 4} > 0$ for all $s>1$.
By a simple inspection, the choice $\nu_{1,3}\geq 0$ gives a divergent spin sum in the  dual objective and therefore no bound on $c_2$. Therefore we take $\nu_{1,3}<0$.

For $\ell=0, 2$ we  have
\be
\mu_0(s)=\frac{1}{\pi s^3} \, , \qquad \mu_2(s)=\frac{1}{\pi s^3}+|\nu_{1,3}| \frac{12}{s^5} \, ,
\ee
which are positive in the whole range of integration of the objective in \reef{dualone}.
Instead for $\ell \geq 4$, the quantity 
 $\bar\mu_\ell(s)$
is negative for $1<s<s_0(\ell)$,
where $s_0(\ell)= \sqrt{\pi \ell (\ell+1)(\ell^2+\ell-8) |\nu_{1,3}|}$, and   solves the equation $\bar\mu_\ell(s_0)=0$.
Therefore
\be
\mu_{\ell\geq 4}(s)=\left(\frac{1}{\pi s^3}-|\nu_{1,3}|F_\ell^{(1,3)}(s)\right) \theta(s-s_0(\ell))\, ,
\ee
The contribution of each spin $\ell \geq 4$ to the dual objective is give by
\be
\int_{s_0(\ell)}^\infty \mu_\ell(s)ds=\frac{1}{4\ell(\ell+1)(\ell^2+\ell-8)\pi^2 |\nu_{1,3}|} \, .
\ee

All in all,  the dual objective with the $(1,3)$ constraint is given by
\be
\sum_{\ell=0}^\infty  \int_{1}^\infty n_\ell^{(4)} 2 \mu_\ell(s)ds=96(1-5\pi \nu_{1,3})+\frac{1}{6\pi \nu_{1,3}}\left(6\log{2}-11+3(H_{\tfrac{5}{4}-\tfrac{\sqrt{33}}{4}}+H_{\tfrac{5}{4}+\tfrac{\sqrt{33}}{4}})\right) \, , 
\ee
where $H_n$ are the Harmonic numbers.
Taking the minimum with respect to  $\nu_{1,3}$ yields 
\be
\max{c_2 }\leq \frac{ 8}{\Lambda^4}\left( 12+\sqrt{5\left(6\log{2}-11+3\left(H_{\tfrac{5}{4}-\tfrac{\sqrt{33}}{4}}+H_{\tfrac{5}{4}+\tfrac{\sqrt{33}}{4}}\right)\right)}\right)\approx 0.7937 \frac{(4\pi)^2}{\Lambda^4}
\label{analytic_dual_bound}
\ee
 for $\nu_{1,3}\sim -0.009728 \Lambda^4$, where we have  reintroduced $\Lambda^2$ to stress that the bound is in $\Lambda^2$ units.

In the language of the previous section, this bound is specially relevant for  theories where $c_2$ is UV dominated. 
Next we generalise this derivation by considering a massive particle, 
so that we can compare the dual bounds with the   results that we derived in the previous section.

\subsubsection{A numerical bound on $c_2$ in presence of the mass gap}
\label{insufficiency_mass_gap}

Here we solve the optimization problem discussed in the previous section for the scattering of gapped particles in $3+1$ dimensions.
Likewise we assume a gap in the imaginary part, i.e. $\im M(s,t=4/3)=0$ for $4<s<\Lambda^2$. For this derivation we  work in units   $m^2=1$.
The problem of interest is 
\be
\max \, c_2, \quad \text{subject to} \quad \mathcal{F}^{(n,m)}=0\ , \quad 0\leq \im f_\ell(s) \leq \frac{2}{\rho^2(s)}\, ,
\label{c2dualgapped}
\ee
where
\be
c_2=\frac{1}{\pi}\int_{\Lambda^2}^\infty dz \frac{M_z(z,4/3)}{(z-4/3)^3}\, , \quad \text{and} \quad \mathcal{F}^{(n,m)}=\int_{\Lambda^2}^\infty dz \sum_{\ell=2}^\infty n_\ell^{(4)}\im f_\ell(z) F^{(n,m)}_\ell(z)\, ,
\ee
with $\rho^2(s)=\sqrt{s-4}/\sqrt{s}$ the two-particle phase space.
The problem \eqref{c2dualgapped} is formally identical to \eqref{c2dual_gapless}, therefore we can write down immediately the Lagrangian
\be
\mathcal{L}=\int_{\Lambda^2}^\infty ds \sum_\ell \left[n_\ell^{(4)} \im f_\ell(s)\left(\frac{P_\ell(1+\tfrac{8}{3(s-4)})}{\pi (s-4/3)^3} +\lambda_\ell-\mu_\ell + \sum_{n,m}\nu_{n,m} F_\ell^{(n,m)}(s)  \right) +
  \frac{2}{\rho^2(s)}n_\ell^{(4)} \mu_\ell(s)\right]ds
\label{lagm}
\ee
and the  dual formulation
\be
\begin{aligned}
& \min\, \sum_{\ell=0}^\infty  \int_{\Lambda^2}^\infty \frac{2}{\rho^2(s)}n_\ell^{(4)} \mu_\ell(s)ds, \label{dual_objective_gapped}\\
\text{subject to} \quad &\frac{P_\ell(1+\tfrac{8}{3(s-4)})}{\pi (s-4/3)^3} +\lambda_\ell-\mu_\ell + \sum_{n,m}\nu_{n,m} F_\ell^{(n,m)}(s) = 0.
\end{aligned}
\ee
Once more, the dual problem admits the  formal solution
\be
\mu_\ell(s)=\bar \mu_\ell(s)\, \texttt{HeavisideTheta}  [\bar \mu_\ell(s)] \quad \text{where}\quad \bar \mu_\ell= \left(\frac{P_\ell(1+\tfrac{8}{3(s-4)})}{\pi (s-4/3)^3} + \sum_{n,m}\nu_{n,m} F_\ell^{(n,m)}\right) \, .  
\ee
which we should plug in  \reef{dual_objective_gapped} and minimise over the $\nu_{n,m}$'s.

Consider first the problem with just one crossing constraint $(n,m)=(1,3)$.
The steps we follow are similar to the analytic solution we discussed in section~\ref{analytic_c2bound}.
 For $\ell=2$ we take  $\bar\mu_2(s)>0$ for all $s>\Lambda^2$, which is true when 
$\nu_{1,3} \leq \frac{3\Lambda^4-16}{18 \pi}$. 
Then, for $\ell\geq 4$ we have that $\bar\mu_{\ell}(s)<0$ for $\Lambda^2<s<s_0(\ell)$, and positive   otherwise. 

Unlike the gapless case, $s_0(\ell)$ here is the root of a generic polynomial of degree $\ell$. 
We solve this problem numerically: for each value of $\nu_{1,3}$ we determine the point $s_0(\ell)$, which we then use  to compute the dual objective \eqref{dual_objective_gapped}. The optimal dual bound is obtained by minimizing numerically with respect to  $\nu_{1,3}$ .
The dual objective \eqref{dual_objective_gapped} contains an infinite sum, and in practice we must introduce a numerical cutoff $L$, and minimize the quantity
\be
d^L_{\Lambda^2}(\nu_{1,3}) = \sum_{\ell=0}^L \int_{\Lambda^2}^\infty \frac{2}{\rho^2(s)}n_\ell^{(4)} \bar \mu_\ell[\nu_{1,3}](s)\theta(\bar \mu_\ell[\nu_{1,3}](s))ds,
\label{num_objective}
\ee
with $\bar \mu_\ell[\nu_{13}](s)=\frac{P_\ell(1+\tfrac{8}{3(s-4)})}{\pi (s-4/3)^3} + \sum_{n,m}\nu_{n,m} F_\ell^{(n,m)}$.
To obtain a  dual bound we have to minimize $d^L_{\Lambda^2}(\nu_{1,3})$ for different values of $L$, and then extrapolate.

\begin{figure}[t]
        \centering
        \includegraphics[width=0.8\textwidth]{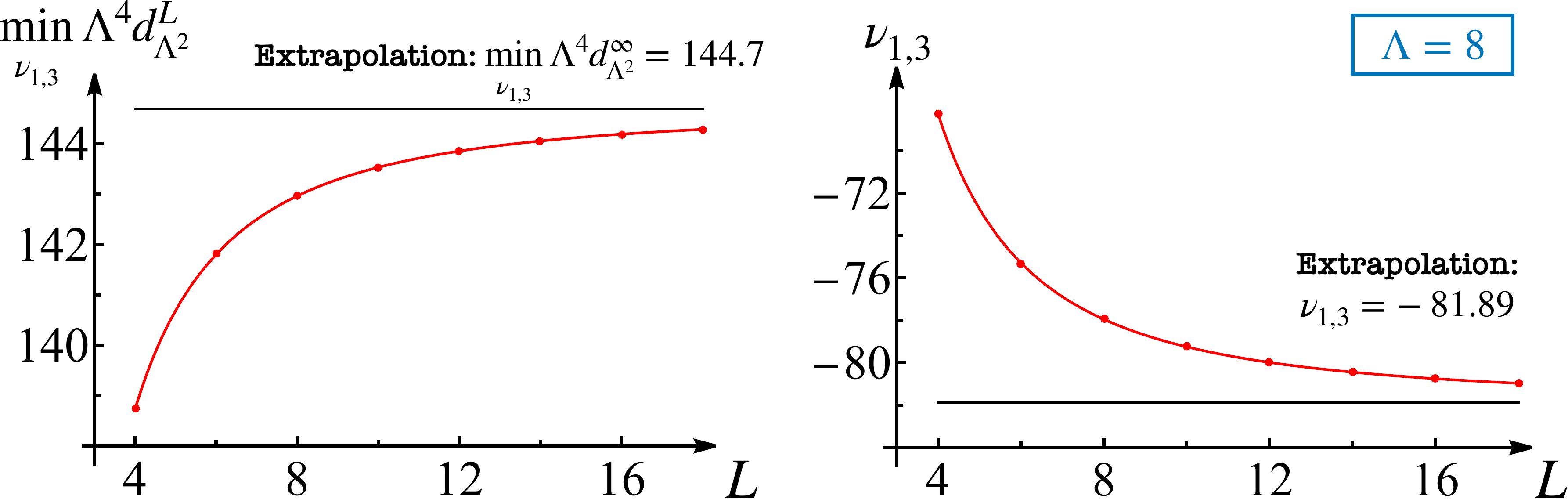}
    \caption{Numerical solution of \reef{dual_objective_gapped} ($\Lambda/m=8$) with a single crossing constraint $(1,3)$ as a function of the spin cutoff (left). 
    Corresponding optimal value  of $\nu_{13}$ as a function of the spin cutoff (right). Black lines represent the extrapolated values for $L\to\infty$.  }
    \label{fig_dual_1}
\end{figure}

In Fig.~\ref{fig_dual_1} we summarize the results of our numerical procedure for $\Lambda=8$.
On the left  we plot the $\min_{\nu_{1,3}}d^L_{\Lambda^2}$ as a function of $L$, and on the right  the corresponding optimal value of $\nu_{1,3}$ (denoted by the red dots).
We extrapolate to $L\to \infty$ using a simple power law ansatz $f(L)=a+\frac{b}{L^c}$ (the solid red line), and report the extrapolated value of the bound $\min_{\nu_{1,3}} d^\infty_{\Lambda^2}$ and the optimal $\nu_{1,3}$ with a black line.
Both quantities show a very smooth behavior in $L$, and the extrapolation is stable.

\begin{figure}[t]
        \centering
        \includegraphics[width=1\textwidth]{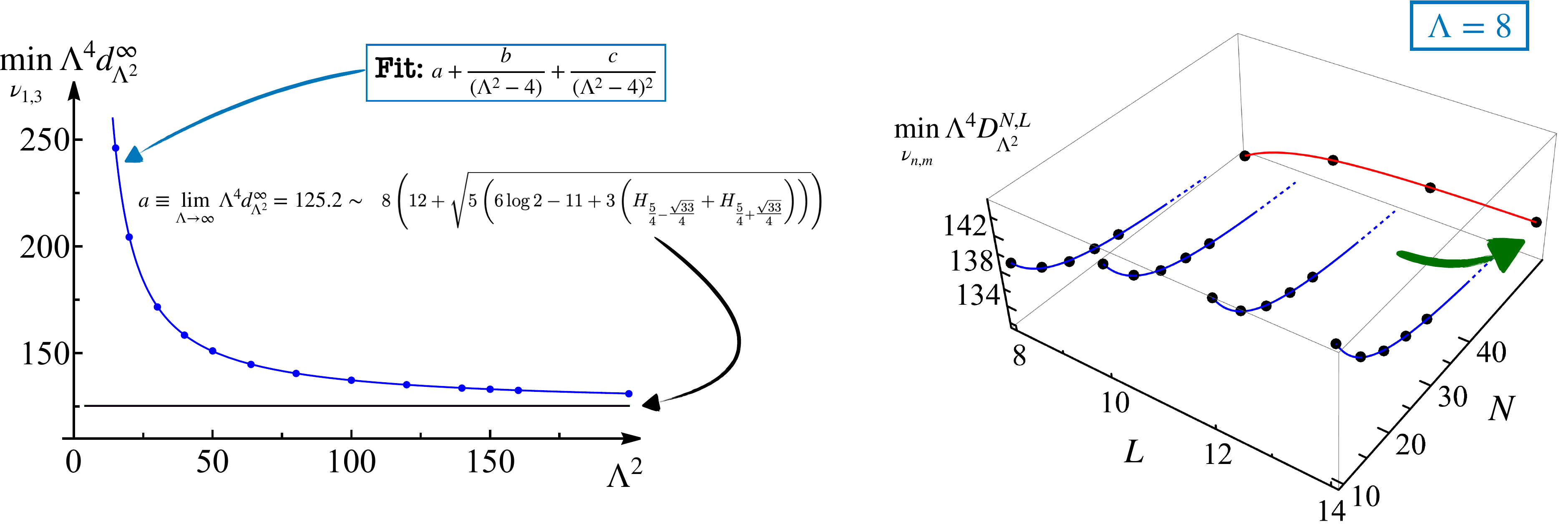}
    \caption{ On the left plot we show the dual bound as a function of the cutoff $\Lambda^2$ with one crossing constraint (blue dots). The solid blue line is our fit of the data. The black line is the extrapolation of the bound for $\Lambda^2\to\infty$ agreeing with the bound obtained in the massless limit at fixed cutoff.
    On the right plot we show the double extrapolation of the relaxed problem in the spin cutoff $L$ and the truncation cutoff $N$ at fixed $\Lambda/m=8$ (the green arrow). The black dots are the data, the blue curves the fit used to extrapolate the data for $N\to\infty$, the red curve is the final extrapolation for $L\to\infty$. }
    \label{fig_dual_2}
\end{figure}

Repeating the same exercise for different values of $\Lambda$, we obtain the left plot in Fig.~\ref{fig_dual_2}.
The data can be nicely fit by a simple function of the form $a+\frac{b}{\Lambda^2-4}+\frac{c}{(\Lambda^2-4)^2}$. 
When $\Lambda/m \to 2$ the dual objective is unbounded: the constraints of linearized unitarity are not strong enough to bound any theory with a cut starting from threshold at $s=4$. 
When $\Lambda/m\to \infty$ we recover precisely the analytic bound in \eqref{analytic_dual_bound}.
This limit is equivalent to sending the mass to zero fixing the cutoff $\Lambda$.
Indeed, in the limit of $m\to 0$ in units of $\Lambda$  the two problems are  equivalent.

We conclude this section by exploring the effect of multiple crossing constraints on the bound on $c_2$.
In principle, we could still try to employ the ansatz method that works for one single constraint. In practice, however, the method is not very stable when we have more than one dual variable: depending on the values of $\nu_{n,m}$ the root $s_0(\ell)$ can become complex, and the value of the function $d_L^{\Lambda^2}(\{\nu_{n,m}\})$ suddenly jumps.

To overcome this issue, we propose a simple \emph{relaxed} version of the original problem, that can be easily solved using linear optimization.
At fixed $L$, we introduce a set of auxiliary variables $x_\ell(s)$ for any $\ell \leq L$, and replace the objective \eqref{num_objective}
with
\be
\begin{aligned}
&D^{L,N}_{\Lambda^2} =\sum_{\ell=0}^L \int_{\Lambda^2}^\infty \frac{2}{\rho^2(s)}n_\ell^{(4)} x_\ell(s)ds,\\
\text{subject to} \quad & x_\ell(s)\geq 0, \quad \text{and} \quad x_\ell(s)-\bar\mu_\ell(s) \geq 0, \quad \text{for}\quad s\geq \Lambda^2,
\end{aligned}
\label{relaxed_D}
\ee
where $N$ is the number of terms in the ansatz for $x_\ell(s)$.~\footnote{We expand $x_\ell(s)$ in a basis of Chebyschev polynomials, and truncate  the basis to degree $N$.}
The new objective satisfies the inequality $D^{L,N}_{\Lambda^2} \geq d^L_{\Lambda^2}$ that is saturated when $x_\ell(s)=\bar\mu_\ell(s)\theta(\bar\mu_\ell(s))$.
To obtain dual rigorous bound we have to minimize $D^{L,N}_{\Lambda^2}$ and perform a double extrapolation -- see right plot in Fig.~\ref{fig_dual_2}: for fixed $L$ we extrapolate in $N$, and only after that we can safely extrapolate in $L$. 
In this example we use all crossing constraints $\mathcal{F}^{(n,m)}$ up to $n+m\leq 7$.

We compare the effect of multiple crossing constraints at $\Lambda=8$
\be
\Lambda^4 c_2 \leq 144.7 \quad n_\text{constraints}=1, \quad \Lambda^4 c_2 \leq 136.3 \quad n_\text{constraints}=7.
\ee
We do not expect the bounds with infinitely many crossing constraints  to change significantly, but it would be worth improving upon our analysis.

\subsection{Dual EFT bounds}

So far in this section, we have assumed a gap in the imaginary part: $\im M(s,4/3)=0$ for $4<s<\Lambda^2$.
Next we will show that  it is possible to obtain dual bounds when we remove the gap, provided that in the IR region we impose additional constraints on the imaginary part. In particular we shall impose the   {\bf M1} and {\bf M2} constraints discussed in section~\ref{singeft}.

\subsubsection{A dual bound on $c_2$ in presence of the M2 constraint}
\label{M2_dual_sec}

\begin{figure}[t]
        \centering
        \includegraphics[width=0.7\textwidth]{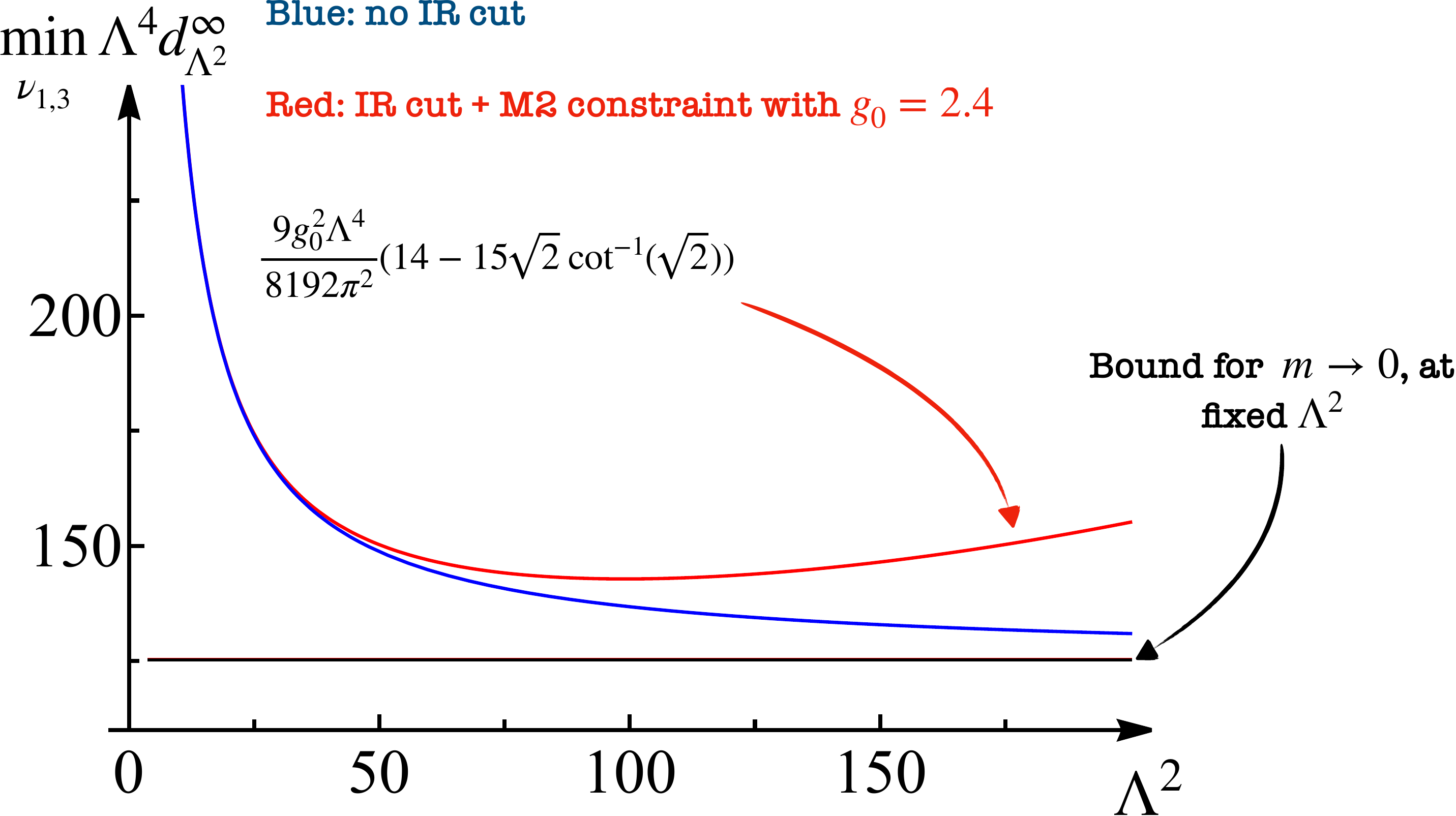}
    \caption{Dual bound $c_2\leq \min_{\nu_{1,3}}\Lambda^4 d_{\Lambda^4}^\infty$ as a function of the cutoff $\Lambda^2$ with the gap in the imaginary part and using linearized unitarity (solid blue). The red solid line represent the dual bound in presence of the IR cut bounded by the {\bf M2} constraint~\eqref{cooo}. The horizontal black line is the bound for the scattering of gapless particles in \eqref{analytic_dual_bound}. }
    \label{fig_dual_3}
\end{figure}

We first solve the problem of maximizing $c_2$ with the extra constraint   on the IR cut given  by the {\bf M2} bounds explained in  section~\ref{secm2},
\be
\im f_0(s)\leq \frac{1}{2}\sqrt{\frac{s-4m^2}{s}}\left(\frac{g_0}{16\pi}\right)^2 \equiv \im f_0^\text{EFT}(s), \quad \im f_{\ell \geq 2}(s)=0,\quad  \text{for} \quad 4m^2<s<\Lambda^2 \, . 
\label{M2_constraints_all}
\ee
The problem is identical to the one we solved in the previous section, except for adding few new terms to the Lagrangian. Namely the  Lagrangian is 
${\cal L}^\text{\bf M2}={\cal L}+\Delta\mathcal{L}^{\textbf{M2}} $ where ${\cal L}$ is given in \reef{lagm} and 
\be
\begin{aligned}
\Delta\mathcal{L}^{\textbf{M2}}&=\frac{1}{\pi}\int_4^{\Lambda^2} \frac{M_z(s,4/3)}{(z-4/3)^3}dz+\int_4^{\Lambda^2} \sigma_0(z)(\im f_0^\text{EFT}(z)-\im f_0(z))dz=\nonumber\\
&=\int_4^{\Lambda^2} \sigma_0(z) \im f_0^\text{EFT}(z)+\int_4^{\Lambda^2} \im f_0(z) \left( \frac{16}{(z-4/3)^3}-\sigma_0(z)\right),
\end{aligned}
\ee
where $\sigma_0(z)\geq 0$.~\footnote{In general one should add the IR cuts for the crossing and unitarity constraints. In our set up however the one-loop bound automatically satisfies linearised unitarity;  and  the crossing constraints are trivially satisfied  $\sum_{\ell=0} n_\ell^{(4)} \im f_\ell(s) F_\ell^{n,m}=0$,    because $F_0^{n,m} ,\, f_{\ell >0}=0$.}

After taking $\sigma_0(z)=16/(z-4/3)^3>0$,  we see that the only effect of {\bf M2} is to add a constant contribution to the dual objective of
the previous problem \reef{dual_objective_gapped}!
All in all, the  solution of this dual problem is given by 
\be
\max  c_2 \leq (\min_{\nu_{1,3}} d_{\Lambda^2}^\infty (\nu_{1,3}))+d_{\Lambda^2}^\textbf{M2}=(\min_{\nu_{1,3}} d_{\Lambda^2}^\infty (\nu_{1,3}))+\int_4^{\Lambda^2}  \frac{16}{(s-4/3)^3}\im f_0^{EFT}(s)ds.
\label{M2_dual_obj}
\ee
In Fig.~\ref{fig_dual_3}, in red we plot the bound \eqref{M2_dual_obj} for $g_0\sim 2.4$, in blue the bound obtained in the previous Section with no IR cut. It is worth noticing that adding the IR cut does not improve the bound obtained with the gap in the imaginary part. 
This is expected, and the larger is the IR contribution the worst the bound becomes. 

At $\Lambda^2=64$, the best dual bound we obtain with linearized unitarity is $c_2 \Lambda^4 \leq 138.8$ to be compared with the primal estimate $\max c_2\Lambda^4  \lesssim 8$.

\subsubsection{A dual bound on $c_2$ in presence of the M1 constraint}

Maximising $c_2$ in presence of the bound {\bf M1} in the IR requires more effort than the corresponding {\bf M2} dual problem.
We consider the constraint in the  Lagrangian form 
\be
\Delta \mathcal{L}^{\textbf{M1}}=-\sum_{n\geq 2} \sigma_n \Delta a_n^\text{IR}=\sum_{n\geq 2} \sigma_n \int_4^{\Lambda^2} \frac{M^\text{EFT}_z(z,4/3)}{(z-4/3)^{n+1}}dz-\sum_{n\geq 2}  \sigma_n \int_4^{\Lambda^2} \frac{M_z(z,4/3)}{(z-4/3)^{n+1}}dz,
\ee
with $\sigma_n \geq 0$. 
The Lagrangian for the full problem then reads
\be
\mathcal{L}^\text{\bf M1}=c_2+\sum_{n,m} \nu_{n,m} \mathcal{F}^{(n,m)}+\sum_{\ell=0}^\infty  \int_{4}^\infty n_\ell^{(4)} \left(\lambda_\ell(s) \im f_\ell(s)+ \mu_\ell(s) \left(\frac{2}{\rho^2(s)}{-}\im f_\ell(s)\right)\right)ds+\Delta\mathcal{L}^{\textbf{M1}}, \label{lagmone}
\ee
where we reintroduce the IR cuts in the following definitions
\be
c_2=\frac{1}{\pi}\int_{4}^\infty dz \frac{M_z(z,4/3)}{(z-4/3)^3}, \quad \text{and} \quad \mathcal{F}^{(n,m)}=\int_{4}^\infty dz \sum_{\ell=2}^\infty n_\ell^{(4)}\im f_\ell(z) F^{(n,m)}_\ell(z).
\ee

Unlike the {\bf M2} case in section~\ref{M2_dual_sec}, the dual problem we derive from \reef{lagmone} has both a modified objective and modified constraints:
\be
\begin{aligned}
& \min\, \sum_{\ell=0}^\infty \int_{4}^\infty \frac{2}{\rho^2(s)}n_\ell^{(4)} \bar\mu_\ell(s) \theta(\bar\mu_\ell(s))ds+\sum_n  \sigma_n  \int_4^{\Lambda^2} \frac{M^\text{EFT}_z(z,4/3)}{(z-4/3)^{n+1}}dz \\
\text{with} \quad &\bar\mu_\ell(s)=\frac{P_\ell(1+\tfrac{8}{3(s-4)})}{\pi (s-4/3)^3} +\sum_{n,m}\nu_{n,m} F_\ell^{(n,m)}(s)- \sum_{n\geq 2} \sigma_n \frac{P_\ell(1+\tfrac{8}{3(s-4)})}{ (s-4/3)^{n+1}}, \quad \text{for}\quad 4<s<\Lambda^2 \\
\text{and} \quad & \bar\mu_\ell(s)=\frac{P_\ell(1+\tfrac{8}{3(s-4)})}{\pi (s-4/3)^3} +\sum_{n,m}\nu_{n,m} F_\ell^{(n,m)}(s), \quad \text{for}\quad s>\Lambda^2.
\end{aligned}
\ee
Even with one crossing constraint this problem is not simple to solve directly, and we employ here the relaxation trick used in \eqref{relaxed_D}.

\begin{figure}[t]
        \centering
        \includegraphics[width=0.9\textwidth]{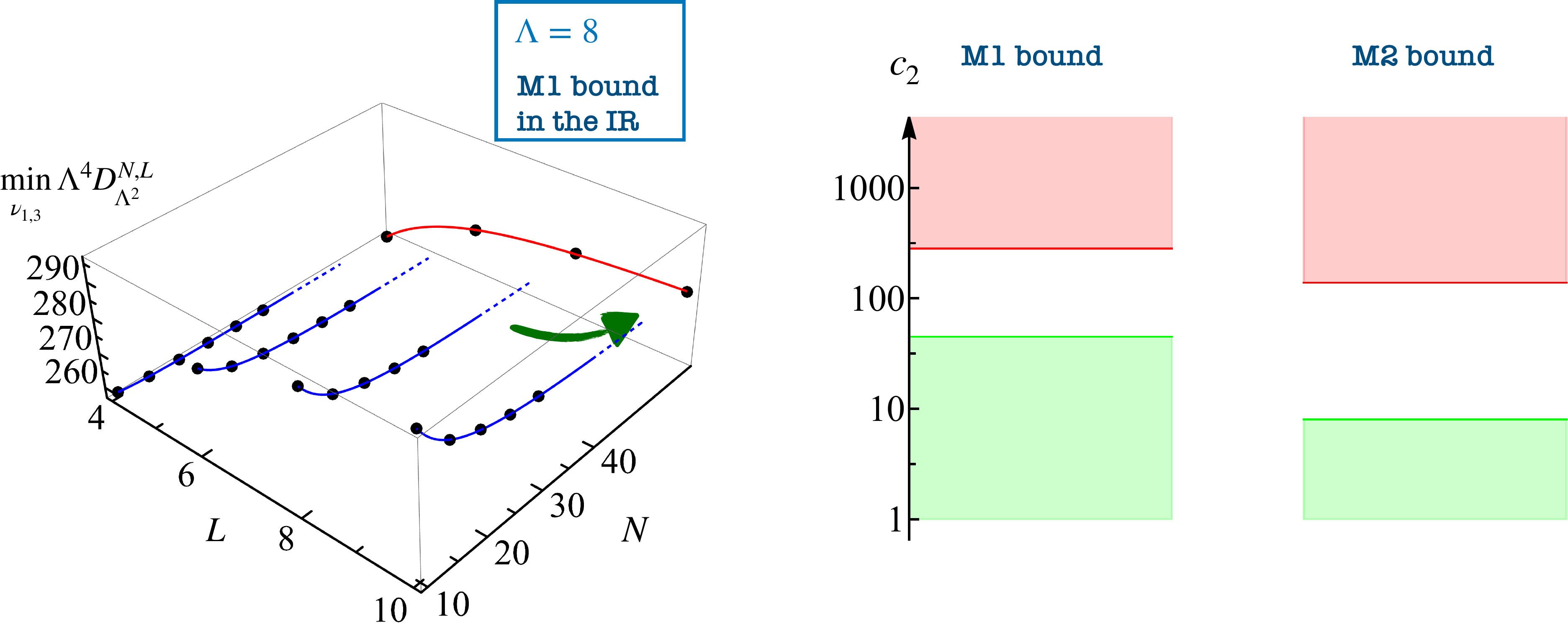}
    \caption{On the left, the double extrapolation procedure to determine the dual bound on $c_2$ in presence of the IR cut and the constraint {\bf M1}. On the right, the comparison between the primal results and the dual exclusion bounds. We expect the gap should close once the constraints on the real parts are included.}
    \label{fig_dual_4}
\end{figure}

We study the bound for $\Lambda=8$, the results are in the left plot of Fig.~\ref{fig_dual_4}.
By performing a double extrapolation we estimate the dual bound for {\bf M1} at this cutoff $ c_2 \Lambda^4 \leq 283.3$ that must be compared with the primal extrapolation $c_2 \Lambda^4 \lesssim 45$.

Finally, we summarise the primal and dual bounds on the $\max c_2$ in the right plot of Fig.~\ref{fig_dual_4}: in green the primal allowed region, in red the dual excluded.
The duality gap  (distance between red and green boundaries in  right plot of Fig.~\ref{fig_dual_4}) was not expected to close because for the dual we are using a subset of the unitarity constraints of the ones used in the primal bootstrap. 
 It would be interesting to efficiently implement the full dual problem in \cite{Guerrieri:2021tak} or \cite{He:2021eqn} and shrink the duality gap.


\section{Dimension-six operators}
\label{dimsix}

In this section we begin the analysis of  the space of theories with $O(n)$ symmetry by 
finding the extremal values of the  two-to-two amplitude -- a thorough analysis is reported in ref.~\cite{upcommingsoon}.
We scatter  two scalars which transform as vectors under the $O(n)$ internal symmetry $a+b\rightarrow c+d$. The  amplitude is 
\be
M_{ab}^{cd}(s,t,u)=  M   (\bar s | \bar t, \bar u) \delta_{ab}\delta^{cd}+
M   (\bar t | \bar u, \bar s) \delta_{a}^{\, c}\delta_{b}^{\, d}+
M   (\bar u | \bar s, \bar t) \delta_{a}^{\, d}\delta_{b}^{\, c} \, .  
\ee
Taylor expanding the  s-channel  amplitude around $(\bar{s},\bar{t},\bar{u}) = (0,0,0)$  we have
\be
M   (\bar s | \bar t, \bar u)  =  - c_0 + c_H  \bar{s} + O(\bar s^2,\bar t^2,\bar u^2)  \, ,
\label{lowamp4dflav}
\ee
A particular interesting aspect of this generalisation   
is  the presence of  dimension-six operators. 

The coefficients in \reef{lowamp4dflav} can be  interpreted in terms of the field theory
\be
\mathcal{L}= \frac{1}{2} \left(  \partial_\mu \vec\phi \cdot \partial^\mu  \vec\phi - m^2  \vec{\phi}\cdot \vec \phi \right)    -\frac{g_0}{4} (\vec{\phi}\cdot \vec \phi)^2  + \frac{1}{4} \frac{g_H}{\Lambda^2}  \partial^\mu( \vec\phi \cdot   \vec\phi)  \partial_\mu(\vec \phi \cdot \vec \phi)  +\dots \label{sm}
\ee 
When the  EFT is weakly coupled we can readily compute 
\be
c_0 = 2 g_0 - 8/3 \,  g_H  \,   m^2/\Lambda^2 + \cdots \  ,  \quad 
c_H  =  2  \, g_H \,  m^2/\Lambda^2 +   \cdots \ ,
\ee
where $\cdots$ involve loop corrections. 
For $n=4$ this particular model is very interesting from the  phenomenological point of view. 
Equation \reef{sm} describes the Standard Model (SM) Higgs sector in the custodial symmetric limit~\footnote{See appendix C of ref.~\cite{Elias-Miro:2013mua} for the transformation properties  of all the SM dimension-six operators under custodial symmetry.  }, after neglecting $SU(2)_L$ interactions.  In a second step it is possible  to relax this assumption and add to  this  set up gauge and Yukawa interactions.

The dimension-six operator 
\be
O_H= \partial^\mu( \vec\phi \cdot   \vec\phi)  \partial_\mu(\vec \phi \cdot \vec \phi)
\ee
is the  leading corrections  to SM Higgs two-to-two scattering.
The current  data from ATLAS and CMS at the LHC  are compatible with the SM and therefore constrain this operator to be  $|g_H \times (5~\text{TeV})^2/\Lambda^2|\lesssim 50$. The exact bound mildly depends on the exact underlying assumptions of the fit or SM EFT, see for instance HEPfit~\cite{DeBlas:2019qco} or SMEFIT~\cite{Ethier:2021bye}.

\begin{figure}[t] 
\centering 
        \includegraphics[width=0.9
       \textwidth]{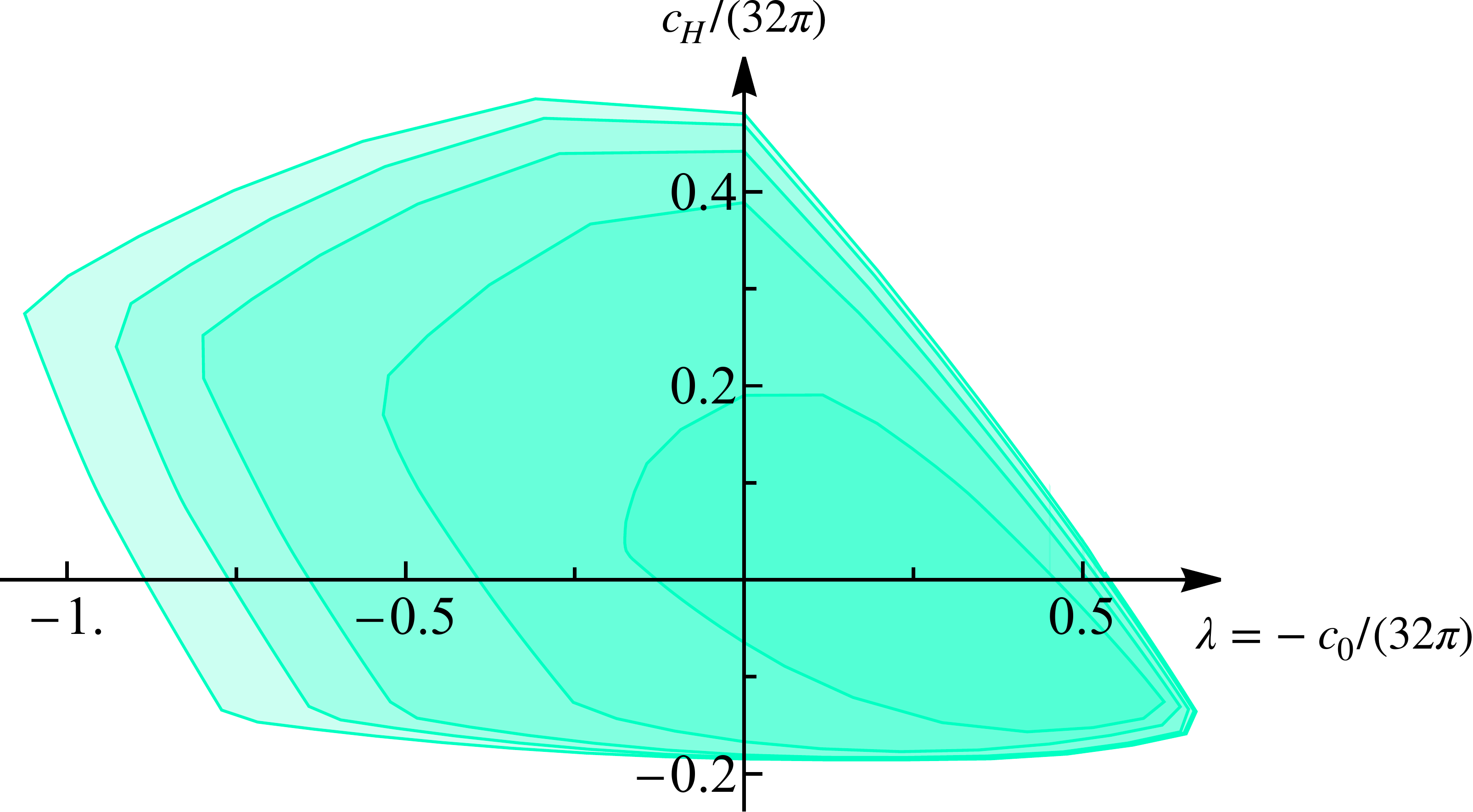} 
        \caption{Allowed space of $(\lambda,c_H)$, for the $O(4)$ model. Different curves correspond to different values of the ansatz cutoff $N$ ranging from $N=2$ (the inner curve) to $N=6$ (the boundary). }
\label{O4_bound}
\end{figure}

In the same spirit as in section~\ref{sec:full_space}, we begin by exploring the space of theories with $O(n)$ symmetry by making no  assumption of 
weak coupling or  reference to a possible EFT description. That is we define couplings  through \reef{lowamp4dflav}
and ask what is their allowed values  compatible with analyticity and unitarity of the S-matrix. 
In Fig.~\ref{O4_bound} we plot the allowed values of  $(c_0,c_H)$, in $m^2=1$ units, and for $n=4$. 

We can  understand the mechanism behind the bound on $c_H$ by means of  a simple sum rule.
The coefficient $c_H$ can be extracted from e.g. the singlet channel  $c_H = \frac{1}{(N-1)} \frac{\partial}{\partial \bar s} M^\text{sing}(\bar s,\bar t)\big |_{\bar s=\bar t=0}$, where $M^\text{sing}=N \, M(\bar s | \bar t , \bar u) + M(\bar t | \bar u , \bar s) + M(\bar u | \bar s , \bar t)$,
and it satisfies the following sum rule 
\be
c_H= C(s,t). \vec M(s,t)+\frac{1}{\pi} \int_{4m^2}^\infty [D_1(z,s,t). \vec M_z(z,t)+ D_2(z,t). \vec M_z(z,4/3) ] dz \, ,  \label{cHsr}
\ee
which  follows from a doubly subtracted fixed-$t$ dispersion relation.~\footnote{The un-substracted dispersion relation for $c_H$ was studied in ref.~\cite{Low:2009di}. }
The row vectors   $C$ and $D_i$ are rational functions  given   in  appendix~\ref{appdispN}.
The key qualitative feature of this sum rule is that it involves both the integral over the absorptive part of the amplitude $M_z(z,t)$, and  the real subtraction constants  $\vec M(s,t)$.
This structure is similar to the sum rule of $c_0$ in \reef{c0sr}. 
Therefore bounding only the imaginary part of the amplitude $M_z(z,t)$ along the unitarity cuts does not produce a bound on $c_H$.
One needs to go beyond the first approximation of linearised unitarity $0\leq \im f_\ell \leq 2/\rho^2(s)$, depicted with the red thick dashed lines below,
\vspace{-.2cm}\be
\begin{minipage}{13cm}\centering
        \includegraphics[width=0.3
       \textwidth]{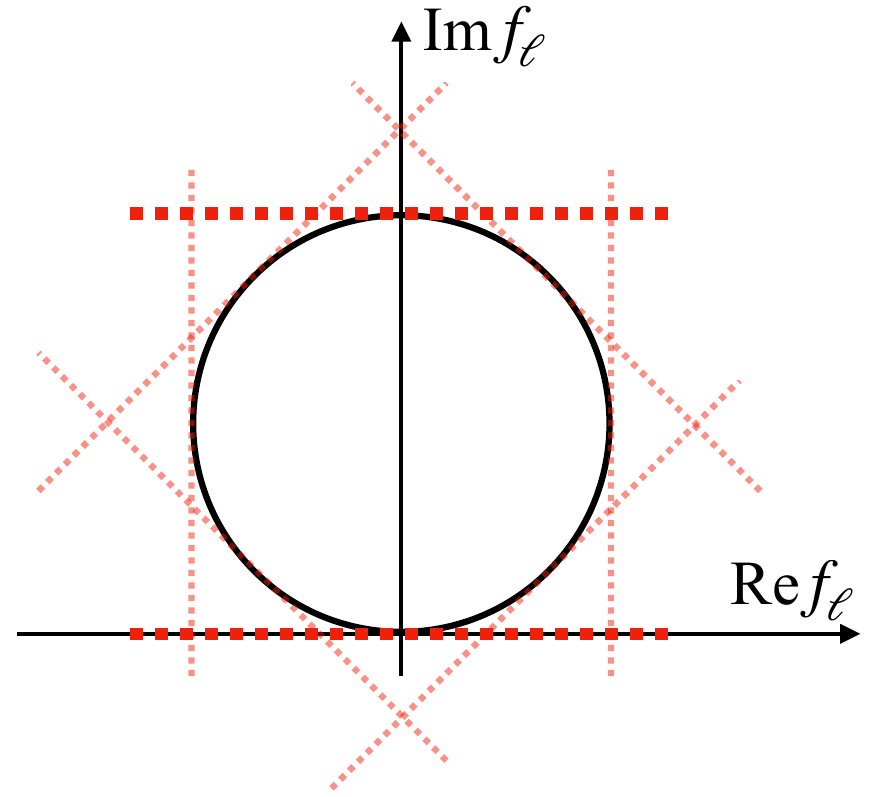} 
  \end{minipage} \nonumber
\vspace{-.2cm}
\ee
and bound the real parts of the partial waves as well. This could be done by including higher order linear approximations to unitarity (depicted above with red thinner dashed lines). It would be interesting in the future to derive dual bounds taking into account these higher order approximations.
In the S-matrix Bootstrap approach the real parts of the amplitude are bounded by means of the  exact two-particle unitarity equation, and  because of the sum rule \reef{cHsr} this strategy leads to a two-sided  bound on $c_H$.
 This is a mechanism similar to the bound on $c_0$ for the singlet case -- see \reef{c0sr} and the discussion there.

The next important point to clarify is the physics along the boundary of the oyster-shaped region in  Fig.~\ref{O4_bound}. 
Two interesting observables are the spin dominance and the $\text{UV}/\text{IR}$ dominance of $c_H$ along the boundary of the oyster~\cite{upcommingsoon}.
From our experience with the singlet in section~\ref{singeft}, we speculate that,  in the scenario of  UV dominance $c_H^\text{UV}/c_H\approx 1$, we can  re-interpret  the maximum/minimum value of $c_H$ in Fig.~\ref{O4_bound}
as the maximum/minimum value of $c_H$ or $g_H$ measured at the cutoff scale $\Lambda$.
In this case the maximum/minimum values of $c_H\sim 2 g_H\eps  $ from Fig.~\ref{O4_bound}, roughly 
  $ g_H \in[-10,20]$,   are of very much physical relevance for the  Higgs precision programme 
at the LHC and  future high energy colliders. 
Perhaps unsurprisingly, this bound  is  compatible with  naive dimensional analysis for  strongly coupled UV completions $O(1)\times |g_H|  \lesssim ( 4 \pi)^2 $. 
We do not know however  the smallest value that $g_H$ can take for  negative values, or  sharp bounds on this quantity.  
The approach  we are taking will establish a precise two sided bound on $g_H$, and  help us understanding better the space of theories and their spectrum for maximally strongly coupled UV completions.

\section{Cutoff dependence, \emph{In medio stat virtus}}
\label{medio}

What is the cutoff dependence of  the  bounds on the $c_i$'s? To answer this question, consider  the  two extreme limits  that one can take
in the {\bf M1}/{\bf M2} methods: $\Lambda^2\rightarrow 4m^2$ or $\Lambda^2/m^2\rightarrow \infty$.

In the first limit, the answer to what is the maximal value of $c_2$ is provided by Fig.~\ref{butterfly_plot}.
The upper branch of that plot is saturated by amplitudes with a pole at threshold $s=4m^2$, i.e.  with  non vanishing  residue $\alpha_\text{th}$ (see left plot in Fig.~\ref{threshold_and_ratios}).
These amplitudes correspond to  theories maximally strongly coupled all the way to the IR $s=4m^2$.

The other limit  $\Lambda^2/m^2\gg 1$, also admits an interesting interpretation,  in terms of the EFT Lagrangian \reef{lagone}. 
Recall that the coefficient $c_2$ is related to the Wilson coefficient $g_2$ through   $c_2=  g_2 m^4/\Lambda^4+ \beta_0 g_0^2  + \dots  $, where $\beta_0$ is a calculable IR  one-loop contribution. 
If $g_0$ is sent to zero first and then the $m^2\rightarrow 0$ limit is taken, we are left with a single scale $\Lambda^2$ in the problem. It is then natural to normalize $c_i$'s in terms of the new cutoff scale $\Lambda$.
This is the limit of exact UV dominance  $c_2^\text{UV}/c_2=1$. 
In this limit the scale   $\Lambda^2$  sets the units of the problem and the bound on $c_2$ is  trivially independent of $\Lambda^2$.
The limit $\Lambda^2\rightarrow\infty$ with $g_0\lesssim O(1)$ and $m^2$ held fixed, is described by the IR domination regime $c^\text{UV}_2/c_2=0$, 
and the single-scale in the problem is $m^2$. See for instance the region $\th\in[\pi,2\pi]$ of Figs.~\ref{M1_ratios},\ref{M2_ratio} where the IR  $s\sim O(m^2)$ is described by the  $g_0\phi^4$ theory.

As for the UV domination regime when $\Lambda^2\gg m^2$, see for instance Fig.~\ref{fig_dual_3}: the blue curve shows that as the cutoff is increased (in units of $m^2=1$) we approach the massless limit bound of the  single-scale $\Lambda^2$ problem, shown with a black horizontal line. 
In red,  the M2 dual bound shows that if   IR physics (controlled by $g_0^2$) is non-negligible, then the bound is loosened; while if  IR physics  goes  to zero faster than $m^2/\Lambda^2$, then we reach the UV domination scenario of the  blue curve -- i.e. the minimum of the blue  curve moves to higher  values of $\Lambda^2$  as $g_0^2$ is decreased in comparison to $m^2/\Lambda^2$. 
Note however that for phenomenologically  reasonable cutoffs like for instance $\Lambda/m\gtrsim O(10)$ and moderately weak IR couplings $g_0\lesssim O(1)$, the bounds for $c_2$ are 
close to the  asymptotic regime $m^2\rightarrow 0$
and show a mild cutoff dependence.

The analyses  that we presented in this work \emph{bridge} between these two limits. That is the construction that we presented  allows to ask precise questions for EFTs that, on one hand feature non-negligible IR physics, but on the other hand are maximally   strongly coupled at energies above a physical cutoff $\Lambda$.

\section{Conclusions}
\label{conc}

We have studied in detail the space of QFTs in $3+1$ dimensions with unprecedented precision due to a number of novel numerical improvements that we have introduced.
In addition we have constructed  a new observable to measure the low spin dominance, and we have found that along the boundary of the $(c_0,c_2)$ space  there exist an arc around the free theory point where spin-0 dominance is realized. 

We have also  introduced a simple formulation to obtain positivity bounds and applied it to the ratio of  coefficients $c_3/c_2$, and compared them with the numerical S-matrix results, finding a nice agreement -- and observed that  in the low spin dominance region the coefficient $c_3>0$.

The notion of the cutoff $\Lambda^2$ in an EFT is tightly related to the presence of resonances in the low energy spectrum. Ideally, it would be enough to impose some analyticity in the second sheet of the Mandelstam $s$-plane to make sure that no resonances would appear for energies below the cutoff $s<\Lambda^2$.
In two dimensional theories there have been already promising explorations in this direction \cite{Kruczenski:2020ujw}, and it would be interesting to find a generalization to higher dimensions.~\footnote{This property seems also be related to real analytic functions studied with the Geometric Function Theory \cite{Haldar:2021rri}.} 

Here we have taken a direct approach. We have  introduced additional constraints on the absorptive part of the amplitude. The intuition comes from dispersion relations (or the Cauchy theorem): once we specify the crossing and analytic structure, the amplitude depends only on its discontinuities.
Introducing suitable physical constraints on the discontinuities is enough to carve out a region in the amplitude space that behave as  expected for the  EFT  amplitude space!

We have  proposed two methods to introduce such constraints.

Firstly, using the definition of arc variables in \cite{Bellazzini:2020cot}, we bound arcs of the amplitude in the IR region. 
We study the effect of a single arc constraint and determine the space of QFTs consistent with it. These  
theories define a region in the $(c_0,c_2)$ place which we call {\bf M1} region. 
We introduce the notion of UV dominance and show that the tree-level positivity bounds are valid when UV dominance is realized, but violated otherwise.

Secondly,  we bound point-wise the imaginary part of the amplitude using a one-loop EFT computation up to the cutoff scale.
We call the region in the $(c_0,c_2)$ obtain through this method {\bf M2} region. 
The region defined by this stronger set of constraints nicely agrees  inside the $M_1$ region, and its boundary overlap when there is IR dominance.
By increasing the number of arc constraints we do expect the {\bf M1} region to collapse on the {\bf M2} region (this is  also suggested by two-dimensional toy models).
We have also  employed linearised positivity to place dual bounds on $c_2$ and compared those with the S-matrix Bootstrap approach.

Our construction is general and applies to other theories such as $O(n)$ scalar field theory. 
In this context we have shown that the S-matrix Bootstrap is able to bound dimension-six operators of EFTs, and leave for ref.~\cite{upcommingsoon}
a more detailed study of the physics.

All in all the construction that we have presented  opens up new avenues to study accurately   EFTs 
with  extremal values of the Wilson coefficients due to UV physics above the  EFT cutoff scale.


\section*{Acknowledgements}

We thank Paolo Creminelli,  James Ingoldby,  Markus Luty, Harish Murali,  Yaron Oz, Jo\~ao Penedones, Riccardo Rattazzi, Marco Serone, Amit Sever, Alessandro Vichi, Pedro Vieira, Shimon Yankielowicz for useful remarks and discussions. 
We thank Brando Bellazzini, James Ingoldby,  Jo\~ao Penedones, Marc Riembau and Pedro Vieira for useful comments on the draft.
J.E.M. thanks the Aspen Center for Physics  for  hospitality  while part of this project was completed. AG is supported by the European Union - NextGenerationEU, under the programme Seal of Excellence@UNIPD, project acronym CluEs.
A.G. was also supported by The Israel Science Foundation (grant number 2289/18). This research was supported in part by the National Science Foundation under Grant No. NSF PHY-1748958.


\appendix

\section{The numerical setup}
\label{numerics}

In this appendix we explain in detail the numerical setup used to obtain the results of this paper. 

\subsection{The wavelet ansatz}
\label{Appendix:wavelet}
The problem of designing ansatzes for the scattering amplitude that would span efficiently the amplitude space, thus improving the $N$ convergence, was first addressed in the pioneering paper \cite{Paulos:2017fhb}.
There it was shown that adding all allowed singularities compatible with unitarity would improve the convergence to the optimal bound in one specific example. 
Sometimes, however, the problem is more subtle, and the convergence rate in $N$ is related to the appearance of resonances (or zeros) in the complex plane, and in particular to their position in the plane.
In fact, a long lived resonance, would show up as a zero relatively close to the physical cut, therefore close to the boundary of the convergence region in the $\rho$ plane.

Intuitively, one should expect that inputting degrees of freedom locally in the resonance region would improve the convergence.
A simple way to do it is to tune the centering parameter of the $\rho$ map.
The conformal map $\rho$ from the cut plane to the unit disk depends on a scale parameter $\sigma$ defined as the point where $\rho_\sigma(\sigma)=i$, or equivalently $\rho_\sigma(8m^2-\sigma)=0$
\be
\rho_s(\sigma)=\frac{\sqrt{\sigma-4m^2}-\sqrt{4m^2-s}}{\sqrt{\sigma-4m^2}+\sqrt{4m^2-s}}.
\ee
All features of the amplitude that appear at scales of order $s\sim \sigma$ will be well captured by an ansatz centered at $\sigma$. 
A priori, however, it is hard to predict where resonances or virtual particles will appear as a result of the bootstrap.
To overcome this problem, we propose an alternative ansatz, where we scatter $\rho_\sigma$ variables around trying to capture particles wherever they appear in the complex Mandelstam $s$ plane.

Our proposal is the following: choose a set of scale parameters  $\sigma \in \Sigma_N \subset [4,\infty)$, where $N$ denotes a measure of the number of points, and write down the ansatz
\bea
\bar M(s,t,u)=\alpha_0+M^{(1)}(s,t,u)+M^{(2)}(s,t,u)=\alpha_0+\sum_{\sigma\in \Sigma_N} \alpha_\sigma (\rho_s (\sigma)+\rho_t(\sigma)+\rho_u(\sigma))+\nonumber\\
+\sum_{(\sigma,\tau) \in \Sigma_N^2} \alpha_{\sigma,\tau} ((\rho_s(\sigma)\rho_t(\tau)+\rho_s(\tau)\rho_t(\sigma)) + ( t \leftrightarrow u)  + (s \leftrightarrow u) ).
\label{wavelet_ansatz_appendix}
\eea
The two functions $M^{(1)}$, and $M^{(2)}$ contain terms that only contribute to the single discontinuity, and terms that contribute to the double discontinuity. This separation makes more manifest the equivalence of this ansatz with the double-subtracted Mandelstam representation in \eqref{Mandelstam_rep}.

\begin{figure}[ht] 
\centering 
        \includegraphics[width=0.8
       \textwidth]{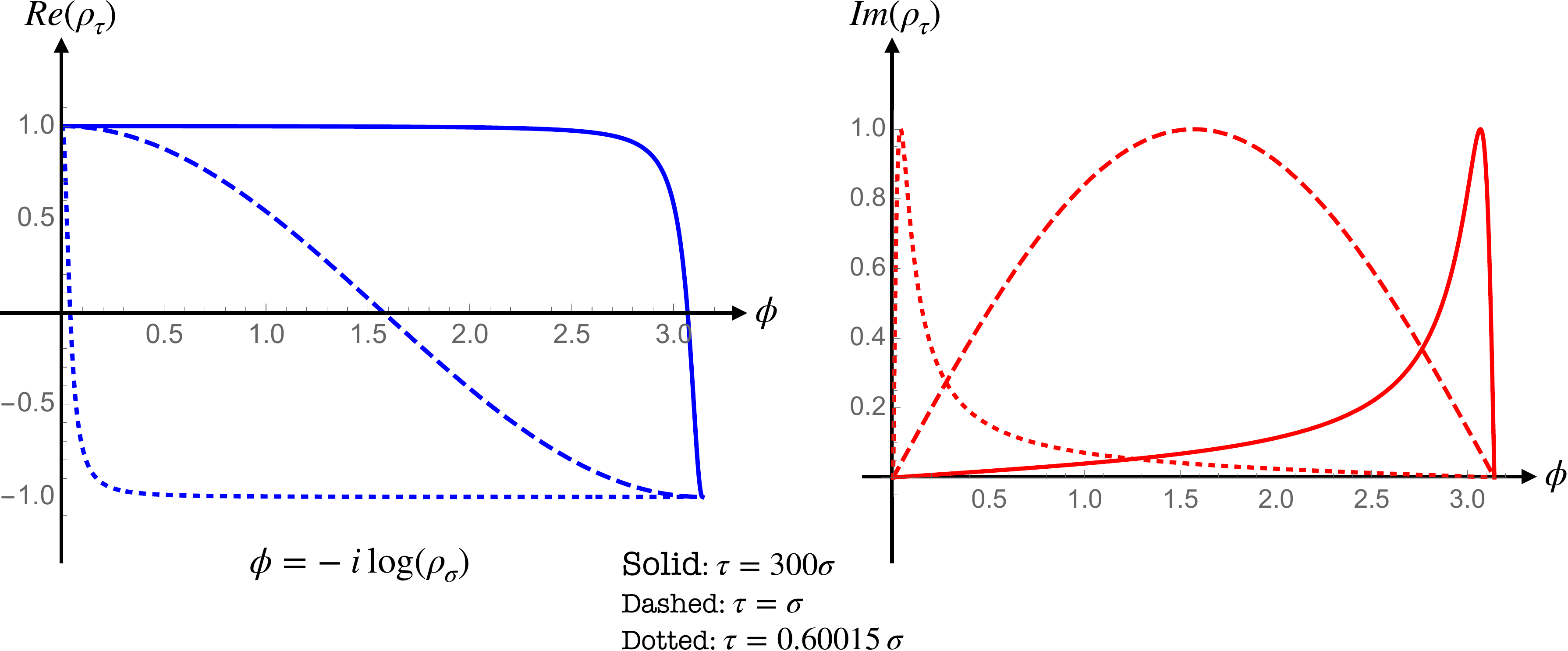} 
        \caption{Real part (on the right) and imaginary part (on the left) of the $\rho_\tau$ map as a function of $\phi=\arg(\rho_\sigma)$. From the figure it is clear that tuning the scale parameter we can zoom in different region of the boundary of the disk. }
\label{mobius}
\end{figure}

We refer to this ansatz with the term \emph{wavelet} ansatz, where the analogy comes from the following simple idea. We can view the $\rho_\sigma$ map as function of the scale parameter $\sigma$. If we map the cut $s$-plane onto the unit disk using some map $\rho_{\sigma}(s)$, then the transformation $\rho_\tau$ can be written as 
\be
\rho_\tau(\rho_\sigma)=\frac{\rho_\sigma-w}{1-\rho_\sigma w}, \quad w=\frac{\sqrt{\sigma-4}-\sqrt{\tau-4}}{\sqrt{\sigma-4}+\sqrt{\tau-4}}<1, \quad \text{for}\quad \sigma,\sigma_0>4,
\ee
in other words $\rho_\tau$ is a M\"obius transformation mapping the disk into itself. 
Inspired by the \emph{wavelet} representation of a signal (though we will not attempt any rigorous connection), we can sample the space of analytic functions in the unit disk using the basis of all possible M\"obius transformations.
In the limit of infinite $N$ the ansatz \reef{generic_ansatz} and the wavelet one are equivalent since $\rho_\tau$ has a convergent Taylor expansion in $\rho_\sigma$, but at finite $N$ we do expect it might improve convergence. We find experimentally that either improves convergence, or it works as well as the usual $\rho$ expansion.~\footnote{On a more technical level: if the projections of the ansatz onto partial waves are computed analytically, then one needs to compute only the projection of $\rho_\sigma(t)$ and $\rho_\sigma(t)\rho_\tau(u)$ for each spin.}

In practice, a critical point is the choice of a grid for the scaling parameters $\Sigma$.
In our numerics we employ the following procedure.
First, we choose a \emph{mother} wavelet: the map we will use to tune all other wavelets.
For a cut starting at 4 we choose the value $\sigma^*=\tfrac{20}{3}$ (corresponding to a $\rho$ map centered at the crossing symmetric point).
At each fixed value of $n$, we define a Chebyschev grid
\be
\vec \phi=\left\{\frac{\pi}{2}(1+\cos\left(\tfrac{k \pi}{n+1}\right)) | k=1,\dots, n\right\},
\label{cheby_grid}
\ee
then using the inverse map $\rho^{-1}_{\sigma^*}=s_{\sigma^*}$, we take $s_{\sigma^*}(e^{i \vec \phi})$ and obtain a set of points for $s>4$ that we call $\Sigma_n$.
The grid at cutoff $N$ is defined by taking the union of all these sets of points with $n \leq N$
\be
\Sigma_N=\bigcup_{n=1}^N \Sigma_n.
\ee
The number of terms in $M^{(1)}$ scales like $N^2$, while the number of terms in $M^{(2)}$ like $N^3$. For instance, when $N=14$, as in many examples in this paper, the number of free variables is 551.~\footnote{We have tested the possibility of distributing the scale parameters in the complex plane, but we haven't observed any improvement. }

\subsection{Unitarity and positivity constraints}

In the S-matrix Bootstrap unitarity is imposed at the level of the partial waves $2\im f_\ell(s) \geq \rho^2(s)|f_\ell(s)|^2$.
This inequality can be written in an equivalent semi-definite positive condition introducing the matrix
\be
\mathcal{U}_\ell=\begin{pmatrix}
1-\frac{\rho^2}{2}\im f_\ell & \rho \re f_\ell \\
\rho \re f_\ell & 2 \im f_\ell
\end{pmatrix}  \succeq 0.
\ee
We impose unitarity $\mathcal{U}_\ell(s) \succeq 0$ for each spin up to some cutoff $\ell \leq L$, and for a set of points in the interval $(4,\infty)$.
In our numerics we choose for each spin a fixed number of points $n_\text{pts}=300$, distributed on a Chebyschev grid on the boundary of the unit disk~\eqref{cheby_grid} of the standard $\rho_{\sigma^*}$ map with scale parameter $\sigma^*=20/3$.

Computing the partial wave projection of the ansatz for large values of $L$ and for many points $s$ is a computationally expensive task. However, there is a simple consequence of unitarity, i.e. positivity of the imaginary part, that can be imposed at the level of the ansatz without projecting.
Usually, it is written as 
\be
\im M(s,t=0) =16\pi \sum_{\ell=0}^\infty (2\ell+1) \im f_\ell(s)\geq 0,
\label{positivity_cross_section}
\ee
that is positive because separately each partial wave is positive $\im f_\ell\geq 0$.
In the numerical bootstrap with fixed cutoff $L$ this sum is not automatically positive, unless we take $L$ ``large enough".
By adding positivity constraints as~\eqref{positivity_cross_section} we bound this sum from below, though it can still be negative
\be
\sum_{\ell=L+2}^\infty(2\ell+1) \im f_\ell(s)\geq -\sum_{\ell=0}^L(2\ell+1) \im f_\ell(s)\geq-(L+1)(2L+1)\frac{\sqrt{s}}{\sqrt{s-4}}.
\label{remainer}
\ee

Although this condition has proved to help convergence in the massless case~\cite{Guerrieri:2021ivu}, in the gapped case we can and we need to do better.
The reason is that by looking at the large spin expansion of the partial wave projections using the Froissart-Gribov representation
\be
\im f_\ell(s)=\int_{t_0(s)}^\infty Q_\ell (1+2\tfrac{t}{s-4}) \,  \im M_t(s,t)dt,
\ee
the integral for large $\ell$ is dominated by the region close to the boundary $t_0(s)$, that in the usual S-matrix Bootstrap ansatz is $t_0(s)=4$.

One should expect that imposing constraints in the region close to $t=4$ will have the most effect on the higher spins $\ell>L$. 
Indeed, another consequence of the positivity of the imaginary part of the partial waves is that
\be
\im M(s,0\leq t<4) = 16 \pi \sum_{J=0}^\infty \im f_J(s)P_\ell(1+2\tfrac{t}{s-4})\geq 0,
\label{tpositivity}
\ee
that follows from the fact that $P_\ell(x)>0$ for $x>1$.
Therefore, by considering different values of $0\leq t<4$ we have access to a number of positive sum rules that help bounding from below the sum over the imaginary parts for $\ell>L$, and by taking $t\sim 4$ we are sure these sum rules are dominated by asymptotically large values of $\ell \gg L$.

Since we are bounding numerically all partial waves for $\ell\leq L$, we can recycle the same numerical integrals to improve further the positivity constraints \eqref{tpositivity}
\be
\im M(s,0\leq t<4) -16\pi\sum_{\ell=0}^L \im f_\ell(s)P_\ell(1+2\tfrac{t}{s-4})\equiv \sum_{\ell=L+2}^\infty \im f_\ell(s)P_\ell(1+2\tfrac{t}{s-4})\geq 0.
\label{improvedtpositivity}
\ee
imposing also positivity of the sum of the remaining spins in eq. \reef{remainer}.

In our numerics we impose the constraint \eqref{improvedtpositivity} for any $s$ of the unitarity grid. In $t$ we choose ten points between $0\leq t <4$, and twenty points for $3.999<t<4$. Indeed, we observe these constraints become more important as we approach $t\to 4$.

We conclude this section with two final observations. In the fixed-$t$ dual formulation \cite{Guerrieri:2021tak} it has been shown that it is possible to obtain dual rigorous bounds by imposing nonlinear unitarity up to some spin $\ell\leq L$, and positivity of the imaginary part for any $\ell$.  In this paper, we observe that the improvements obtained by imposing the positivity constraints at fixed cutoff $L$ are consistent with the existence of such fixed $L$ bounds. It would be interesting to further explore this direction and prove it starting also from the Mandelstam representation~\cite{He:2021eqn}.
Finally, note that so far we have exploited the discontinuity  in $s$ to boost  convergence. However it is been shown in \cite{Correia:2020xtr} that discontinuity  in $ t$ and its threshold expansion puts further constraints on the problem, we leave it for future investigations.

\subsection{High energy improvement}
\label{Appendix:HighEnergy}

As we will see in the next section, the assumptions on the high energy behaviour of the amplitude can have big impact on the bounds on physical observables.
In the original S-matrix Bootstrap formulation, the ansatz proposed has the property that 
\be
\lim_{|s|\to\infty}M(s,t)=M_\infty,
\ee
compatible with a subtracted dispersion relation.
As observed already in \cite{Paulos:2017fhb}, allowing for a more general high energy behaviour could improve the convergence of the bounds.

Our simple proposal is to add few terms of the form \footnote{The growing terms we chose to add to our numerics are redundant in the physical region where we impose the constraints. Increasing the number of terms would introduce zero modes in the ansatz and make the numerics unstable. Therefore, we keep few growing terms (to specify) to limit this issue. We observe anyway an improvement in the convergence.}
\bea
M_{\infty}^\text{ans}(s,t,u)=\sum_\sigma \alpha^\prime_\sigma (\rho_\sigma(s)(4-s)^{3/2}+\rho_\sigma(t)(4-t)^{3/2}+\rho_\sigma(u)(4-u)^{3/2}) \nonumber\\
+\sum_{\sigma,\tau}\beta^\prime_{(\sigma,\tau)} (((4-s)^{3/2}+(4-t)^{3/2})\rho_{\sigma}(s)\rho_{\tau}(t)+(s\leftrightarrow u)+(t\leftrightarrow u)).
\eea
Naively, $M_{\infty}^\text{ans}(s,t(s,x),u(s,x))\sim s^{3/2}$ for large $s$ in the physical region. 
However, it is possible to expand at large $s$ and for fixed spin the partial wave projection of $M_{\infty}^\text{ans}$ and impose a finite number of linear constraints among the $\alpha^\prime$ and $\beta^\prime$ coefficients such that for large $s$
\be
\int_{-1}^{1} P_\ell(x) M_{\infty}^\text{ans}(s,t(s,x),u(s,x))dx \sim O(1)
\ee
Furthermore, we need to cancel the leading $\mathcal{O}(s^{3/2})$ growth for $t=0$ as it would violate the Froissart-Martin bound  $\im M(s,t=0) \leq c s \log^2(s)$, and therefore violate unitarity for asymptotically large energies. After imposing these conditions the ansatz will still grow as $M^{\infty}(s,t,4-s-t)\sim s^{3/2}$ at fixed-$t$, in particular at $t=4/3$, allowing for a more flexible high energy behaviour. 

In many contexts the S-matrix Bootstrap has reached a stage where the high energy behavior of the ansatz has become relevant as in the case of gravitational interactions, or even for scalar amplitudes in dimensions $d>4$. 
It would be important to study this problem more systematically as it might allow to construct an ansatz compatible with the Froissart bound, or an ansatz able to accommodate the exchange of higher spin particles. We leave this problem to future explorations.

\subsection{Minimum coupling cusp in more details}
\label{minimum_coupling_numerics}

\begin{figure}[t] 
\centering 
        \includegraphics[width=1
       \textwidth]{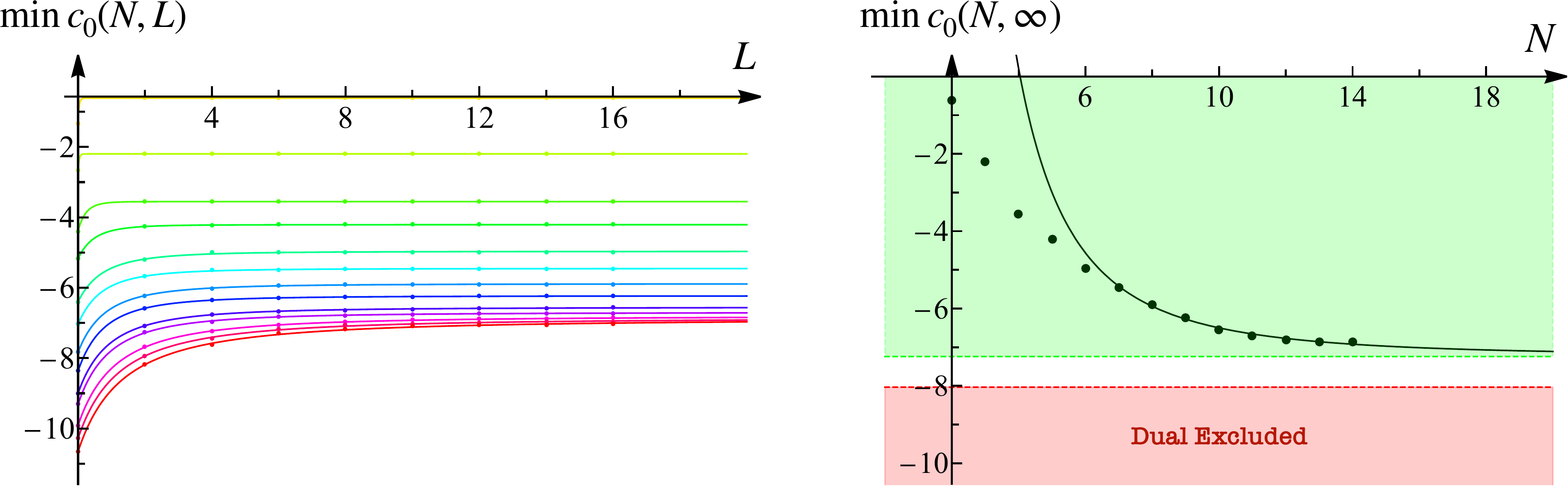} 
        \caption{On the left, set of data points for the problem $\min c_0$ as a function of $L$. Different colors correspond to different values $N$ ranging from $N=2$ (light green) to $N=14$ (red). On the right, asymptotic values of $\min c_0$ (black dots) after spin extrapolation as function of $N$. The red region is excluded by the dual problem~\cite{Guerrieri:2021tak}, the solid black line is a power law fit starting at $N=6$, and the green region is the allowed region given our extrapolation.}
\label{minimum_plots}
\end{figure}

In this section we study the hardest problem along the boundary of Fig.~\ref{butterfly_plot}, namely the minimum coupling cusp.
In Fig.~\ref{minimum_plots} on the left we show the data obtained for $\min c_0(N,L)$ at fixed $N$ as a function of $L$.
The different colors correspond to the values of $N$ ranging from $N=2$ (light green) to $N=14$ (red). The dots are our numerical data, the solid colored curves our fits used to extrapolate for $L=\infty$. Although it is clear that for $L=16$ even for the highest $N$ we consider we are almost asymptotic, and the extrapolation does not improve the bound much. We stress the power of the positivity constraints we impose since the number of free variables for $N=14$ is 552 and the plateau appears already at low spins.

In Fig.~\ref{minimum_plots} on the right we show the extrapolated values of  $\min c_0(N,\infty)$ that we try to fit using a power law. However, this fit must be taken as indicative only, since the behavior for small $N$ can be hardly taken into account with a simple function like a power law, and because the result changes slightly by changing the starting point of the fit from $N=5$ and above (starting at lower values of $N$ the fit is hard).

Taking into account the extrapolation, we determine a primal estimate of $c_0 \geq -7.24$, still far from the dual bound, but reasonable looking at the rate of convergence in $N$. This number should be compared with the asymptotic value for $N=14$ which is instead $\min c_0(14,\infty)=-7.0$. Therefore, in the worst case scenario, we should think that the blue boundary of Fig.~\ref{butterfly_plot} has a maximum uncertainty of order $\sim 0.2$ on the position of the $B$ cusp.

\section{Review of Mandelstam representation and partial waves}
\label{manrep}
The useful Mandelstam representation expresses the amplitude as a function of its double discontinuity.
In general, due to the Froissart-Martin bound, we have to consider double-subtracted dispersion relations
\be
M(s,t,u)=c_0+\int_4^\infty \sigma(x)K_1(x;s,t,u)+\iint_\mathcal{D} \rho(x,y)K_2(x,y;s,t,u),
\label{Mandelstam_rep}
\ee
where
\be
K_1=\frac{s-s_0}{(x-s)(x-s_0)}+\frac{t-t_0}{(x-t)(x-t_0)}+\frac{u-u_0}{(x-u)(x-u_0)},
\ee
and
\be
K_2=\frac{(s-s_0)(t-t_0)}{(x{-}s)(x{-}s_0)(y{-}t)(y{-}t_0)}+\frac{(s-s_0)(u-u_0)}{(x{-}s)(x{-}s_0)(y{-}u)(y{-}u_0)}+\frac{(t-t_0)(u-u_0)}{(x{-}t)(x{-}t_0)(y{-}u)(y{-}u_0)}.
\ee
That is the reason why in the above representation depends on a subtraction constant, and also on the single discontinuity.

The imaginary part in the $s$-channel of the amplitude can be read off by using the standard identity $\tfrac{1}{x-s-i\epsilon}=\mathcal{P}\frac{1}{x-s}-i \pi \delta(x-s)$ yielding
\be
\im M=\pi \sigma(s)+\pi\int_{\bar y(s)}^\infty \rho(s,y) \left(\frac{t-t_0}{(y-t)(y-t_0)}+\frac{u-u_0}{(y-u)(y-u_0)}\right)dy,
\ee
where $\bar y(s)$ is the starting point of the support of the double discontinuity$\rho(s,y)$ at fixed-$s$ (for physical amplitudes is a point on the Karplus curve).
Projecting this equation into partial waves we obtain the expression
\be
\im f_J(s)=\frac{\sigma(s)}{16}\delta_{J,0}+\frac{1}{32}\int_{\bar y(s)}^\infty \rho(s,y)\left(\frac{8}{s-4}Q_J(1+\tfrac{2y}{s-4})-2\delta_{J,0}(\tfrac{1}{y-t_0}+\tfrac{1}{y-u_0})\right)dy,
\ee
where the $Q_J$ are the usual Legendre functions of second kind such that $\tfrac{1}{2i}\text{disc}_x Q_J(x)=P_J(x)$ for $-1<x<1$.
The above relation for $J=0$ allows us to replace the single discontinuity $\sigma$ with
\be
\sigma(s)=16\im f_0(s)-\int_{\bar y(s)}^\infty \rho(s,y)\left(\frac{4}{s-4}Q_0(1+\tfrac{2y}{s-4})-(\tfrac{1}{y-t_0}+\tfrac{1}{y-u_0})\right)dy.
\label{single_disc}
\ee
The higher spins, on the other hand, depend only on the double discontinuity 
\be
\im f_{J\geq 2}(s)=\frac{1}{4}\int_{\bar y(s)}^\infty \rho(s,y)\frac{Q_J(1+\tfrac{2y}{s-4})}{s-4}dy\equiv \int_{\bar y(s)}^\infty \rho(s,y)K_J^I(s,y)dy.
\label{higher_spins_equation}
\ee

For completeness we write down also the projection onto partial waves of the real part of the amplitude
\be
\re f_J(s)=\frac{c_0}{16\pi}\delta_{J,0}+\int_4^\infty \im f_0(x)\mathcal K_J^1(x;s)dx+\iint_\mathcal{D} \rho(x,y)\mathcal K^2_J(x,y;s)dxdy,
\ee
where
\be
\mathcal K_J^1(x;s)=\frac{\delta_{J,0}}{\pi}\left(\frac{1}{x-s}-\frac{1}{x-s_0}-\frac{1}{x-t_0}-\frac{1}{x-u_0}\right)+\frac{4}{\pi(s-4)}Q_J(1+\tfrac{2x}{s-4}),
\ee
and
\begin{align}
&\mathcal{K}_J^2(x,y;s)=\frac{1}{32\pi}\frac{s-s_0}{(x{-}s)(x{-}s_0)}\left( \frac{8}{s-4}Q_J(1+\tfrac{2y}{s-4})+2\delta_{J,0} \left( \frac{1}{(x{-}t_0)(y{-}u_0)}-\frac{1}{y{-}t_0}-\frac{1}{y{-}u_0} \right) \right)\nonumber\\
&+\frac{1}{32\pi}\frac{4}{s-4}\left( \frac{1}{s{-}4{+}x{+}y}(Q_J(1+\tfrac{2x}{s-4})+Q_J(1+\tfrac{2y}{s-4})) -\frac{1}{x{-}t_0}Q_J(1+\tfrac{2y}{s-4})-\frac{1}{y{-}u_0}Q_J(1+\tfrac{2x}{s-4})\right)\nonumber\\
&-\frac{1}{32 \pi}\left(\frac{4}{x-4}Q_0(1+\tfrac{2y}{x-4}) - \frac{1}{y-t_0}-\frac{1}{y-u_0} \right)(2\pi\mathcal{K}_J^1(x;s)).
\end{align}

All in all, the double-subtracted Mandelstam representation depending only on the spin zero imaginary part and the double discontinuity takes the form
\bea
&M(s,t,u)=c_0+16\int_4^\infty \im f_0(x)K_1(x;s,t,u)+\nonumber\\
&+\iint_\mathcal{D}\rho(x,y)\left(K_2(x,y;s,t,u)-\left(\frac{4}{x-4}Q_0(1+\tfrac{2y}{x-4})-\left(\tfrac{1}{y-t_0}+\tfrac{1}{y-u_0}\right)\right)\right)dxdy.
\label{dispHDfinal}
\eea

\section{$g_0 \phi^4$ theory perturbatively}
It is possible to compute the imaginary part of the amplitude, $\text{Im } M$, up to $O(g_0^3)$ perturbatively in $g_0 \phi^4$ theory. In this section we are going to review this computation. Finally we extract contributions into $c_2$ and $c_3$  coefficients up to $O(g_0^3)$.

\subsection{One-loop contributions}
Remember that the one-loop imaginary part of the amplitude can be obtained by a phase space integration of tree-level amplitudes. At leading order, it is given by
\bea
M_z(z, t) &= \frac{1}{2} \frac{g_0^2}{16\pi} \sqrt{\frac{z-4}{z}} \, , \label{phi4oneloop}
\eea
in $m^2=1$ units. Plugging this into dispersion integrals \reef{c2def} and \reef{c3sr} enables us to compute the contributions to the  coefficients $c_2$ and $c_3$ up to $O(g_0^2)$.
\bea
c_2^{O(g_0^2)} &= \frac{9 g_0^2}{8192 \pi^2} \left(14 - 15 \sqrt{2} \, \text{arctan} \frac{1}{\sqrt{2}}\right) \quad \approx 1.05 \cdot 10^{-4} \, g_0^2 \, ,\label{c2_1loop}\\
c_3^{O(g_0^2)} &= \frac{27 g_0^2}{131072 \pi^2} \left(- 130 + 153 \sqrt{2} \, \text{arctan} \frac{1}{\sqrt{2}}\right) \quad \approx 0.66 \cdot 10^{-4} \, g_0^2 \, .
\eea
We can also calulate the ratio $c_3/c_2$ through these numbers, giving us \reef{phi4ratio}.

\subsection{Two-loop contributions}

Analogously, the two-loop imaginary part can be obtained by a phase space integration involving tree-level and one-loop amplitudues. For this reason, we need not just the imaginary part, but also the full one-loop amplitude $M_\text{1-loop}(s,t)$.

Fortunately, fixed-$t$ dispersion relation \reef{fixed-t} enables us to obtain the full amplitude from the knowledge of one-loop discontinuity and a subtraction constant $M(s_0,t_0)$. Plugging in \reef{phi4oneloop} and choosing $M(4/3,4/3)=-g_0$ gives us
\bea
M_\text{1-loop}(s,t) &= \frac{g_0^2}{16 \pi^2} \Big[ f(s) + f(t) + f(4-s-t) \Big] \label{oneloopphi4} \\
\text{where} \quad f(s) &=  \left( \sqrt{2} \, \arctan \frac{1}{\sqrt{2}} - \sqrt{\frac{4-s}{s}}  \, \arctan \sqrt{\frac{s}{4-s}} \right) \, .
\eea
A separate computation with the Feynman diagrams confirms above result.

Notice that $f(s>4)$ has a non-zero imaginary part as expected and $M_\text{1-loop}(4/3,4/3)=0$ such that constant piece of the total amplitude corresponds to $-g_0$. 

Now it is time to use the unitarity equation to get the two-loop imaginary part. We consider the two particle unitarity cuts of two-loop Feynman diagrams\footnote{There exists a three particle cut diagram, but its support on the phase space for physical external particles is zero and therefore we don't consider it.} and we express the equation in terms of partial-wave coefficients $f_\ell$
\bea
2 \, \text{Im} \, M(s,t) &= \sum_\ell n_\ell^{(4)} 2 \im f_\ell(s) P_\ell(1+\tfrac{2t}{s-4})=\rho^2(s) \, \sum_\ell n_\ell^{(4)} \, |f_\ell(s)|^2  P_\ell(1+\tfrac{2t}{s-4})+O(g_0^4) \nonumber \\
&= \rho^2(s) \left( \frac{g_0^2}{16 \pi} - g_0 \, 2 \text{Re} f^\text{1-loop}_0 + O(g_0^4) \right)
\eea
where $\rho^2(s) = \sqrt{s-4}/\sqrt{s}$ is two-particle phase space factor, and $n_\ell^{(4)}=16 \pi (2\ell+1)$.
The first term gives us the leading contribution as mentioned in \reef{phi4oneloop}, and the second term is what we need at $O(g_0^3)$, so plug in the following:
\bea
\text{Re } f^\text{1-loop}_0 = \frac{1}{32\pi} \int_{-1}^{1} dx \, \frac{g_0^2}{16 \pi^2} \Big[ \text{Re } f(s) + f(t') + f(u') \Big] \\
t' = (s-4)(-1+x)/2 \quad , \quad \quad u' = (s-4)(-1-x)/2
\eea
Notice that $f(t')$ and $f(u')$ are purely real functions in the region $s>4$, which is the same interval as in the dispersion integrals for $c_2$ and $c_3$. For the real part of the first term, we used the following nice formula: $\arctan y = 1/2i \log (1+iy)/(1-iy) $. \\
After analytically obtaining $\text{Im }M$ given above, we can plug it in the dispersion integrals \reef{c2def} and \reef{c3sr}, then evaluate them numerically. This will give us
\be
c_2^{O(g_0^3)} \approx 1.16 \cdot 10^{-7} \, g_0^3\quad  , \quad  \quad 
c_3^{O(g_0^3)} \approx -1.25 \cdot 10^{-7} \, g_0^3 \ . 
\label{c2_2loop}
\ee
  

\section{Subtracted dispersion relations for the $O(n)$ theory}
\label{appdispN}

In this appendix we provide further details for the derivation of the sum rule \reef{cHsr}.
The starting point the the fixed-$t$ doubly-subtracted dispersion relation of Roy~\cite{Roy:1971tc} adapted to global  $O(n)$ symmetry.
After blowing up the contour $\frac{s^2}{2\pi i}\oint dz\frac{1}{z^2}\frac{1}{z-s} \overrightarrow{ M} (z,t)$, we are lead to
\be
\overrightarrow{ M} (s,t)= C_{st}. \left[  \vec{c}(t) + (s-u)  \vec d(t) \right] + \frac{1}{\pi} \int_4^\infty dzK(z;s,u). \overrightarrow{M}_z(z,t) \label{dispN}
\ee
where the kernel is a matrix given  by
\be
K(z;s,u)=\frac{1}{z^2} \left(  \frac{s^2}{z-s}  \mathds{1} +  \frac{u^2}{z-u} C_{su}\right) \, . 
\ee
and recall that we are working in $m^2=1$ units, and  that $ M_z(z,t) \equiv \text{disc}_z M(z,t)/(2 i )$. 
The vector $\overrightarrow M=(M^\text{sing}, M^\text{sym},  M^\text{anti})$ is given by the irreps
\bea
M^\text{sing} &=n \, M(\bar s | \bar t , \bar u) + M(\bar t | \bar u , \bar s) + M(\bar u | \bar s , \bar t)  \, , \\[.2cm]
M^\text{sym}&= M(\bar t | \bar u , \bar s) + M(\bar u | \bar s , \bar t) \, ,   \\[.2cm]
M^\text{anti }&= M(\bar t | \bar u , \bar s) - M(\bar u | \bar s , \bar t) \, . 
\eea
The crossing matrices are given by
\be
C_{st} =
 \left(
\begin{array}{ccc} \frac{1}{n} & \frac{n^2+n-2}{2 n} & \frac{n-1}{2} \\ \frac{1}{n} & \frac{1}{2}-\frac{1}{n} & -\frac{1}{2} \\ \frac{1}{n} & -\frac{n+2}{2 n} & \frac{1}{2} \\\end{array}
\right)   \quad , \quad \quad 
C_{su}=  \left(
\begin{array}{ccc}
 \frac{1}{n} & \frac{n^2+n-2}{2 n} & \frac{1-n}{2} \\
 \frac{1}{n} & \frac{1}{2}-\frac{1}{n} & \frac{1}{2} \\
 -\frac{1}{n} & \frac{1}{n}+\frac{1}{2} & \frac{1}{2} \\
\end{array}
\right) \, , 
\ee
and satisfy  $C_{st}^2=C_{su}^2=1$. 
The dispersion relation in \reef{dispN} is composed of the integral over  the discontinuity  $\overrightarrow{M}_z(z,t) $
and the subtraction functions 
\be
\vec{c}(t)= \left(\begin{matrix} c^\text{sing}(t) \\ c^\text{sym}(t) \\ 0 \end{matrix} \right)   \quad , \quad \quad  
\vec{d}(t)=\left(\begin{matrix} 0 \\ 0 \\ d^\text{anti}(t) \end{matrix} \right)  \label{vectors}
\, . 
\ee 
This form of the vectors are dictated by the explicit $s \leftrightarrow u$ symmetry property of \reef{dispN}.

Next we would like to eliminate these subtraction functions in \reef{dispN} in favour  of the amplitude $\overrightarrow{M}(s_0,t_0)$ and the integrals over the discontinuities $\overrightarrow{M}_z(z,t) $.
In order to do so we follow Roy's strategy~\cite{Roy:1971tc}.
Using  first 
 crossing-symmetry  $\overrightarrow{M}(t_0,t)= C_{st}. \overrightarrow{M}(t,t_0)$  we have 
 \be
 C_{st}. \left[ \vec c(t) +(2t_0-4+t) \, \vec d(t)  \right]= \left[ \vec c(t_0) +(2t-4+t_0) \, \vec d(t_0)  \right] + \int \text{\emph{absorptive pieces}} \label{sk1}
 \ee
 the second term schematically denotes an integral of the absortive part of the amplitude $M_z(z,t)$ or  $M_z(z,t_0)$ against various kernels.
 The second piece  $ \vec c(t_0) +(-4+2t+t_0) \vec d(t_0) $ can  be expressed in terms of absorptive pieces and $\vec M(s_0,t_0)$ using \reef{dispN}
 \be
  \vec c(t_0) +(t_0-4+2s_0) \vec d(t_0)  =  \overrightarrow M(t_0,s_0) + \int \text{\emph{absorptive pieces}}
 \ee
 Because the components of the vectors  \reef{vectors} do not mix we can easily solve for them. 
 Namely,  
 we plug the last equation into \reef{sk1}   leading to    $ c^{\text{sing}},  c^{\text{sym}}, d^{\text{anti}}\sim \overrightarrow M(t_0,s_0) + \int \text{\emph{absorptive pieces}}$.
 Plugging this solution in \reef{dispN} we arrive to a dispersion relation expressed in terms of an arbitrary subtraction point $M(s_0,t_0)$,
 i.e. $\overrightarrow M(s,t) \sim  C'. \overrightarrow{ M} (s_0,t_0) +  \int \text{\emph{absorptive pieces}}$, where $C^\prime(s,t;s_0,t_0)$ is a matrix easy to determine.
 
 Having outlined the logic, it is now a matter of algebra to derive   \reef{cHsr} from
  $\vec M(s,t) \sim  C'. \overrightarrow{ M} (s_0,t_0) +  \int \text{\emph{absorptive pieces}}$ and $c_H = \frac{1}{(n-1)} \frac{\partial}{\partial \bar s} M^\text{sing}(\bar s,\bar t)\big |_{\bar s=\bar t=0}$.
  We find  that the vectors and integration kernels in \reef{cHsr} are given by
  \bea
C(s,t)& = \frac{1}{2n} \frac{(3 s-4)}{(3 t-4)} \frac{1}{(s-u)} \left(-1 \, , \, \frac{n+2}{2} \, , \, \frac{n (9 s+6 t-20)}{2 (3 s-4)}\right)   \,  , \\[.2cm]
D_1(z,s,t) &= \frac{1}{2n} \frac{(3s-4)}{(3t-4)} \frac{(3 s+3 t-8)}{(3 t+3 z-8)} \frac{1}{(z-s)(z-u)} \, \left(1 , \frac{-n-2}{2} , -\frac{n (6 t+9 z-20)}{2 (3 z-4)}\right) , \\[.2cm]
D_2(z,t)   &=  \frac{3}{n} \frac{(3 t-4)^2}{(3 z-4)^2} \frac{1}{(t-z) (3 t+3 z-8)} \left( 1 \, , \, \frac{-n-2}{2} \, , \, \frac{n}{2}\right) \, . 
  \eea
 In terms of crossing-symmetry matrices we find that  the $c_H$  sum rule is given by   following vector-like dispersion relation
 \be
c_H \left( \begin{matrix}n-1\\ -1 \\ 1\end{matrix}\right)=F_2(t).F_1(s,t). \overrightarrow{M}(s,t)  + \int_4^\infty \frac{dz}{\pi} \left[K_1(z,t). \overrightarrow M_z(z,t)+  K_2(z,t). \overrightarrow M_z(z,4/3)   \right] \ , 
\ee
where 
\bea
K_1(z,s,t)& =  F_2(t). \left[ K(z;4/3,8/3-t)-F_1(s,t).K(z;4-s-t)\right]   \, ,    \nonumber\\[.2cm]
  K_2(z,t) & =   K^\prime(z) -F_2(t).C_{st}.K(z;t,8/3-t)  \, , \nonumber
 \eea
 and  $ K^\prime(z)= \partial_s K(z;s,4-t-s)\big|_{{t,s=4/3}}$  and
 \be
 F_1 (s,t)=\mathds{1}+\frac{4/3-s}{s-u}(\mathds{1}-C_{su}) \quad  , \quad \quad  
 F_2(t)=  \frac{1}{2}\frac{1}{t-4/3}C_{st}.(\mathds{1}-C_{su}) \, . 
 \ee


\section{A toy model}

In this section we develop the ideas presented in the main text  in a simplified setting where we scatter massive particles in a line, i.e. in $d=1+1$ spacetime dimensions. 
The theory is invariant under a spacetime parity transformation and there is no scattering angle, therefore 
there is a single kinematical invariant given by the Mandelstam variable $s=(p_1+p_2)^2$.
We assume a $Z_2$ symmetry under which the particle we scatter is odd
, thus forbidding the triple vertex 
 interaction.

\begin{figure}[t] \centering
 \includegraphics[scale=.25]{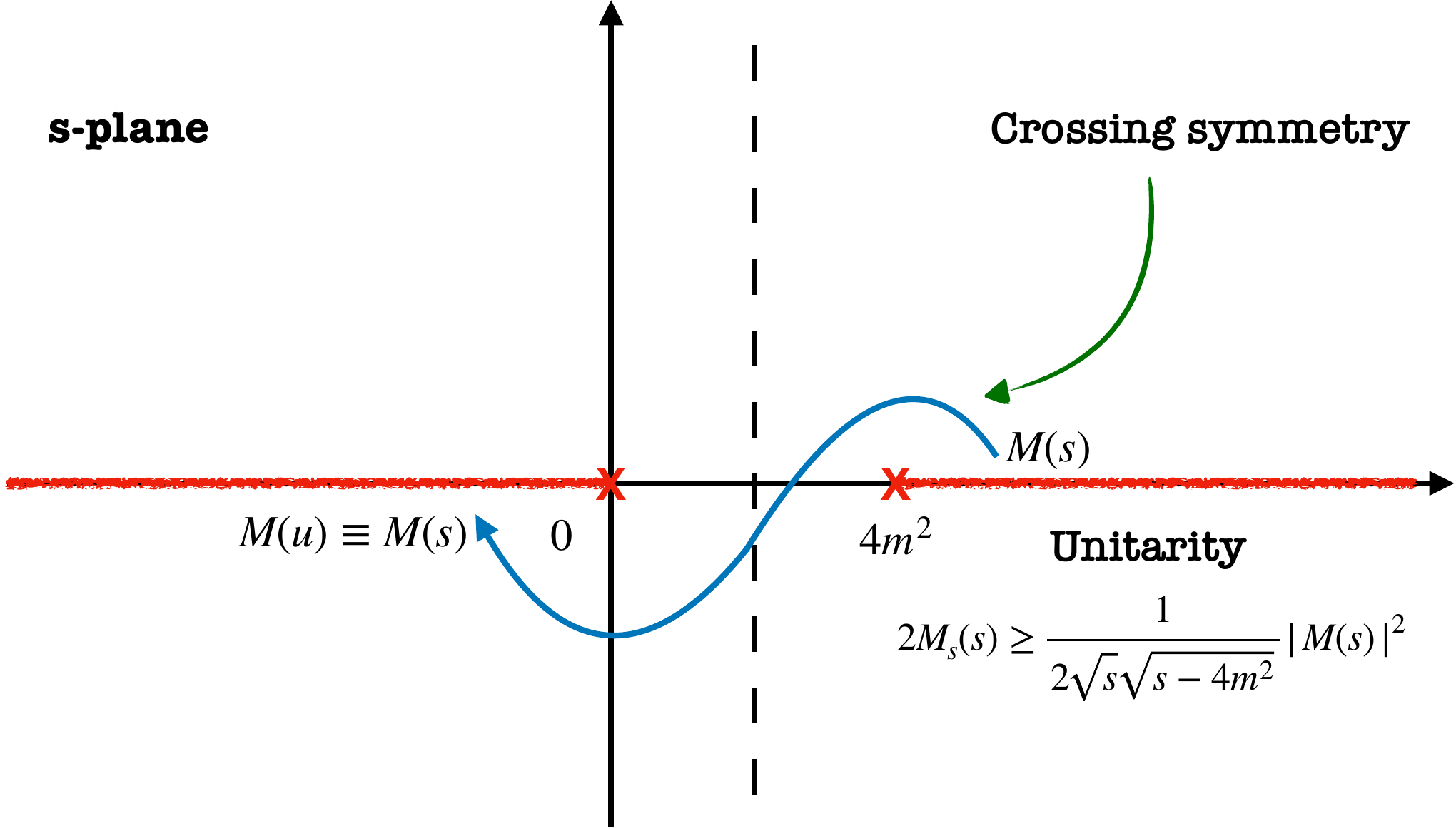} 
 \caption{ Sketch of the analytical structure of the two-dimensional amplitude. }
  \label{2d_complex_plane} \end{figure}

 The analytic structure of $M$ is sketched in Fig.~\ref{2d_complex_plane}. 
  Crossing symmetry implies $M(s-2m^2) = M(2m^2-s)$, and there is a unitary branch-cut starting at $s=(2m)^2$ (and its crossed starting at $s=0)$ and extending all the way to $s=\infty$ (and $s=-\infty)$. 
 The scattering matrix element  $M$ is assumed to be analytic everywhere in the complex plane away from the real axis and in the segment $s\in (0,4m^2)$. In particular, it is analytic in the crossing-symmetric point  $s - 2m^2 = 0$, and thus
 we can represent the amplitude in terms of a series around this point. Due to crossing-symmetry, this series involves even powers of $s-2m^2$ only. 
It is convenient to define  
\be
2m^2 \bar{s} \equiv s-2m^2  \quad \text{and} \quad  \eps \equiv \frac{2m^2}{\Lambda^2} \, , \label{defs}
\ee
with $\Lambda$ an additional scale.  
With these two definitions,  we can represent the scattering  amplitude     by 
\be
M(\bar{s}) = m^2 \left( -c_0 + c_2 ( \eps \hs)^2 + c_4 (\eps \hs )^4 +  c_6 (\eps \hs )^6 + \dots \right)  \, .
\label{lowampnoflav}
\ee
where we have factored out   $m^2$ to account for the dimensions --  recall that the   two-to-two  scattering amplitude has  energy dimension $4-d$.  The coefficients $c_i$ are in general functions $c_i(\epsilon)$; and we have factored out suitable powers of $\epsilon$ to match the power of $\bar s$, $(\eps \bar s)^n$. As we show below, this power counting is useful in order to interpret the amplitude \reef{lowampnoflav} in terms of an Effective Field Theory.

The S-matrix~\footnote{The S-matrix operator is given by 
$\hat S(s)=\mathds{1}+(2\pi)^d\delta^{(d)}(p_1+p_2-p_3-p_4)M(s)$, where as usual $\mathds{1}= \langle p_1,p_2|p_3,p_4\rangle=\prod_{i=1,2}((2\pi)^{(d-1)} 2E_i)(\delta^{(d-1)}(\vec k_1- \vec k_3)\delta^{(d-1)}(\vec k_2- \vec k_4)+ [\vec k_1 \leftrightarrow \vec k_2])$.
In $d=2$, using the relation $\delta(x)\delta(y)=\delta([x+y]/2)\delta([x-y]/2)$, we can relate the total momentum conservation delta function $\delta^{(d)}(p_1+p_2-p_3-p_4)$ to the identity operator $\mathds{1}$ and obtain $\hat S(s)= \mathds{1}S(s)= \mathds{1}(1+  \frac{1}{2\sqrt{s}\sqrt{s-4m^2}} iM(s))$.}  is given by
 $
S(\bar{s}) 
= 1 + \frac{1}{4 \sqrt{1-\bar{s}^2}}  \left[ \,  -c_0 + \sum^{\infty}_{i=1} c_{2i} (\eps \bar{s})^{2i} \, \right]  $.
In order to encode the analytic structure of $M(s)$ it is convenient to 
map the complex $s$-plane into the unit disk using the following map
\be
\rho(\bar{s}) = \frac{ i - \sqrt{\bar{s}+1}\sqrt{\bar{s}-1} }{i + \sqrt{\bar{s}+1}\sqrt{\bar{s}-1}}  \, . \label{map}
\ee
The map is crossing-symmetric ($\hs\rightarrow -\hs$), and thus maps only the upper half plane into the unit disk. 
This transformation opens the $s$-plane branch cuts and map them into the boundary of the unit disk, 
it has the fixed point  $\rho(\bar{s} = 0)=0$, and  the thresholds of physical  scattering  energy map to $\rho(\bar{s} = \pm 1) = 1$.
Now, upon expanding the S-matrix in the unit disk around $\rho=0$ we have
\be
\begin{aligned}
S(\rho) &= (1 - c_0/4) + \left( c_2 \eps^{2} - c_0/2 \right) \rho +  \left( 4 c_4 \eps^{4} - c_0/2 \right) \rho^2 \\
&+ \left( c_2 \eps^{2} - 8 c_4 \eps^{4} + 16 c_6 \eps^{6} - c_0/2 \right) \rho^3 + \mathcal{O}(\rho^4)  \, .  \label{smat1}
\end{aligned}
\ee
Note that  the expansion in (\ref{smat1}) has  even and odd powers in $\rho$,  because we have solved for crossing-symmetry through the map \reef{map}. 
Unitarity of the S-matrix implies
\be
|S|^2\leq 1  \label{unit1}
\ee
for physical energies $s\in[ 4m^2,\infty)$; or equivalently  in  the $\rho$-plane  for $\arg \rho \in [0,\pi)$ and $|\rho|=1$.

The amplitude $M(s)$ looks like   a higher dimensional amplitude in the  forward limit  ($t=0$). 
This is true regarding the analytic and crossing-symmetry properties of $M$. However, unitarity is much simpler in the $d=1+1$ setting, see \reef{unit1}. The large simplification of this toy model can be summarised by noting that the system of equations we are studying is as if we had a single partial wave \reef{unit1}. In higher dimensions the presence of infinitely many partial waves~\footnote{Often defined by projecting with Legendre polynomials $P_\ell$,  $S_\ell\equiv 1+i \sqrt{\frac{s-4}{s}}\int_{-1}^1 dx P_\ell(x) M(s,t)|_{t\rightarrow (s/2-2)(x-1)} $.}    makes  crossing symmetry and the corresponding unitarity  equations --   $|S_\ell(s)|^2\leq 1$ with $\ell=0,1,2,\dots$ -- much harder to analyse.

\subsection{Bounds on LECs}

In this appendix, we  shall call the parameters $c_i$ in \reef{smat1} \emph{low energy constants} (LECs). 
Clearly, in perturbation theory the LECs  are  identified with couplings and Wilson coefficients of a putative Effective Field Theory Lagrangian, we comment more on this identification later on in section~\ref{eft2}.

The low energy couplings $c_i$ in \reef{smat1} can  be optimally bounded as in ref.~\cite{EliasMiro:2019kyf}.~\footnote{We emphasise that ref.~\cite{EliasMiro:2019kyf} considered the EFT of massless goldstones (associated to transverse fluctuation of a relativistic flux tube) while here we are interested in the EFT of massive particles.}
The  idea is to exploit the nice rigidly-smooth shape that holomorphic functions  have, which we will now briefly review.

The   function $S(\rho)$ is holomorphic in the unit disk $\rho \in \mathds{D}$, therefore it is equal to the average of neighbouring points $S(z)=\oint_{|w-z|=\eps}\frac{dw}{2\pi i}\frac{S(w)}{z-w}=\int_0^{2\pi}\frac{d\th}{2\pi}S(z+\eps e^{i\th})$, for $z\in \mathds{D}$.
Thus, the modulus $|S(z)|$ is  bounded  in the region of analyticity by the value of the function at the boundary $\partial \mathds{D}$. 
This is the content of the Maximum Modulus Principle. In our case, the modulus of $S(z)$ is bounded by one at the boundary of the unit disk  \reef{unit1} and therefore  $S(z)$ is bounded everywhere inside the disk by one. 
In particular it implies
\be
|S(\rho=0)| =|1-c_0/4|\leq 1 \, .  \label{b1}
\ee
for our S-matrix \reef{smat1}. The bound in \reef{b1} is often interpreted as a bound on the maximal coupling, defined precisely as the value of $M$ at the crossing-symmetric point. A similar analysis can be done in the presence of bound states, in order to constrain the maximal residue at the pole~\cite{Creutz:1972ikj,Paulos:2016but}.

It turns out that the derivatives  $|\partial_\rho^n S(\rho)|_{\rho=0}$  are also bounded as a consequence of the  
Schwarz-Pick multi-point lemmas. The basic logic goes as follows. 
Consider the function of $z$
\be
S^{[1]}(z|w)=\frac{S(z)-S(w)}{1-S^*(w)S(z)} \left(\frac{z-w}{1-w^*z}\right)^{-1}
\ee
with  $w\in\mathds{D}$. Clearly,  $S^{[1]}(z|w)$ is a holomorphic function of $z\in\mathds{D}$.
As a function of $z$ (and for  $w\in\mathds{D}$), it is bounded   $|S^{[1]}(e^{i \phi}|w)|^2\leq 1$ in the boundary $\partial\mathds{D}$. 
Therefore we can apply the Maximum Modulus Principle, and conclude   $|S^{[1]}(z|w)|^2\leq 1$  for  $z\in\mathds{D}$.
Upon taking the  limit $w\rightarrow z \rightarrow 0$  on the last inequality we get
\be
\Big| \, \frac{ c_2 \eps^{2} - c_0/2}{1 -  (1 - c_0/4)^2} \, \Big|  \leq 1 \, , \label{sps1}
\ee
for the S-matrix in \reef{smat1}. 
Clearly one can now recurse over this construction and define the function 
$S^{[n]}(z| w_1, \dots, w_{n} )$ out of  $S^{[n-1]}(z|w_1, \dots, w_{n-1})$, which is a holomorphic function of $z\in\mathds{D}$ (with $w_i\in \mathds{D}$) and bounded in the boundary $|S^{[n-1]}(e^{i\phi}|w_1, \dots, w_{n-1})|\leq 1$.
The content of the multi-point Schwarz-Pick lemma is the bound on  $|S^{[n]}|$, which 
implies a  bound on the  $n$-th derivative of  $S(z)$.  
See section~\ref{SPapp} for a summary of the Schwarz-Pick formulas we  use in this section. Further details on the theory of Schwarz-Pick applied to S-matrices can be found in  ref.~\cite{EliasMiro:2019kyf}.

\begin{figure}[t] \centering
 \includegraphics[scale=.69]{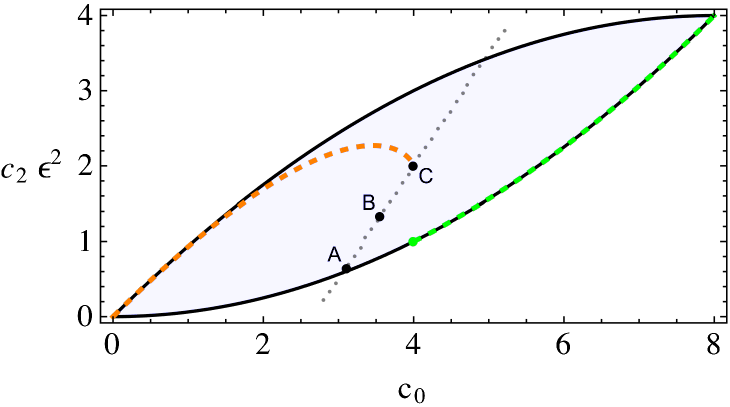} 
 \includegraphics[scale=.71]{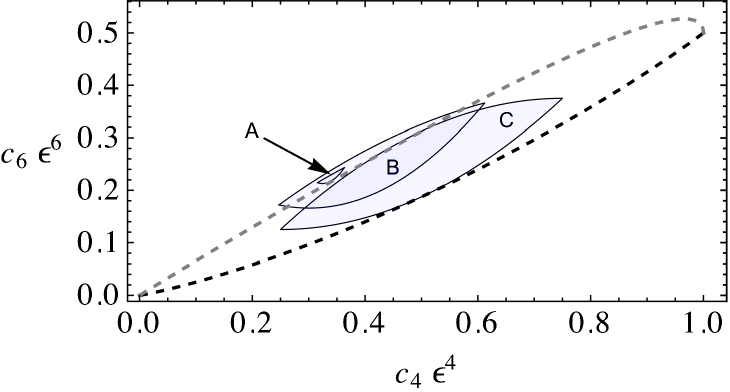} 
 \caption{Left plot: In blue the allowed region in the  $c_0 - c_2 \eps^2$ plane. 
 In orange and green we plot the values of the $T\bar T$ S-matrix in \reef{TTbar} and the \emph{Roaming Trajectories} S-matrix, respectively, as they approach the EFTs of a massive free boson and massive free fermion on the lower and upper cusp.   Right plot:  Regions of allowed values for $\{c_4\eps^4,\, c_6 \eps^6  \}$.  }
  \label{c0c2} \end{figure}

In figure~\ref{c0c2}, left plot, we show the allowed region of the first two LECs $(c_0, c_2 \eps^2)$.
 The boundary of the blue region is described by two parabolas and saturates  \reef{sps1}.
At the cusps (0,0) and (8,4) both of inequalities \reef{b1} and \reef{sps1} are saturated. 
The boundary in figure~\ref{c0c2} is saturated by  the LECs of the following functions
\be
S_\text{up}(\rho, \rho_0) = \frac{\rho - \rho_0}{1 - \rho \rho_0} \quad \text{and} \quad S_\text{low}(\rho, \rho_0) = - \frac{\rho + \rho_0}{1 + \rho \rho_0}, \label{exactsmat1}
\ee
with $\rho_0\in [-1,1]$, parametrising the upper and lower branch respectively.~\footnote{ 
These functions are  called CDD-factors (for Castillejo-Dalitz-Dyson) in the two-dimensional S-matrix literature  or Blaschke products in the context of the Schwarz-Pick theorems. }
Indeed, after matching the lowest LEC of our S-matrix \reef{smat1} with   $\rho_0= (c_0/4-1 )$, we have 
\be
S_\text{up}(\rho, c_0/4-1)=\left(1-c_0/4\right)+    \big(\underbrace{1-\left(c_0/4-1\right)^2+c_0/2}_{c_2\eps^2}-c_0/2\big) \rho + O(\rho^2) \, , 
\ee
and similarly for the lower branch. 
The upper branch has a resonance at $S_\text{up}(\rho_0,\rho_0)=0$, while the lower branch at $S_\text{low}(-\rho_0,\rho_0)=0$.~\footnote{See for instance  ref.~\cite{Doroud:2018szp} for a discussion of resonances. }
Varying  $\rho_0$ parametrises  the edges connecting  the two cusps along the upper and lower branch. 
The cusp at the origin $S_\text{up/low}(\rho, -1) = 1$  is a free bosonic theory, and the other one $S_\text{up/low}(\rho, 1) = -1$ is a free fermionic theory.

Approaching the lower cusp at $(0,0)$ through the upper branch  can be described  by an Effective Field Theory of a single massive scalar because,  for a  small positive $\delta$,
\be
S_\text{up}(0, \delta-1) =1-\delta + O(\delta)  \, , 
\ee
the resonance of $S_\text{up}$ is at $\rho=\delta-1$, which corresponds to a resonance at $s=\infty$ as $\delta\rightarrow 0$.
 If instead we approach the cusp at the origin from the lower branch 
 $S_\text{low}(0,\delta-1)=1+O(\delta) $, the resonance  is at $\rho=1-\delta$,  thus at threshold $\rho(\hat s=1)=1$ as $\delta\rightarrow 1$.
A particular Effective Field Theory example touching the cusp at the origin is provided by the $T\bar T$ deformation of a free massive boson~\cite{Dubovsky:2012wk,Dubovsky:2013ira,Cavaglia:2016oda}. In terms of the $\rho$ variable, the S-matrix is given by 
 \be
 S(\rho)=e^{-\frac{\eps}{4}\sqrt{\frac{(1-\rho)^2}{(1+\rho)^2}}}  \, .  \label{TTbar}
 \ee
 The coefficients $c_0(\eps)$ and $c_2(\eps)\eps^2$ can be read from the last equation after performing the Taylor series around $\rho=0$
 and matching with equation \reef{smat1}.  
 In figure~\ref{c0c2}, left, we plot with a dashed orange curve  the values of $\{c_0(\eps),c_2(\eps)\eps^2\}$ as we vary the parameter $\eps\in[0,\infty)$. When the cutoff is increased (i.e. $\eps$ decreased) the values of $\{c_0(\eps),c_2(\eps)\eps^2\}$  get close to the origin monotonically. 
 
A similar analysis holds for the upper cusp (8,4): if approached through the lower branch it is described by an EFT of a massive fermion, while if approached from the upper branch the scattering of the almost free fermions features a resonance at threshold $\hat s =(s-2m^2)/(2m^2)=1$. 
 We can identify a theory approaching the $(8,4)$ cusp from the lower branch: the \emph{Roaming Trajectories} S-matrix~\cite{Zamolodchikov:1992ulx}, which is in fact given by $S_\text{up}(\rho, \rho_0) $ with $\rho_0$ restricted to the range $\rho_0\in(-1,0]$.
 For these values of the parameter $\rho_0$, the S-matrix features a resonance at purely imaginary values, and thus  $\rho_0$ increases monotonically as $\eps$ decreases. In figure~\ref{c0c2}, left,  we plot with a dashed green curve  the values of $\{c_0(\eps),c_2(\eps)\eps^2\}$ for this S-matrix as $\eps$ is varied.  
The values $\rho_0\in[0,1)$ corresponds instead to a resonance in the range  $[0,4m^2)$ in the Mandelstam $s$-plane, and thus it is  identified with the Sinh-Gordon S-matrix.

In the right plot of  figure~\ref{c0c2}, we show allowed values of $\{c_4\eps^4, c_6 \eps^6\}$ for fixed values of $c_0$ and $c_2\eps^2$. 
Each region A, B and C of the right plot corresponds to the    points A, B and C shown on the left plot.  
As we approach the boundary of the $c_0 - c_2\eps^2$ allowed region, i.e. as we move along the line $C\rightarrow B \rightarrow A$, the region shrinks, and reduces to a point when hitting the boundary. At the boundary of the allowed region in the $c_0 - c_2\eps^2$ plane, the values of     $\{c_4\eps^4, c_6 \eps^6\}$ are fixed by the S-matrices in \reef{exactsmat1}.
These values are shown by the gray and black curve, which correspond to  $S_\text{low}$ and $S_\text{up}$, respectively.

\subsection{Positivity constraints and the space of Effective Field Theories}
\label{eft2}

\begin{figure}[t]
        \centering
        \includegraphics[width=.4\textwidth]{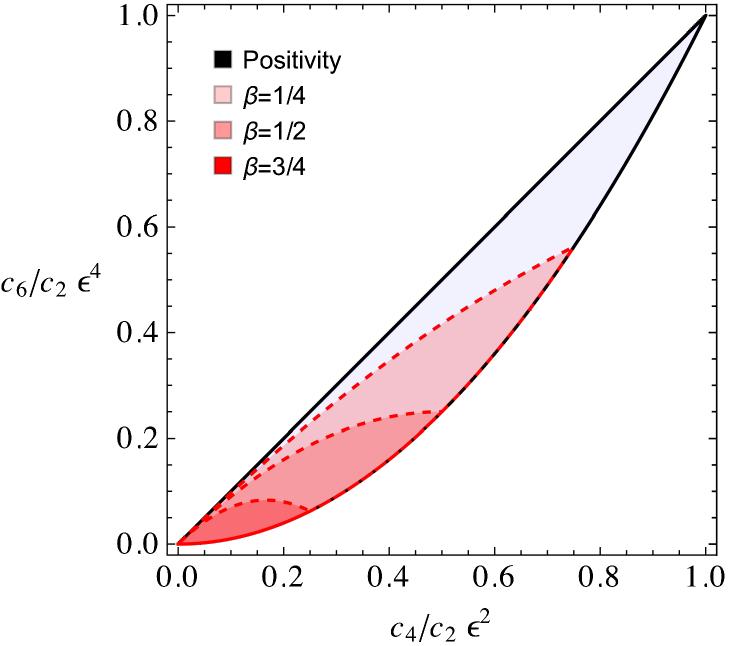}
    \caption{We show in black the bounds from the positivity constraint \reef{pos1}, while with dashed lines we show the bounds that follow from the exact  $2\rightarrow 2$ unitarity constraints   \reef{unit1}, a.k.a. S-matrix bootsrap. }
    \label{figpos}
\end{figure}

Next we compare the optimal bounds  on the LECs just obtained with  positivity. 
We use the 
 so-called arc variables~\cite{Bellazzini:2020cot}  
 \be
a_n(x) \equiv \frac{1}{m^2} \frac{2}{\pi} \int_{x}^{\infty} du \frac{\text{Im} M(u)}{(2m^2 u)^{2n+3}} \quad , \quad n \geq 0 \, , \label{arc1}
\ee
and define $a_n\equiv a_n(1)$. The following optimal constraints are satisfied
\be
\left(\begin{array}{cc} a_0 & a_1 \\ a_1 & a_2 \end{array} \right) \geq 0 \, , \ \ \ a_1\geq 0 \, , \ \ \  a_0 \geq \hat s^2 a_1 \, , \  \ \ a_1 \geq \hat s^2 a_2 \, ,
\label{cpos}
\ee
as a consequence of \emph{positivity}
\be
\text{Im}M \geq 0 \, . \label{pos1}
\ee
The last equation  follows from the $2\rightarrow 2$ S-matrix unitarity equation \reef{unit1}.
Next we  modify the contour of integration and relate the arc variables to the LECs 
\be
a_n = c_{2n+2} (\eps/(2m^2))^{2n+2}
\ee
where recall that $\eps/(2m^2)=1/\Lambda^2$.
In $d=3+1$ the convergence of \reef{arc1}  at high energies is justified thanks to the Froissart-Martin bound, proved for theories with a mass gap. 
In the set up we are discussing, $d=1+1$,  the convergence of \reef{arc1} is a simple consequence of the unitary equation  \reef{unit1}, and is  valid in the massless limit. 
We  summarise the positivity bounds \reef{cpos}  in figure~\ref{figpos}: all theories consistent with the positivity constraint must take values inside the solid black region (with $a_0,\, a_1>0$). 

An interesting question is to  compare  these constraints with the optimal bounds that follow from the exact $2\rightarrow 2$ unitary equation that we showed  in figure~\ref{c0c2}.
Clearly the bootstrap bounds in figure~\ref{c0c2} are in general more constraining than the positivity bounds \reef{cpos},  and thus   the allowed values in figure~\ref{c0c2} lie inside the black solid line of figure~\ref{figpos}.

We also note that   deep inside the region of validity of the EFT the positivity bounds are not saturated, and  the theory obeys stronger bounds, namely the bootstrap bounds. 
In order to show this, it is convenient to define 
\be
C_4\equiv \frac{c_4 \eps^4}{c_2\eps^2}= \frac{a_1}{a_0} (2m^2)^2 \quad \text{and}  \quad C_6\equiv  \frac{c_6 \eps^6}{c_2\eps^2}= \frac{a_2}{a_0} (2m^2)^4  \, . 
\ee
These variables  are singular as we take the weak coupling limit  $c_2 \eps^2 \xrightarrow{} 0$. In this limit, we approach the origin in the $c_4 \eps^4 -  c_6 \eps^6$ plane. 
The regions inside the positivity plot of figure~\ref{figpos} depend on the angle at which we approach the origin of the $c_0 - c_2 \eps^2$ plane, i.e. the free theory. 
We define the angle by the   $\beta \equiv c_2 \eps^2 / c_0$. 
In figure~\ref{figpos} we plot with dashed lines the allowed regions  for different values of $\beta$.
Interesting cases are the extremal values: $\beta=0$ filling completely the banana region and $\beta=1$ shrinking to the origin. Both cases can be understood  by analysing   the  Schwarz-Pick  inequalities in  \reef{sp2} and \reef{sp3}.
For $\beta \xrightarrow{} 1$ these inequalities are given by 
\be
1 \geq |1 + 2 C_4/(\beta-1)+O(c_0,\beta-1)| \quad \text{and} \quad 1 \geq |-1+O(c_0,\beta-1)| \ , \label{eftlimit}
\ee
which implies $C_4=0$ and thus $C_6 = 0$; instead,   the limit $\beta \xrightarrow{} 0$ leads to 
\be
1 \geq |1 - 2  C_4+O(c_0,\beta)| \quad \text{and}\quad 
 1 \geq |(C_4 + C_4^2 - 2 C_6)/(C_4^2 - C_4)+O(c_0,\beta)| \, , \label{secondlimit}
\ee
which are equivalent to the arc inequalities $C_4> C_6$ and $C_4^2 < C_6$, and therefore to the region plotted in black in figure~\ref{figpos}.

Our observations provide a new interpretation of \reef{cpos} in terms of $d=1+1$ S-matrices.  
These equations were   analysed in ref.~\cite{Bellazzini:2020cot} in the context of $d=3+1$.
On one hand, there it was observed that theories in the far IR  flow to the origin of the $c_0 - c_2\eps^2$ plane along trajectories compatible with the positivity bounds (i.e. by flowing in the reverse direction pointed by the arrows of figure~2, left, in  ref.~\cite{Bellazzini:2020cot}). 
This corresponds to our limit in \reef{eftlimit}. Interestingly as we approach the free theory  along various directions in the $c_0-c_2\eps^2$ plane we can place tighter bounds than  the positivity bounds. 
On the other hand it was observed that tree-level theories with resonances at threshold of the lower limit of the arc integrations do saturate the bounds. 
This corresponds to our second limit in \reef{secondlimit}. Here however we are not working at tree-level, but with exact S-matrices. 
As we approach the free theory though the lower branch of the left plot in figure~\ref{c0c2}, the S-matrix $S_\text{low}$ in \reef{exactsmat1}
features a bound state that goes to threshold, giving $\text{Im}M_\text{low}\propto  \,  \delta(s-4m^2+0^+)$.

\subsection{Schwarz-Pick inequalities}
\label{SPapp}
Let $f(z)$ be a real analytic function defined inside the unit disk $|z| \leq 1$ with real Taylor coefficients $\alpha_i \in \mathbb{R}$.
\be
f(z) = \alpha_0 + \alpha_1 z + \alpha_2 z^2 + \dots
\ee
If $f(z)$ is bounded along the boundary of the disk i.e. $|f(z)| \leq 1$ for $|z|=1$, its Taylor coefficients obey the following inequalities
\begin{align}
1 &\geq | \alpha_0 | \label{sp0} \\
1 &\geq \Big| \, \frac{\alpha_1}{1 - \alpha_0^2} \, \Big| \label{sp1} \\
1 &\geq \Big| \, \frac{\alpha_0 \alpha_1^2 + \alpha_2 - \alpha_0^2 \alpha_2}{1 - 2 \alpha_0^2 + \alpha_0^4 - \alpha_1^2} \, \Big|  \label{sp2} \\
1 &\geq \Big| \frac{2 \alpha_0 \alpha_1 \alpha_2 - 2 \alpha_0^3 \alpha_1 \alpha_2 + \alpha_1 \alpha_2^2 + 
     \alpha_0^2 (\alpha_1^3 - 2 \alpha_3) + \alpha_3 + \alpha_0^4 \alpha_3 - 
     \alpha_1^2 \alpha_3}{-1 - 3 \alpha_0^4 + \alpha_0^6 + 2 \alpha_1^2 - \alpha_1^4 + 
     2 \alpha_0 \alpha_1^2 \alpha_2 + \alpha_2^2 - \alpha_0^2 (-3 + 2 \alpha_1^2 + \alpha_2^2)} \Big|  \label{sp3} 
\end{align}
These are obtained from  the Maximum Modulus Principle,  and the  1st, 2nd, 3rd Schwarz-Pick inequalities,  respectively.



\small

\bibliography{biblio}

\providecommand{\href}[2]{#2}\begingroup\raggedright\begin{thebibliography}{10}

\bibitem{Adams:2006sv}
A.~Adams, N.~Arkani-Hamed, S.~Dubovsky, A.~Nicolis, and R.~Rattazzi,
  ``{Causality, analyticity and an IR obstruction to UV completion},''
  \href{http://dx.doi.org/10.1088/1126-6708/2006/10/014}{{\em JHEP} {\bfseries
  10} (2006) 014}, \href{http://arxiv.org/abs/hep-th/0602178}{{\ttfamily
  arXiv:hep-th/0602178}}.

\bibitem{Pham:1985cr}
T.~N. Pham and T.~N. Truong, ``{Evaluation of the Derivative Quartic Terms of
  the Meson Chiral Lagrangian From Forward Dispersion Relation},''
  \href{http://dx.doi.org/10.1103/PhysRevD.31.3027}{{\em Phys. Rev. D}
  {\bfseries 31} (1985) 3027}.

\bibitem{Pennington:1994kc}
M.~R. Pennington and J.~Portoles, ``{The Chiral Lagrangian parameters, l1, l2,
  are determined by the rho resonance},''
  \href{http://dx.doi.org/10.1016/0370-2693(94)01551-M}{{\em Phys. Lett. B}
  {\bfseries 344} (1995) 399--406},
  \href{http://arxiv.org/abs/hep-ph/9409426}{{\ttfamily arXiv:hep-ph/9409426}}.

\bibitem{Ananthanarayan:1994hf}
B.~Ananthanarayan, D.~Toublan, and G.~Wanders, ``{Consistency of the chiral
  pion pion scattering amplitudes with axiomatic constraints},''
  \href{http://dx.doi.org/10.1103/PhysRevD.51.1093}{{\em Phys. Rev. D}
  {\bfseries 51} (1995) 1093--1100},
  \href{http://arxiv.org/abs/hep-ph/9410302}{{\ttfamily arXiv:hep-ph/9410302}}.

\bibitem{Manohar:2008tc}
A.~V. Manohar and V.~Mateu, ``{Dispersion Relation Bounds for pi pi
  Scattering},'' \href{http://dx.doi.org/10.1103/PhysRevD.77.094019}{{\em Phys.
  Rev. D} {\bfseries 77} (2008) 094019},
  \href{http://arxiv.org/abs/0801.3222}{{\ttfamily arXiv:0801.3222 [hep-ph]}}.

\bibitem{Low:2009di}
I.~Low, R.~Rattazzi, and A.~Vichi, ``{Theoretical Constraints on the Higgs
  Effective Couplings},'' \href{http://dx.doi.org/10.1007/JHEP04(2010)126}{{\em
  JHEP} {\bfseries 04} (2010) 126},
  \href{http://arxiv.org/abs/0907.5413}{{\ttfamily arXiv:0907.5413 [hep-ph]}}.

\bibitem{Komargodski:2011vj}
Z.~Komargodski and A.~Schwimmer, ``{On Renormalization Group Flows in Four
  Dimensions},'' \href{http://dx.doi.org/10.1007/JHEP12(2011)099}{{\em JHEP}
  {\bfseries 12} (2011) 099}, \href{http://arxiv.org/abs/1107.3987}{{\ttfamily
  arXiv:1107.3987 [hep-th]}}.

\bibitem{Luty:2012ww}
M.~A. Luty, J.~Polchinski, and R.~Rattazzi, ``{The $a$-theorem and the
  Asymptotics of 4D Quantum Field Theory},''
  \href{http://dx.doi.org/10.1007/JHEP01(2013)152}{{\em JHEP} {\bfseries 01}
  (2013) 152}, \href{http://arxiv.org/abs/1204.5221}{{\ttfamily arXiv:1204.5221
  [hep-th]}}.

\bibitem{Bellazzini:2016xrt}
B.~Bellazzini, ``{Softness and amplitudes\textquoteright{} positivity for
  spinning particles},'' \href{http://dx.doi.org/10.1007/JHEP02(2017)034}{{\em
  JHEP} {\bfseries 02} (2017) 034},
  \href{http://arxiv.org/abs/1605.06111}{{\ttfamily arXiv:1605.06111
  [hep-th]}}.

\bibitem{Cheung:2016yqr}
C.~Cheung and G.~N. Remmen, ``{Positive Signs in Massive Gravity},''
  \href{http://dx.doi.org/10.1007/JHEP04(2016)002}{{\em JHEP} {\bfseries 04}
  (2016) 002}, \href{http://arxiv.org/abs/1601.04068}{{\ttfamily
  arXiv:1601.04068 [hep-th]}}.

\bibitem{Distler:2006if}
J.~Distler, B.~Grinstein, R.~A. Porto, and I.~Z. Rothstein, ``{Falsifying
  Models of New Physics via WW Scattering},''
  \href{http://dx.doi.org/10.1103/PhysRevLett.98.041601}{{\em Phys. Rev. Lett.}
  {\bfseries 98} (2007) 041601},
  \href{http://arxiv.org/abs/hep-ph/0604255}{{\ttfamily arXiv:hep-ph/0604255}}.

\bibitem{Englert:2019zmt}
C.~Englert, G.~F. Giudice, A.~Greljo, and M.~Mccullough, ``{The
  $\hat{H}$-Parameter: An Oblique Higgs View},''
  \href{http://dx.doi.org/10.1007/JHEP09(2019)041}{{\em JHEP} {\bfseries 09}
  (2019) 041}, \href{http://arxiv.org/abs/1903.07725}{{\ttfamily
  arXiv:1903.07725 [hep-ph]}}.

\bibitem{Bellazzini:2017fep}
B.~Bellazzini, F.~Riva, J.~Serra, and F.~Sgarlata, ``{Beyond Positivity Bounds
  and the Fate of Massive Gravity},''
  \href{http://dx.doi.org/10.1103/PhysRevLett.120.161101}{{\em Phys. Rev.
  Lett.} {\bfseries 120} no.~16, (2018) 161101},
  \href{http://arxiv.org/abs/1710.02539}{{\ttfamily arXiv:1710.02539
  [hep-th]}}.

\bibitem{Alberte:2020bdz}
L.~Alberte, C.~de~Rham, S.~Jaitly, and A.~J. Tolley, ``{QED positivity
  bounds},'' \href{http://arxiv.org/abs/2012.05798}{{\ttfamily arXiv:2012.05798
  [hep-th]}}.

\bibitem{Bellazzini:2019bzh}
B.~Bellazzini, F.~Riva, J.~Serra, and F.~Sgarlata, ``{Massive Higher Spins:
  Effective Theory and Consistency},''
  \href{http://dx.doi.org/10.1007/JHEP10(2019)189}{{\em JHEP} {\bfseries 10}
  (2019) 189}, \href{http://arxiv.org/abs/1903.08664}{{\ttfamily
  arXiv:1903.08664 [hep-th]}}.

\bibitem{Gu:2020ldn}
J.~Gu, L.-T. Wang, and C.~Zhang, ``{An unambiguous test of positivity at lepton
  colliders},'' \href{http://arxiv.org/abs/2011.03055}{{\ttfamily
  arXiv:2011.03055 [hep-ph]}}.

\bibitem{deRham:2018qqo}
C.~de~Rham, S.~Melville, A.~J. Tolley, and S.-Y. Zhou, ``{Positivity Bounds for
  Massive Spin-1 and Spin-2 Fields},''
  \href{http://dx.doi.org/10.1007/JHEP03(2019)182}{{\em JHEP} {\bfseries 03}
  (2019) 182}, \href{http://arxiv.org/abs/1804.10624}{{\ttfamily
  arXiv:1804.10624 [hep-th]}}.

\bibitem{Arkani-Hamed:2020blm}
N.~Arkani-Hamed, T.-C. Huang, and Y.-T. Huang, ``{The EFT-Hedron},''
  \href{http://arxiv.org/abs/2012.15849}{{\ttfamily arXiv:2012.15849
  [hep-th]}}.

\bibitem{Green:2019tpt}
M.~B. Green and C.~Wen, ``{Superstring amplitudes, unitarily, and Hankel
  determinants of multiple zeta values},''
  \href{http://dx.doi.org/10.1007/JHEP11(2019)079}{{\em JHEP} {\bfseries 11}
  (2019) 079}, \href{http://arxiv.org/abs/1908.08426}{{\ttfamily
  arXiv:1908.08426 [hep-th]}}.

\bibitem{Bellazzini:2020cot}
B.~Bellazzini, J.~Elias~Mir\'o, R.~Rattazzi, M.~Riembau, and F.~Riva,
  ``{Positive Moments for Scattering Amplitudes},''
  \href{http://arxiv.org/abs/2011.00037}{{\ttfamily arXiv:2011.00037
  [hep-th]}}.

\bibitem{Bellazzini:2021oaj}
B.~Bellazzini, M.~Riembau, and F.~Riva, ``{The IR-Side of Positivity Bounds},''
  \href{http://arxiv.org/abs/2112.12561}{{\ttfamily arXiv:2112.12561
  [hep-th]}}.

\bibitem{Bellazzini:2021shn}
B.~Bellazzini, G.~Isabella, M.~Lewandowski, and F.~Sgarlata, ``{Gravitational
  causality and the self-stress of photons},''
  \href{http://dx.doi.org/10.1007/JHEP05(2022)154}{{\em JHEP} {\bfseries 05}
  (2022) 154}, \href{http://arxiv.org/abs/2108.05896}{{\ttfamily
  arXiv:2108.05896 [hep-th]}}.

\bibitem{Tolley:2020gtv}
A.~J. Tolley, Z.-Y. Wang, and S.-Y. Zhou, ``{New positivity bounds from full
  crossing symmetry},'' \href{http://arxiv.org/abs/2011.02400}{{\ttfamily
  arXiv:2011.02400 [hep-th]}}.

\bibitem{Caron-Huot:2020cmc}
S.~Caron-Huot and V.~Van~Duong, ``{Extremal Effective Field Theories},''
  \href{http://arxiv.org/abs/2011.02957}{{\ttfamily arXiv:2011.02957
  [hep-th]}}.

\bibitem{Komatsu:2020sag}
S.~Komatsu, M.~F. Paulos, B.~C. Van~Rees, and X.~Zhao, ``{Landau diagrams in
  AdS and S-matrices from conformal correlators},''
  \href{http://dx.doi.org/10.1007/JHEP11(2020)046}{{\em JHEP} {\bfseries 11}
  (2020) 046}, \href{http://arxiv.org/abs/2007.13745}{{\ttfamily
  arXiv:2007.13745 [hep-th]}}.

\bibitem{Caron-Huot:2021rmr}
S.~Caron-Huot, D.~Mazac, L.~Rastelli, and D.~Simmons-Duffin, ``{Sharp
  Boundaries for the Swampland},''
  \href{http://arxiv.org/abs/2102.08951}{{\ttfamily arXiv:2102.08951
  [hep-th]}}.

\bibitem{Bern:2021ppb}
Z.~Bern, D.~Kosmopoulos, and A.~Zhiboedov, ``{Gravitational Effective Field
  Theory Islands, Low-Spin Dominance, and the Four-Graviton Amplitude},''
  \href{http://arxiv.org/abs/2103.12728}{{\ttfamily arXiv:2103.12728
  [hep-th]}}.

\bibitem{Alberte:2021dnj}
L.~Alberte, C.~de~Rham, S.~Jaitly, and A.~J. Tolley, ``{Reverse Bootstrapping:
  IR Lessons for UV Physics},''
  \href{http://dx.doi.org/10.1103/PhysRevLett.128.051602}{{\em Phys. Rev.
  Lett.} {\bfseries 128} no.~5, (2022) 051602},
  \href{http://arxiv.org/abs/2111.09226}{{\ttfamily arXiv:2111.09226
  [hep-th]}}.

\bibitem{Chiang:2021ziz}
L.-Y. Chiang, Y.-t. Huang, W.~Li, L.~Rodina, and H.-C. Weng, ``{Into the
  EFThedron and UV constraints from IR consistency},''
  \href{http://arxiv.org/abs/2105.02862}{{\ttfamily arXiv:2105.02862
  [hep-th]}}.

\bibitem{Henriksson:2021ymi}
J.~Henriksson, B.~McPeak, F.~Russo, and A.~Vichi, ``{Rigorous bounds on
  light-by-light scattering},''
  \href{http://dx.doi.org/10.1007/JHEP06(2022)158}{{\em JHEP} {\bfseries 06}
  (2022) 158}, \href{http://arxiv.org/abs/2107.13009}{{\ttfamily
  arXiv:2107.13009 [hep-th]}}.

\bibitem{Sinha:2020win}
A.~Sinha and A.~Zahed, ``{Crossing Symmetric Dispersion Relations in Quantum
  Field Theories},''
  \href{http://dx.doi.org/10.1103/PhysRevLett.126.181601}{{\em Phys. Rev.
  Lett.} {\bfseries 126} no.~18, (2021) 181601},
  \href{http://arxiv.org/abs/2012.04877}{{\ttfamily arXiv:2012.04877
  [hep-th]}}.

\bibitem{Knop:2022viy}
W.~Knop and D.~Mazac, ``{Dispersive Sum Rules in AdS$_2$},''
  \href{http://arxiv.org/abs/2203.11170}{{\ttfamily arXiv:2203.11170
  [hep-th]}}.

\bibitem{Creminelli:2022onn}
P.~Creminelli, O.~Janssen, and L.~Senatore, ``{Positivity bounds on effective
  field theories with spontaneously broken Lorentz invariance},''
  \href{http://arxiv.org/abs/2207.14224}{{\ttfamily arXiv:2207.14224
  [hep-th]}}.

\bibitem{Haring:2022cyf}
K.~H\"aring and A.~Zhiboedov, ``{Gravitational Regge bounds},''
  \href{http://arxiv.org/abs/2202.08280}{{\ttfamily arXiv:2202.08280
  [hep-th]}}.

\bibitem{Li:2022rag}
X.~Li, K.~Mimasu, K.~Yamashita, C.~Yang, C.~Zhang, and S.-Y. Zhou, ``{Moments
  for positivity: using Drell-Yan data to test positivity bounds and
  reverse-engineer new physics},''
  \href{http://dx.doi.org/10.1007/JHEP10(2022)107}{{\em JHEP} {\bfseries 10}
  (2022) 107}, \href{http://arxiv.org/abs/2204.13121}{{\ttfamily
  arXiv:2204.13121 [hep-ph]}}.

\bibitem{Paulos:2016fap}
M.~F. Paulos, J.~Penedones, J.~Toledo, B.~C. van Rees, and P.~Vieira, ``{The
  S-matrix bootstrap. Part I: QFT in AdS},''
  \href{http://dx.doi.org/10.1007/JHEP11(2017)133}{{\em JHEP} {\bfseries 11}
  (2017) 133}, \href{http://arxiv.org/abs/1607.06109}{{\ttfamily
  arXiv:1607.06109 [hep-th]}}.

\bibitem{Paulos:2016but}
M.~F. Paulos, J.~Penedones, J.~Toledo, B.~C. van Rees, and P.~Vieira, ``{The
  S-matrix bootstrap II: two dimensional amplitudes},''
  \href{http://dx.doi.org/10.1007/JHEP11(2017)143}{{\em JHEP} {\bfseries 11}
  (2017) 143}, \href{http://arxiv.org/abs/1607.06110}{{\ttfamily
  arXiv:1607.06110 [hep-th]}}.

\bibitem{Paulos:2017fhb}
M.~F. Paulos, J.~Penedones, J.~Toledo, B.~C. van Rees, and P.~Vieira, ``{The
  S-matrix bootstrap. Part III: higher dimensional amplitudes},''
  \href{http://dx.doi.org/10.1007/JHEP12(2019)040}{{\em JHEP} {\bfseries 12}
  (2019) 040}, \href{http://arxiv.org/abs/1708.06765}{{\ttfamily
  arXiv:1708.06765 [hep-th]}}.

\bibitem{Homrich:2019cbt}
A.~Homrich, J.~a. Penedones, J.~Toledo, B.~C. van Rees, and P.~Vieira, ``{The
  S-matrix Bootstrap IV: Multiple Amplitudes},''
  \href{http://dx.doi.org/10.1007/JHEP11(2019)076}{{\em JHEP} {\bfseries 11}
  (2019) 076}, \href{http://arxiv.org/abs/1905.06905}{{\ttfamily
  arXiv:1905.06905 [hep-th]}}.

\bibitem{EliasMiro:2019kyf}
J.~Elias~Mir\'o, A.~L. Guerrieri, A.~Hebbar, J.~a. Penedones, and P.~Vieira,
  ``{Flux Tube S-matrix Bootstrap},''
  \href{http://dx.doi.org/10.1103/PhysRevLett.123.221602}{{\em Phys. Rev.
  Lett.} {\bfseries 123} no.~22, (2019) 221602},
  \href{http://arxiv.org/abs/1906.08098}{{\ttfamily arXiv:1906.08098
  [hep-th]}}.

\bibitem{EliasMiro:2021nul}
J.~Elias~Mir\'o and A.~Guerrieri, ``{Dual EFT bootstrap: QCD flux tubes},''
  \href{http://dx.doi.org/10.1007/JHEP10(2021)126}{{\em JHEP} {\bfseries 10}
  (2021) 126}, \href{http://arxiv.org/abs/2106.07957}{{\ttfamily
  arXiv:2106.07957 [hep-th]}}.

\bibitem{Guerrieri:2018uew}
A.~L. Guerrieri, J.~Penedones, and P.~Vieira, ``{Bootstrapping QCD Using Pion
  Scattering Amplitudes},''
  \href{http://dx.doi.org/10.1103/PhysRevLett.122.241604}{{\em Phys. Rev.
  Lett.} {\bfseries 122} no.~24, (2019) 241604},
  \href{http://arxiv.org/abs/1810.12849}{{\ttfamily arXiv:1810.12849
  [hep-th]}}.

\bibitem{Guerrieri:2020bto}
A.~L. Guerrieri, J.~Penedones, and P.~Vieira, ``{S-matrix bootstrap for
  effective field theories: massless pions},''
  \href{http://dx.doi.org/10.1007/JHEP06(2021)088}{{\em JHEP} {\bfseries 06}
  (2021) 088}, \href{http://arxiv.org/abs/2011.02802}{{\ttfamily
  arXiv:2011.02802 [hep-th]}}.

\bibitem{Hebbar:2020ukp}
A.~Hebbar, D.~Karateev, and J.~Penedones, ``{Spinning S-matrix bootstrap in
  4d},'' \href{http://dx.doi.org/10.1007/JHEP01(2022)060}{{\em JHEP} {\bfseries
  01} (2022) 060}, \href{http://arxiv.org/abs/2011.11708}{{\ttfamily
  arXiv:2011.11708 [hep-th]}}.

\bibitem{Guerrieri:2021ivu}
A.~Guerrieri, J.~Penedones, and P.~Vieira, ``{Where is String Theory?},''
  \href{http://arxiv.org/abs/2102.02847}{{\ttfamily arXiv:2102.02847
  [hep-th]}}.

\bibitem{WhereIsMTheory}
A.~Guerrieri, H.~Murali, J.~Penedones, and P.~Vieira. To appear soon.

\bibitem{Correia:2020xtr}
M.~Correia, A.~Sever, and A.~Zhiboedov, ``{An Analytical Toolkit for the
  S-matrix Bootstrap},'' \href{http://arxiv.org/abs/2006.08221}{{\ttfamily
  arXiv:2006.08221 [hep-th]}}.

\bibitem{Correia:2021etg}
M.~Correia, A.~Sever, and A.~Zhiboedov, ``{Probing multi-particle unitarity
  with the Landau equations},''
  \href{http://arxiv.org/abs/2111.12100}{{\ttfamily arXiv:2111.12100
  [hep-th]}}.

\bibitem{Doroud:2018szp}
N.~Doroud and J.~Elias~Mir\'o, ``{S-matrix bootstrap for resonances},''
  \href{http://dx.doi.org/10.1007/JHEP09(2018)052}{{\em JHEP} {\bfseries 09}
  (2018) 052}, \href{http://arxiv.org/abs/1804.04376}{{\ttfamily
  arXiv:1804.04376 [hep-th]}}.

\bibitem{Paulos:2018fym}
M.~F. Paulos and Z.~Zheng, ``{Bounding scattering of charged particles in $1+1$
  dimensions},'' \href{http://dx.doi.org/10.1007/JHEP05(2020)145}{{\em JHEP}
  {\bfseries 05} (2020) 145}, \href{http://arxiv.org/abs/1805.11429}{{\ttfamily
  arXiv:1805.11429 [hep-th]}}.

\bibitem{He:2018uxa}
Y.~He, A.~Irrgang, and M.~Kruczenski, ``{A note on the S-matrix bootstrap for
  the 2d O(N) bosonic model},''
  \href{http://dx.doi.org/10.1007/JHEP11(2018)093}{{\em JHEP} {\bfseries 11}
  (2018) 093}, \href{http://arxiv.org/abs/1805.02812}{{\ttfamily
  arXiv:1805.02812 [hep-th]}}.

\bibitem{Cordova:2018uop}
L.~C\'ordova and P.~Vieira, ``{Adding flavour to the S-matrix bootstrap},''
  \href{http://dx.doi.org/10.1007/JHEP12(2018)063}{{\em JHEP} {\bfseries 12}
  (2018) 063}, \href{http://arxiv.org/abs/1805.11143}{{\ttfamily
  arXiv:1805.11143 [hep-th]}}.

\bibitem{Karateev:2019ymz}
D.~Karateev, S.~Kuhn, and J.~a. Penedones, ``{Bootstrapping Massive Quantum
  Field Theories},'' \href{http://dx.doi.org/10.1007/JHEP07(2020)035}{{\em
  JHEP} {\bfseries 07} (2020) 035},
  \href{http://arxiv.org/abs/1912.08940}{{\ttfamily arXiv:1912.08940
  [hep-th]}}.

\bibitem{Chen:2021pgx}
H.~Chen, A.~L. Fitzpatrick, and D.~Karateev, ``{Bootstrapping 2d
  \ensuremath{\phi}$^{4}$ theory with Hamiltonian truncation data},''
  \href{http://dx.doi.org/10.1007/JHEP02(2022)146}{{\em JHEP} {\bfseries 02}
  (2022) 146}, \href{http://arxiv.org/abs/2107.10286}{{\ttfamily
  arXiv:2107.10286 [hep-th]}}.

\bibitem{Bose:2020shm}
A.~Bose, P.~Haldar, A.~Sinha, P.~Sinha, and S.~S. Tiwari, ``{Relative entropy
  in scattering and the S-matrix bootstrap},''
  \href{http://dx.doi.org/10.21468/SciPostPhys.9.5.081}{{\em SciPost Phys.}
  {\bfseries 9} (2020) 081}, \href{http://arxiv.org/abs/2006.12213}{{\ttfamily
  arXiv:2006.12213 [hep-th]}}.

\bibitem{Bose:2020cod}
A.~Bose, A.~Sinha, and S.~S. Tiwari, ``{Selection rules for the S-Matrix
  bootstrap},'' \href{http://arxiv.org/abs/2011.07944}{{\ttfamily
  arXiv:2011.07944 [hep-th]}}.

\bibitem{Karateev:2020axc}
D.~Karateev, ``{Two-point functions and bootstrap applications in quantum field
  theories},'' \href{http://dx.doi.org/10.1007/JHEP02(2022)186}{{\em JHEP}
  {\bfseries 02} (2022) 186}, \href{http://arxiv.org/abs/2012.08538}{{\ttfamily
  arXiv:2012.08538 [hep-th]}}.

\bibitem{Karateev:2022jdb}
D.~Karateev, J.~Marucha, J.~a. Penedones, and B.~Sahoo, ``{Bootstrapping the
  $a$-anomaly in $4d$ QFTs},''
  \href{http://arxiv.org/abs/2204.01786}{{\ttfamily arXiv:2204.01786
  [hep-th]}}.

\bibitem{Chen:2022nym}
H.~Chen, A.~L. Fitzpatrick, and D.~Karateev, ``{Nonperturbative Bounds on
  Scattering of Massive Scalar Particles in $d \geq 2$},''
  \href{http://arxiv.org/abs/2207.12448}{{\ttfamily arXiv:2207.12448
  [hep-th]}}.

\bibitem{Gabai:2019ryw}
B.~Gabai and X.~Yin, ``{On The S-Matrix of Ising Field Theory in Two
  Dimensions},'' \href{http://arxiv.org/abs/1905.00710}{{\ttfamily
  arXiv:1905.00710 [hep-th]}}.

\bibitem{Tourkine:2021fqh}
P.~Tourkine and A.~Zhiboedov, ``{Scattering from production in 2d},''
  \href{http://dx.doi.org/10.1007/JHEP07(2021)228}{{\em JHEP} {\bfseries 07}
  (2021) 228}, \href{http://arxiv.org/abs/2101.05211}{{\ttfamily
  arXiv:2101.05211 [hep-th]}}.

\bibitem{Cordova:2019lot}
L.~C\'ordova, Y.~He, M.~Kruczenski, and P.~Vieira, ``{The O(N) S-matrix
  Monolith},'' \href{http://dx.doi.org/10.1007/JHEP04(2020)142}{{\em JHEP}
  {\bfseries 04} (2020) 142}, \href{http://arxiv.org/abs/1909.06495}{{\ttfamily
  arXiv:1909.06495 [hep-th]}}.

\bibitem{Guerrieri:2020kcs}
A.~L. Guerrieri, A.~Homrich, and P.~Vieira, ``{Dual S-matrix bootstrap. Part I.
  2D theory},'' \href{http://dx.doi.org/10.1007/JHEP11(2020)084}{{\em JHEP}
  {\bfseries 11} (2020) 084}, \href{http://arxiv.org/abs/2008.02770}{{\ttfamily
  arXiv:2008.02770 [hep-th]}}.

\bibitem{Kruczenski:2020ujw}
M.~Kruczenski and H.~Murali, ``{The R-matrix bootstrap for the 2d O(N) bosonic
  model with a boundary},''
  \href{http://dx.doi.org/10.1007/JHEP04(2021)097}{{\em JHEP} {\bfseries 04}
  (2021) 097}, \href{http://arxiv.org/abs/2012.15576}{{\ttfamily
  arXiv:2012.15576 [hep-th]}}.

\bibitem{He:2021eqn}
Y.~He and M.~Kruczenski, ``{S-matrix bootstrap in 3+1 dimensions:
  regularization and dual convex problem},''
  \href{http://arxiv.org/abs/2103.11484}{{\ttfamily arXiv:2103.11484
  [hep-th]}}.

\bibitem{Guerrieri:2021tak}
A.~Guerrieri and A.~Sever, ``{Rigorous Bounds on the Analytic S Matrix},''
  \href{http://dx.doi.org/10.1103/PhysRevLett.127.251601}{{\em Phys. Rev.
  Lett.} {\bfseries 127} no.~25, (2021) 251601},
  \href{http://arxiv.org/abs/2106.10257}{{\ttfamily arXiv:2106.10257
  [hep-th]}}.

\bibitem{Martin:1965jj}
A.~Martin, ``{Extension of the axiomatic analyticity domain of scattering
  amplitudes by unitarity. 1.},''
  \href{http://dx.doi.org/10.1007/BF02720568}{{\em Nuovo Cim. A} {\bfseries 42}
  (1965) 930--953}.

\bibitem{Lopez:1974cq}
C.~Lopez, ``{A Lower Bound to the pi0 pi0 S-Wave Scattering Length},''
  \href{http://dx.doi.org/10.1016/0550-3213(75)90287-4}{{\em Nucl. Phys. B}
  {\bfseries 88} (1975) 358--364}.

\bibitem{Lopez:1975wf}
C.~Lopez, ``{Rigorous Lower Bounds for the pi pi p-Wave Scattering Length},''
  \href{http://dx.doi.org/10.1007/BF02753880}{{\em Lett. Nuovo Cim.} {\bfseries
  13} (1975) 69}.

\bibitem{Lopez:1975ca}
C.~Lopez and G.~Mennessier, ``{A New Absolute Bound on the pi0 pi0 S-Wave
  Scattering Length},''
  \href{http://dx.doi.org/10.1016/0370-2693(75)90583-3}{{\em Phys. Lett. B}
  {\bfseries 58} (1975) 437--441}.

\bibitem{Bonnier:1975jz}
B.~Bonnier, C.~Lopez, and G.~Mennessier, ``{Improved Absolute Bounds on the pi0
  pi0 Amplitude},'' \href{http://dx.doi.org/10.1016/0370-2693(75)90528-6}{{\em
  Phys. Lett. B} {\bfseries 60} (1975) 63--66}.

\bibitem{Lopez:1976zs}
C.~Lopez and G.~Mennessier, ``{Bounds on the pi0 pi0 Amplitude},''
  \href{http://dx.doi.org/10.1016/0550-3213(77)90237-1}{{\em Nucl. Phys. B}
  {\bfseries 118} (1977) 426--444}.

\bibitem{wavelet}
 See the \href{https://en.wikipedia.org/wiki/Wavelet}{Wikipedia page}.

\bibitem{Froissart:1961ux}
M.~Froissart, ``{Asymptotic behavior and subtractions in the Mandelstam
  representation},'' \href{http://dx.doi.org/10.1103/PhysRev.123.1053}{{\em
  Phys. Rev.} {\bfseries 123} (1961) 1053--1057}.

\bibitem{Roy:1971tc}
S.~M. Roy, ``{Exact integral equation for pion pion scattering involving only
  physical region partial waves},''
  \href{http://dx.doi.org/10.1016/0370-2693(71)90724-6}{{\em Phys. Lett. B}
  {\bfseries 36} (1971) 353--356}.

\bibitem{Rattazzi:2008pe}
R.~Rattazzi, V.~S. Rychkov, E.~Tonni, and A.~Vichi, ``{Bounding scalar operator
  dimensions in 4D CFT},''
  \href{http://dx.doi.org/10.1088/1126-6708/2008/12/031}{{\em JHEP} {\bfseries
  12} (2008) 031}, \href{http://arxiv.org/abs/0807.0004}{{\ttfamily
  arXiv:0807.0004 [hep-th]}}.

\bibitem{Poland:2018epd}
D.~Poland, S.~Rychkov, and A.~Vichi, ``{The Conformal Bootstrap: Theory,
  Numerical Techniques, and Applications},''
  \href{http://dx.doi.org/10.1103/RevModPhys.91.015002}{{\em Rev. Mod. Phys.}
  {\bfseries 91} (2019) 015002},
  \href{http://arxiv.org/abs/1805.04405}{{\ttfamily arXiv:1805.04405
  [hep-th]}}.

\bibitem{Haldar:2021rri}
P.~Haldar, A.~Sinha, and A.~Zahed, ``{Quantum field theory and the Bieberbach
  conjecture},'' \href{http://dx.doi.org/10.21468/SciPostPhys.11.1.002}{{\em
  SciPost Phys.} {\bfseries 11} (2021) 002},
  \href{http://arxiv.org/abs/2103.12108}{{\ttfamily arXiv:2103.12108
  [hep-th]}}.

\bibitem{Zahed:2021fkp}
A.~Zahed, ``{Positivity and geometric function theory constraints on pion
  scattering},'' \href{http://dx.doi.org/10.1007/JHEP12(2021)036}{{\em JHEP}
  {\bfseries 12} (2021) 036}, \href{http://arxiv.org/abs/2108.10355}{{\ttfamily
  arXiv:2108.10355 [hep-th]}}.

\bibitem{upcommingsoon}
J.~Elias-Miro, A.~Guerrieri, and M.~Gumus, ``{to appear soon},''.

\bibitem{Elias-Miro:2013mua}
J.~Elias-Miro, J.~R. Espinosa, E.~Masso, and A.~Pomarol, ``{Higgs windows to
  new physics through d=6 operators: constraints and one-loop anomalous
  dimensions},'' \href{http://dx.doi.org/10.1007/JHEP11(2013)066}{{\em JHEP}
  {\bfseries 11} (2013) 066}, \href{http://arxiv.org/abs/1308.1879}{{\ttfamily
  arXiv:1308.1879 [hep-ph]}}.

\bibitem{DeBlas:2019qco}
J.~De~Blas, G.~Durieux, C.~Grojean, J.~Gu, and A.~Paul, ``{On the future of
  Higgs, electroweak and diboson measurements at lepton colliders},''
  \href{http://dx.doi.org/10.1007/JHEP12(2019)117}{{\em JHEP} {\bfseries 12}
  (2019) 117}, \href{http://arxiv.org/abs/1907.04311}{{\ttfamily
  arXiv:1907.04311 [hep-ph]}}.

\bibitem{Ethier:2021bye}
{\bfseries SMEFiT} Collaboration, J.~J. Ethier, G.~Magni, F.~Maltoni,
  L.~Mantani, E.~R. Nocera, J.~Rojo, E.~Slade, E.~Vryonidou, and C.~Zhang,
  ``{Combined SMEFT interpretation of Higgs, diboson, and top quark data from
  the LHC},'' \href{http://dx.doi.org/10.1007/JHEP11(2021)089}{{\em JHEP}
  {\bfseries 11} (2021) 089}, \href{http://arxiv.org/abs/2105.00006}{{\ttfamily
  arXiv:2105.00006 [hep-ph]}}.

\bibitem{Creutz:1972ikj}
M.~Creutz, ``{Rigorous bounds on coupling constants in two-dimensional field
  theories},'' \href{http://dx.doi.org/10.1103/PhysRevD.6.2763}{{\em Phys. Rev.
  D} {\bfseries 6} (1972) 2763--2765}.

\bibitem{Dubovsky:2012wk}
S.~Dubovsky, R.~Flauger, and V.~Gorbenko, ``{Solving the Simplest Theory of
  Quantum Gravity},'' \href{http://dx.doi.org/10.1007/JHEP09(2012)133}{{\em
  JHEP} {\bfseries 09} (2012) 133},
  \href{http://arxiv.org/abs/1205.6805}{{\ttfamily arXiv:1205.6805 [hep-th]}}.

\bibitem{Dubovsky:2013ira}
S.~Dubovsky, V.~Gorbenko, and M.~Mirbabayi, ``{Natural Tuning: Towards A Proof
  of Concept},'' \href{http://dx.doi.org/10.1007/JHEP09(2013)045}{{\em JHEP}
  {\bfseries 09} (2013) 045}, \href{http://arxiv.org/abs/1305.6939}{{\ttfamily
  arXiv:1305.6939 [hep-th]}}.

\bibitem{Cavaglia:2016oda}
A.~Cavagli\`a, S.~Negro, I.~M. Sz\'ecs\'enyi, and R.~Tateo, ``{$T
  \bar{T}$-deformed 2D Quantum Field Theories},''
  \href{http://dx.doi.org/10.1007/JHEP10(2016)112}{{\em JHEP} {\bfseries 10}
  (2016) 112}, \href{http://arxiv.org/abs/1608.05534}{{\ttfamily
  arXiv:1608.05534 [hep-th]}}.

\bibitem{Zamolodchikov:1992ulx}
A.~B. Zamolodchikov, ``{Resonance factorized scattering and roaming
  trajectories},'' \href{http://dx.doi.org/10.1088/0305-4470/39/41/S08}{{\em J.
  Phys. A} {\bfseries 39} (2006) 12847--12862}.

\end{thebibliography}\endgroup
\bibliographystyle{utphys}

\end{document}